\documentstyle{mn}

\title[SCUBA mapping of prestellar cores]{The
initial conditions of isolated star formation - VI.
SCUBA mapping of prestellar cores.}

\author[Kirk, Ward-Thompson and Andr\'e]{J. M. Kirk$^{1,2}$,
D. Ward-Thompson$^1$ and P. Andr\'e$^3$ \\
$^1$ Department of Physics and Astronomy, Cardiff University,
5 The Parade, Cardiff, CF24 3YB \\
$^2$ Department of Astronomy, 207 Astronomy Building,
1002 West Green St, Urbana, IL 61801, USA \\
$^3$ CEA, DSM, DAPNIA, Service d'Astrophysique, C.E. Saclay,
F-91191 Gif-sur-Yvette Cedex, France}

\date{Accepted 2005 March 19; received 2005 March 18; in original form
2004 July 8.}

\pagerange{000--000}

\begin{document}

\label{firstpage}

\maketitle

\begin{abstract}
Observations have been carried out
with the submillimetre common-user bolometer array (SCUBA)
at the James Clerk Maxwell Telescope (JCMT)
of regions of comparatively isolated star
formation in molecular cloud cores.
52 starless cores were observed, which are molecular cloud cores that
do not contain any sign of protostellar activity such as infrared sources
or bipolar outflows. These are all therefore candidate prestellar cores,
which are believed to represent the stage of star formation that precedes
the formation of a protostar. 29 of the 52 cores were detected at 850~$\mu$m
at varying levels of signal-to-noise ratio greater than 3$\sigma$ at peak,
while 23 of the cores were observed but not detected. The mean detection
lower limit of the data corresponds roughly  to
an A$_V$$\sim$15 under typical assumptions. The detected cores were
split into `bright' cores and `intermediate' cores, depending on their
peak flux density at 850~$\mu$m.
Those with peak 850-$\mu$m flux densities greater than 170~mJy/beam were
designated `bright' cores. Those with peak 850-$\mu$m
flux densities less than this value were
designated `intermediate' cores. This dividing line corresponds to
a mean detection limit of 10$\sigma$ at peak, and an approximate
A$_V$$\sim$50 under typical assumptions.
13 of the 29 detected cores are found to be bright and 16
are intermediate. The data are combined with our previously published ISO
data, and the physical parameters of the cores, such as density and
temperature, are calculated. The bright cores
are detected with sufficiently high signal-to-noise ratio
to allow their structure to be mapped. Radial flux density profiles of
these show flattened inner regions and sharp boundaries, consistent with
previous observations of prestellar cores.
Detailed fitting of the bright core radial profiles shows that they are not
critical Bonnor-Ebert spheres, in agreement with previous
findings. However, we find that intermediate cores, such as B68
(which has previously been claimed to be a Bonnor-Ebert sphere),
may in fact be consistent with the Bonnor-Ebert criterion, suggesting
perhaps that cores pass through such a phase during their evolution.
We also find that the masses of the bright cores have a mean value of
approximately the same order as their virial masses.
We make rough estimates of core lifetimes
based on the statistics of detections and find that
the lifetime of a prestellar core is roughly
$\sim$3 $\times$ 10$^5$ years, while that of a bright core
is $\sim$1.5 $\times$ 10$^5$ years.
Comparisons with some
models that regulate collapse using either
magnetic fields or turbulence show that no model can match
all of the data. Models that are tuned to fit the
total prestellar core lifetime, do not
predict the relative numbers of cores seen at each stage.
\end{abstract}
\begin{keywords}
stars: formation -- ISM: dust -- infrared: ISM -- submillimetre: ISM
\end{keywords}

\section{Introduction}

Star formation occurs in dense cores within molecular clouds (e.g. Williams,
Blitz \& McKee 2000). Study of such regions was hampered for many
years by their very large optical depths at near-infrared and optical
wavelengths. It is only since the opening up of the far-infrared to
submillimetre regime that astronomers have been able to study molecular
clouds in detail. In this series of papers we report on a programme of
molecular cloud core studies at infrared and submillimetre wavelengths.

Theories of star formation have until now lacked a detailed observational
determination of the initial conditions of the protostellar collapse phase
(e.g. see: Andr\'e, Ward-Thompson \& Barsony 2000 for a review).
The pre-protostellar (or prestellar for short) core phase (Ward-Thompson
et al. 1994 -- hereafter Paper I)
is believed to be the stage of star formation that precedes
the formation of a protostar and hence should represent observationally the
initial conditions of protostellar collapse.

Some recent observations
have even indicated that the Initial Mass Function (IMF)
of stars may be determined at the prestellar core stage
(Motte, Andr\'e \& Neri 1998; Motte et al. 2001).
Furthermore, theory predicts that the core geometry prior to this
stage is critical in determining the manner of collapse
(e.g. Whitworth \& Summers 1985; Foster \& Chevalier 1993;
Whitworth et al. 1996).
For example, a decreasing accretion rate is obtained when the
initial radial density profile
is relatively flat in the centre, and steepens towards the edge
(Foster \& Chevalier 1993; Henriksen, Andr\'e \& Bontemps 1997;
Whitworth \&
Ward-Thompson 2001). Therefore it is of vital importance to
star formation theories to determine
observationally the physical parameters of prestellar cores.

Molecular line surveys of dense cores by Myers and co-workers
identified a significant number of
isolated cores (Myers \& Benson 1983; Myers et al. 1988;
Benson \& Myers 1989). Comparisons of these surveys
with the IRAS point source catalogue
detected a class of core that had no associated infrared
source. The lack of an embedded source led to the
classification of these as `starless' cores
(Beichman et al. 1986).
In this series of papers we have been studying starless cores in some detail.

In Paper I (Ward-Thompson et al. 1994) we found
that the cores all appeared to follow a form of density profile
that was relatively flat in the centre and steeper towards the edge. This is
similar to the profiles
predicted by theory to produce a decreasing accretion rate with
time (Whitworth \& Summers 1985; Foster \& Chevalier 1993).
In Papers II \& III (Andr\'e,
Ward-Thompson \& Motte 1996; Ward-Thompson,
Motte \& Andr\'e 1999) we found that although the radial
density profiles of the cores appear similar to those predicted by
magnetically-regulated models of star formation
(e.g. Mouschovias 1991),
the details of the time-scales required by the models at
different stages did not appear to match exactly
the life-times calculated from
the numbers detected at each evolutionary stage.
The flat inner radial density profiles were also seen in absorption
at 7 \& 15$\mu$m using ISOCAM (Bacmann et al., 2000).

In Paper IV (Jessop \& Ward-Thompson 2001) we
showed that a temperature gradient from
the outside-in was probably required to explain the data. This is
significant because a decrease in temperature could potentially produce
a decrease in submillimetre flux, and hence possibly explain the
apparent flattening observed towards the centre of prestellar cores.

However, in
Paper V (Ward-Thompson, Andr\'e \& Kirk 2002) we showed that
although the
edges of prestellar cores appear warmer than their centres. This is
consistent with external heating of the cores.
The central regions of cores are nevertheless consistent with being
isothermal (at the resolution of ISOPHOT). Hence the flattening
previously observed in the centres of prestellar cores' submillimetre
radial flux density profiles
translates directly into a flattening of their volume density profiles.
A recent claim that this is not the case (Zucconi, Walmsley \& Galli 2001)
does not take account of the fact mentioned above
that we also see the same form of radial
density profile in absorption (Bacmann et al 2000) towards prestellar
cores as we see in emission.
This cannot be explained by a temperature gradient.
A larger ISOPHOT survey of one of our sample, L183, was carried out by
Lehtinen et al., (2003), who found results consistent with our survey.

The significance of the form of the radial profile is that it determines the
subsequent nature of the protostellar collapse. The form of profile
that produces a decreasing accretion rate with time appears consistent
with the observed relative numbers of Class 0 and Class I protostars
(Andr\'e, Ward-Thompson \& Barsony 1993; Andr\'e 1994; Ward-Thompson 1996)
and is also consistent with a decreasing accretion rate extending
throughout the pre-main sequence phase (e.g. Kenyon \& Hartmann 1995;
Safier, McKee \& Stahler 1997).

Recent radiative transfer modelling (Stamatellos \& Whitworth 2003;
Stamatellos et al., 2004)
indicates that submillimetre wavelengths are the best to use for
studying density variations in cold molecular cloud cores.
In this paper we present the results of a SCUBA survey of prestellar
cores at wavelengths of 450 and 850~$\mu$m to study further the detailed
morphologies of prestellar cores.

\section{The Sample}

Our sample of cores is mainly derived from the
molecular line surveys of dense cores by Myers and co-workers
(Myers \& Benson 1983; Myers et al. 1988;
Benson \& Myers 1989),
by choosing only those cores
that have no associated infrared
sources. These cores are known generally
as `starless' cores (Beichman et al. 1986).
As such, our sample is an approximate
subset of the roughly 200 cores catalogued by Jijina et al. (1999).
We have also added a couple of cores that we ourselves found to
be strong submillimetre sources (L1689SMM and L1521SMM). In addition, there
are two cores with nearby star formation that may have affected their
`starless' status -- L43 and L1524.

The sample contains a total of 52 cores, which are listed in Tables 1--3.
The majority of the cores lie in or near the
constellations of Taurus-Auriga (22 cores) and Ophiuchus (21 cores). Of the
remaining cores, five lie in Aquila and four lie in Cepheus.
Some cores form part of more extended molecular cloud
structures, while others are more isolated Bok globules
(e.g. Bok \& Reilly 1947).

Figures 1 \& 2 show the distribution of the cores in the Taurus-Auriga
and Ophiuchus regions respectively.
It can be seen that the cores do not cluster in the centres
of dense molecular clouds.
For example, in Figure 1 the cores can be seen to be scattered
apparently randomly across the Taurus molecular cloud.
Likewise, in Figure 2 it can be seen that some of the cores
lie near to the
cluster-forming $\rho$ Oph main cloud,
but none of them lie directly within it.
All lie either on the fringes of the cloud or along the filamentary
streamers to the north.
We note that not all of our sample of cores were detected in NH$_3$,
and so not all may in fact be dense cores -- some may be simply
sight-lines with high column density. Tables 1--3 indicate whether or
not a core was detected in NH$_3$ (Benson \& Myers 1989).

\begin{figure*}
\setlength{\unitlength}{1mm}
\noindent
\begin{picture}(170,220)
\put(0,10){\includegraphics{fig1.ps}}
\end{picture}
\caption{Image of the IRAS 100$\mu$m emission from the region
of Taurus-Auriga (Wheelock et al. 1994).
Superposed are contours of CO (J=1$\rightarrow$0) intensity
(Dame, Hartman \& Thaddeus 2001).
The positions of the cores in our sample
are marked.}
\end{figure*}

\begin{figure*}
\setlength{\unitlength}{1mm}
\noindent
\begin{picture}(170,220)
\put(0,10){\includegraphics{fig2.ps}}
\end{picture}
\caption{Image of the IRAS 100$\mu$m emission from the region
of Ophiuchus (Wheelock et al. 1994).
Superposed are contours of CO (J=1$\rightarrow$0) intensity
(Dame et al., 2001).
The positions of the cores in our sample
are marked.}
\end{figure*}

\begin{table*}
\caption{Source names, peak positions and 850- \& 450-$\mu$m flux densities
of the 13 `bright' cores detected with an
850-$\mu$m peak flux density $\geq$170mJy/beam at peak.
The peak flux densities $S_{850}^{peak}$ \& $S_{450}^{peak}$
are measured using a beam size FWHM of 14.8 arcsec
and are quoted to 3 sig. figs. in mJy/beam.
The integrated flux densities $S_{850}^{int}$ \& $S_{450}^{int}$
are measured in a 150-arcsec diameter
circular aperture and are quoted to 3 sig. figs. in mJy.
The given errors are the statistical measurement errors
and are quoted to 2 sig. figs. Absolute
calibration errors are $\pm$10\% at 850~$\mu$m and $\pm$25\% at 450~$\mu$m.
The source marked with a $*$ -- L1524 --
appears to have not been fully mapped and
the peak may have been missed, so both the peak flux densities and integrated
flux densities may have been underestimated for this source.
L1696A is also sometimes referred to as Oph D (e.g. Motte et al., 1998).
L1521F has also been referred to as Oni-MC27 (Onishi et al., 1999).
The final column indicates whether or not a NH$_3$ detection is
associated with the source (Benson \& Myers 1989).
A dash indicates the position was not searched for NH$_3$.}
\begin{tabular}{lccccccc}
\hline
Source   &  RA    &  Dec. &  $S_{850}^{peak}$  &  $S_{850}^{int}$
  &  $S_{450}^{peak}$  & $S_{450}^{int}$  & NH$_3$  \\
  &  (2000)     &   (2000)    &    (mJy/beam)  &    (mJy)
  &  (mJy/beam)          &     (mJy)        &   \\ \hline
L1521D   &  04$^{\rm h}$27$^{\rm m}$47.0$^{\rm s}$
  &  $+$26$^\circ$17$^\prime$38$^{\prime\prime}$
  &  206 $\pm$ 14   &  3130 $\pm$ 120
  &  1360 $\pm$ 170  &  4800  $\pm$ 1400 & Y \\
L1521F   &  04$^{\rm h}$28$^{\rm m}$39.2$^{\rm s}$
  &  $+$26$^\circ$51$^\prime$36$^{\prime\prime}$
  &  406 $\pm$ 26    &  3230 $\pm$ 230
  &  3170 $\pm$ 350  &  12400 $\pm$ 2400 & Y \\
L1524$^{*}$ &  04$^{\rm h}$29$^{\rm m}$21.1$^{\rm s}$
  &  $+$24$^\circ$33$^\prime$38$^{\prime\prime}$
  &  227 $\pm$ 13    &  2480 $\pm$ 110
  &  3480 $\pm$ 900  &  26800 $\pm$ 7800 & Y \\
L1517B &  04$^{\rm h}$55$^{\rm m}$17.9$^{\rm s}$
  &    $+$30$^\circ$37$^\prime$47$^{\prime\prime}$
  &    170 $\pm$ 12  &  2630 $\pm$ 100
  &  920 $\pm$ 120 & 12100 $\pm$ 900 & Y \\
L1544    &  05$^{\rm h}$04$^{\rm m}$16.9$^{\rm s}$
  &  $+$25$^\circ$10$^\prime$48$^{\prime\prime}$
  &  319 $\pm$ 26     &  3850 $\pm$ 220
  &  3650 $\pm$ 430  &  22900 $\pm$ 3600 & Y  \\
L1582A &  05$^{\rm h}$32$^{\rm m}$01.0$^{\rm s}$
  &    $+$12$^\circ$30$^\prime$23$^{\prime\prime}$
  &    170 $\pm$ 15  &  1960 $\pm$ 130
  &  1960 $\pm$ 240 & 18200 $\pm$ 2000 & Y \\
L183     &  15$^{\rm h}$54$^{\rm m}$08.8$^{\rm s}$
  &  $-$02$^\circ$52$^\prime$38$^{\prime\prime}$
  &  333 $\pm$ 24   &  4560 $\pm$ 190
  &  1370 $\pm$ 130 &  12800  $\pm$ 1000  & Y \\
L1696A &  16$^{\rm h}$28$^{\rm m}$28.9$^{\rm s}$
  &  $-$24$^\circ$19$^\prime$09$^{\prime\prime}$
  &  271 $\pm$ 19   &  6250 $\pm$ 160
  &  2120 $\pm$ 310   & 17500 $\pm$ 2600  & Y    \\
L1689SMM  &  16$^{\rm h}$31$^{\rm m}$57.0$^{\rm s}$
  &  $-$24$^\circ$57$^\prime$17$^{\prime\prime}$
  &  414 $\pm$ 40    &  5150 $\pm$ 340
  &  3880 $\pm$ 360  &  33900 $\pm$ 3100  & --  \\
L43      &  16$^{\rm h}$34$^{\rm m}$35.7$^{\rm s}$
  &  $-$15$^\circ$47$^\prime$05$^{\prime\prime}$
  &  428 $\pm$ 27   &  5400 $\pm$ 230
  &  3330 $\pm$ 250 &  28600 $\pm$ 1800  & Y \\
L1689B   &  16$^{\rm h}$34$^{\rm m}$48.2$^{\rm s}$
  &  $-$24$^\circ$38$^\prime$04$^{\prime\prime}$
  &  231 $\pm$ 20   &  3050 $\pm$ 170
  &  2230 $\pm$ 390 &  17900 $\pm$ 3200  & Y \\
L63      &  16$^{\rm h}$50$^{\rm m}$14.2$^{\rm s}$
  &  $-$18$^\circ$06$^\prime$17$^{\prime\prime}$
  &  239 $\pm$ 17  &  3570 $\pm$ 140
  &  1460 $\pm$ 220 &  15500 $\pm$ 1800 & Y \\
B133     &  19$^{\rm h}$06$^{\rm m}$08.4$^{\rm s}$
  &  $-$06$^\circ$52$^\prime$52$^{\prime\prime}$
  &  174 $\pm$ 15   &  1720 $\pm$ 120
  &  1970 $\pm$ 630 &  15900 $\pm$ 2900 & Y \\
\hline
\end{tabular}
\end{table*}

For the Taurus-Auriga cores, we adopt a distance of 140$\pm$20~pc (e.g.
Elias 1978; Ungerechts \& Thaddeus 1987; Kenyon, Dobryzcka \& Hartmann 1994).
There is some evidence that the southern cores may be slightly closer than
this mean value at 125~pc (Bertout, Robichon \& Arenou 1999), but we will
not distinguish between them for the purposes of this paper, as the
difference lies within 1~$\sigma$ of the mean value. There are two exceptions
to this: The L1551 cloud is somewhat further away, and we use the recently
found value of 168$\pm$20~pc for this cloud (Bertout et al. 1999); and the
L1582A core, which is visible at the southern edge of Figure 1, is actually
part of the $\lambda$ Orionis Bubble at a distance of 400$\pm$40~pc
(Murdin \& Penston 1977).

For the Ophiuchus cores we adopt a distance of 130$\pm$20~pc, in the light
of Hipparcos measurements of the star $\rho$ Ophiuchi, which was found to
be at a distance of 128$\pm$10~pc (Bertout et al. 1999). There is some
evidence that L1719 and L1721 (see Figure 2)
may be slightly more distant at 150~pc and
associated with the star $\chi$ Ophiuchi (de Geuss, Bronfman \& Thaddeus
1990). Additionally, L63 may be slightly closer, at a distance of 125~pc
(Snell 1981).
However, all of these lie within 1$\sigma$ of our adopted
distance of 130~pc, so we treat the whole Ophiuchus
sample to be at the same distance.

B68 lies between the main Ophiuchus complex and the Galactic Plane, and
was recently claimed to be at a distance of only 80~pc (Hotzel et al. 2002).
However, the evidence for this was based upon assumptions about the stability
of this core and the case for this revision is as yet unproven, so we
follow previous workers in assigning B68 the same distance as the rest of
the Ophiuchus cores (Bok \& McCarthy 1974).
The only exceptions to this distance assignment are the L134/L183 cloud and
L1778. These can be seen in the upper right of Figure~2. They have been
found to be at a distance of 110$\pm$10~pc (Mattila 1979; Franco 1989), and
therefore this is the distance we adopt for these cores.

For the clouds in the Aquila region there is some possible confusion as there
are two complexes on the same line of sight. L530 and L549 have distances of
530~pc (Beichman et al. 1986), and we here assign L547 the same distance by
virtue of its proximity to L530. L581 has been associated with the Aquila
Rift at a distance of 200~pc (Lee \& Myers 1999), as has B133 (Paper III),
and we use the same value here.
We adopt a distance for the Cepheus clouds (L1148 and L1155) of 325$\pm$25~pc
(Straizys et al. 1992).

The selection effects associated with our sample are somewhat complex. The
original sample (Myers, Linke
\& Benson 1983) was selected based upon criteria stating
that the core must be less than 5 arcmin in diameter, must be visible as a
patch of obscuration on the Palomar Observatory Sky Survey (POSS), must be
within 10$^\circ$ of the Galactic Plane, and must be `elongated and with a
condensation' (Myers et al. 1983). The further selection effect for starless
cores requires that the cores must not have an IRAS source within 6 arcmin,
or one core diameter, as defined by the CO (J=1$\rightarrow$0) emission, that
is detected at either 25$\mu$m or both 60 \& 100$\mu$m (Beichman et al. 1986).

A further bias in our sample is that most of the cores lie in previously
known and famous star-forming regions, with by far the majority lying in
either Ophiuchus or Taurus-Auriga. This is because the original CO
observations were primarily directed towards known areas of nearby star
formation (Myers et al. 1983). However, since these two regions are at very
similar distances, they at least provide us with a largely homogeneous
sample of cores. The more distant cores that appear to have similar
brightness must of course have higher luminosities.

\begin{table*}
\caption{Source names, peak positions and 850- \& 450-$\mu$m flux densities
of the sixteen `intermediate' cores detected with an
850-$\mu$m peak flux density $<$170mJy/beam at peak.
The peak flux densities are measured using a beam size FWHM of 14.8 arcsec
and are quoted to an accuracy of 1 mJy/beam
(or 3 sig. figs., whichever is the lower accuracy).
The integrated flux densities are measured in a 150-arcsec diameter
circular aperture and are quoted to 3 sig. figs. in mJy.
The errors and upper limits given are the statistical measurement errors
and are quoted to 2 sig. figs. Absolute
calibration errors are $\pm$10\% at 850~$\mu$m and $\pm$25\% at 450~$\mu$m.
Sources marked with a $*$ appear to have not been fully mapped and
their peaks may have been missed, so both their peak and integrated
flux densities may have been underestimated
(e.g. for L1689A, see Evans et al. 2001).
L1686 is sometimes also referred to as Oph A3 (Motte et al., 1998).
The final column indicates whether or not a NH$_3$ detection is
associated with the source (Benson \& Myers 1989).
A dash indicates the position was not searched for NH$_3$.}
\begin{tabular}{lccccccc}
\hline
Source  &  RA  &  Dec. & $S_{850}^{peak}$ & $S_{850}^{int}$
  & $S_{450}^{peak}$ & $S_{450}^{int}$  & NH$_3$ \\
        &  (2000)  &  (2000) &    (mJy/beam)  &     (mJy)
  &    (mJy/beam)  &     (mJy)   & \\
\hline
L1498  &  04$^{\rm h}$10$^{\rm m}$52.8$^{\rm s}$
  &    $+$25$^\circ$10$^\prime$09$^{\prime\prime}$
  &    116 $\pm$ 15  &  2270 $\pm$ 130
  &    560 $\pm$ 150 &  9450 $\pm$ 1100 & Y \\
L1521SMM &  04$^{\rm h}$21$^{\rm m}$04.0$^{\rm s}$
  &    $+$27$^\circ$02$^\prime$30$^{\prime\prime}$
  &    130 $\pm$ 22  &  1250 $\pm$ 190
  &    1750 $\pm$ 290 & $\leq$6200 & -- \\
L1521E &  04$^{\rm h}$29$^{\rm m}$13.6$^{\rm s}$
  &    $+$26$^\circ$14$^\prime$05$^{\prime\prime}$
  &    132 $\pm$ 17  &  1440 $\pm$ 150
  &    1590 $\pm$120 &  4020 $\pm$ 1000 & Y  \\
L1517A &  04$^{\rm h}$55$^{\rm m}$08.9$^{\rm s}$
  &    $+$30$^\circ$33$^\prime$42$^{\prime\prime}$
  &    72 $\pm$ 13   &  460 $\pm$  110
  &    393 $\pm$ 66  & 4820 $\pm$ 560 & Y \\
L1512  &  05$^{\rm h}$04$^{\rm m}$08.4$^{\rm s}$
  &    $+$32$^\circ$43$^\prime$26$^{\prime\prime}$
  &    113 $\pm$ 17  &  1370 $\pm$ 140
  &    730 $\pm$ 150 &  8000 $\pm$ 1200 & Y \\
L1719D  &  16$^{\rm h}$21$^{\rm m}$09.2$^{\rm s}$
   &  $-$20$^\circ$07$^\prime$10$^{\prime\prime}$
   &  62 $\pm$ 10 &  556 $\pm$ 83
   &  441 $\pm$ 87 & $\leq$1900  & N \\
L1686   &  16$^{\rm h}$25$^{\rm m}$57.2$^{\rm s}$
   &  $-$24$^\circ$19$^\prime$00$^{\prime\prime}$
   &  123 $\pm$ 25  & 1050 $\pm$ 210
   &  10500 $\pm$ 3100 & $\leq$60000 & N \\
L1709A$^{*}$ &  16$^{\rm h}$30$^{\rm m}$51.9$^{\rm s}$
   &  $-$23$^\circ$41$^\prime$52$^{\prime\prime}$
   &  140 $\pm$ 17 &  1530 $\pm$ 140
   &  $\leq$450 & $\leq$3300 & Y \\
L1689A$^{*}$ &  16$^{\rm h}$32$^{\rm m}$13.2$^{\rm s}$
  &    $-$25$^\circ$03$^\prime$45$^{\prime\prime}$
  &    138 $\pm$ 13  &  2430 $\pm$ 110
  &    1440 $\pm$ 380 & 19400 $\pm$ 3200 & Y \\
L234E  &  16$^{\rm h}$48$^{\rm m}$06.9$^{\rm s}$
  &    $-$10$^\circ$57$^\prime$08$^{\prime\prime}$
  &    129 $\pm$ 24  &  2190 $\pm$ 200
  &    $\leq$450 & $\leq$3000 & Y \\
B68    &  17$^{\rm h}$22$^{\rm m}$39.2$^{\rm s}$
  &    $-$23$^\circ$50$^\prime$01$^{\prime\prime}$
  &    70 $\pm$ 11   &  1270 $\pm$ 90
  &    $\leq$1000 & $\leq$8500 & Y \\
L530D  &  18$^{\rm h}$49$^{\rm m}$57.6$^{\rm s}$
  &    $-$04$^\circ$49$^\prime$47$^{\prime\prime}$
  &    81 $\pm$ 14   &  1220 $\pm$ 120
  &    1130 $\pm$ 200 &  5800 $\pm$ 1400 & Y \\
L1148   &  20$^{\rm h}$40$^{\rm m}$32.1$^{\rm s}$
  &   $+$67$^\circ$20$^\prime$45$^{\prime\prime}$
  &   65 $\pm$ 14 & 580 $\pm$ 120
  &   $\leq$720 & $\leq$5100 & Y \\
L1155H$^{*}$ &  20$^{\rm h}$43$^{\rm m}$02.5$^{\rm s}$
  &    $+$67$^\circ$47$^\prime$29$^{\prime\prime}$
  &    81 $\pm$ 10   &  894 $\pm$ 88
  &    $\leq$930 & $\leq$6300 & N \\
L1155C &  20$^{\rm h}$43$^{\rm m}$28.0$^{\rm s}$
  &    $+$67$^\circ$52$^\prime$50$^{\prime\prime}$
  &    100 $\pm$ 13  &  1600 $\pm$ 110
  &    $\leq$2400 & $\leq$20000 & Y \\
L1155D  &  20$^{\rm h}$43$^{\rm m}$52.8$^{\rm s}$
  &    $+$67$^\circ$37$^\prime$32$^{\prime\prime}$
  &    65 $\pm$ 10   & 290 $\pm$ 80
  &    $\leq$930 & $\leq$6300  & N \\
\hline
\end{tabular}
\end{table*}

\section{Observations}

The observations were carried out using the Submillimetre Common User
Bolometer Array (SCUBA) on the James Clerk Maxwell Telescope (JCMT).
Observations in the Right Ascension (RA)
range 15--21 hours were carried out in the evenings of 1997
August 7--13  between 17:30 and 01:30 H.S.T. (03:30--11:30 UT).
Observations in the RA range 04--06 hours were
carried out in the mornings of 1997 October 4--7 from 01:30
to 09:30 H.S.T. (11:30--19:30 UT).

The observations were conducted simultaneously at 850 and 450 $\mu$m
using a 64-point jiggle pattern to create fully sampled maps of 2.3
arcminute diameter fields around the coordinate centres
(Holland et al. 1999).
Time-dependent variations
in the sky emission were removed by chopping the secondary mirror at 7 Hz
with a chop throw of 120 arcsec in azimuth.
The submillimetre zenith opacity at 850 and 450~$\mu$m
was determined using the `skydip' method
and by comparison with the 1.3-mm sky opacity.
The average 850-$\mu$m zenith optical depth was 0.33
$\pm$ 0.09 for the August run and 0.25 $\pm$ 0.09 for the October run,
corresponding to a mean zenith transmission of 72\% and 78\%
respectively. The average 450-$\mu$m optical depth
was 1.81 $\pm$ 0.66 for the August run and 1.55
$\pm$ 0.74 for the October run,
corresponding to a zenith transmission in the range of $\sim$10--30\% and
$\sim$10--45\% respectively.

The data were reduced in the normal way using the SCUBA User
Reduction Facility (Jenness \& Lightfoot 1998).
Calibration was performed using
observations of the planets Mars and Uranus taken during
each shift. We estimate that the absolute calibration uncertainty
is $\pm$10\% at 850 $\mu$m and $\pm$25\% at 450$\mu$m, based on
the consistency and reproducibility of the calibration.
The average beam size full-width at half maximum (FWHM)
was found to be 14.8 arcsec at 850 $\mu$m and 8.2 arcsec at 450 $\mu$m.
There was found to be a negligible error beam at 850 $\mu$m,
but at 450 $\mu$m a significant error beam was detected that
was found to contribute up to 50\% of the flux density. This was taken into
account when calibrating the data. In the final maps it was found that
the average 1$\sigma$ noise (off-source)
was 17 mJy/beam at 850 $\mu$m and 91 mJy/beam at 450 $\mu$m.
To see the morphology of the cores at 450~$\mu$m it was found
necessary to smooth the data to the same resolution as the 850-$\mu$m data.

\section{Results}

We observed 52 starless cores from the catalogue of Myers and
co-workers (Myers \& Benson 1983; Myers et al. 1988;
Benson \& Myers 1989).
We list the core names, positions and
measured flux densities in Tables 1--3. We
detected 29 cores with 850-$\mu$m peak
signal-to-noise levels greater than 3$\sigma$.
Some were detected with significantly higher signal-to-noise ratio than
others because some cores are significantly brighter than others.
Consequently, we divided the detected
cores into two groups: `bright' and `intermediate'.
We based this division on the 850-$\mu$m peak
flux density of the cores, but we note that the 450-$\mu$m flux densities
scale roughly accordingly.

We defined a core as `bright' if its peak 850-$\mu$m flux
density was found to be greater than, or equal to,
170~mJy/beam. This level was chosen for a number of reasons. Firstly,
this value is 10 times the mean 1$\sigma$ level of $\sim$17~mJy/beam
of our observations. In addition,
in terms of the physical parameters of the cores themselves, this lower
limit brightness cutoff represents a lower limit H$_2$ column density
of N(H$_2$)$\sim$5$\times$10$^{22}$cm$^{-2}$, or a value of
A$_V\sim$50, using typical assumptions -- T=10K,
$\kappa_{850}=$0.01cm$^2$g$^{-1}$,
N(H$_2$)/A$_V$ = 9.4 $\times$ 10$^{20}$ cm$^{-2}$mag$^{-1}$
(see section 5.3 below for discussion; see
also e.g. Bohlin, Savage \& Drake 1978; Frerking et al. 1982).
Also, by using reasonable assumptions about
core geometries (see section 5.3 below) we find that this corresponds to an
approximate minimum volume
number density of n(H$_2$)$\sim$3$\times$10$^5$cm$^{-3}$ at the distance
of the majority of our sample. This is discussed in more detail below.

This value also corresponds roughly to cores for which we could
detect the 450-$\mu$m emission sufficiently to produce a map of this
emission in most cases. Finally, the 10-$\sigma$ peak level typically
meant that we could map the detailed structure of most of these cores
at 850~$\mu$m out to reasonable radii (see below).
Table 1 lists the peak and integrated flux densities of the bright cores.
Throughout this paper we list all sources in order of
increasing Right Ascension (RA).
The integrated flux densities were measured in a 150-arcsec diameter
circular aperture. 13 of the 52 cores satisfy the criterion for bright cores.

\begin{table}
\caption{Source names, positions searched and
850- \& 450-$\mu$m upper limits to the flux densities
of the 23 cores which were undetected -- i.e.
their peak flux densities are less than three
times the 1-$\sigma$ level.
The limits are measured using a FWHM beam size of 14.8 arcsec
and are quoted to 2 sig. figs.
The quoted upper limits are based on
the statistical measurement errors. Absolute
calibration errors are $\pm$10\% at 850~$\mu$m and $\pm$25\% at 450~$\mu$m.
Sources marked with a $*$ show some evidence for low level
emission at one edge of the field, so
their peaks may have been missed.
The final column indicates whether or not a NH$_3$ detection is
associated with the source (Benson \& Myers 1989).}
\begin{tabular}{lccccc}
\hline
Source  &  RA & Dec. & $S_{850}^{peak}$ & $S_{450}^{peak}$  \\
   & (2000) & (2000) & (mJy/ & (mJy/ & NH$_3$ \\
   &        &        &  beam) & beam) & \\
\hline
L1495D  &  04$^{\rm h}$14$^{\rm m}$27.5$^{\rm s}$
   &  $+$28$^\circ$14$^\prime$49$^{\prime\prime}$
   &  $\leq$ 48 &   $\leq$ 3300 & N \\
L1495B  &  04$^{\rm h}$15$^{\rm m}$36.5$^{\rm s}$
   &  $+$28$^\circ$46$^\prime$06$^{\prime\prime}$
   &  $\leq$ 36 &  $\leq$ 370 & N \\
L1506   &  04$^{\rm h}$18$^{\rm m}$32.5$^{\rm s}$
   &  $+$25$^\circ$20$^\prime$37$^{\prime\prime}$
   &  $\leq$ 48 &  $\leq$ 1600 & N \\
L1521C  &  04$^{\rm h}$19$^{\rm m}$19.2$^{\rm s}$
   &  $+$27$^\circ$16$^\prime$29$^{\prime\prime}$
   &  $\leq$ 42   &  $\leq$ 270 & Y \\
L1521B$^{*}$  &  04$^{\rm h}$24$^{\rm m}$18.3$^{\rm s}$
   &  $+$26$^\circ$36$^\prime$47$^{\prime\prime}$
   &  $\leq$ 84 &  $\leq$ 250 & N \\
L1521A  &  04$^{\rm h}$26$^{\rm m}$43.1$^{\rm s}$
   &  $+$26$^\circ$16$^\prime$00$^{\prime\prime}$
   &  $\leq$ 27  &  $\leq$ 280 & N \\
L1551C$^{*}$  &  04$^{\rm h}$31$^{\rm m}$25.9$^{\rm s}$
   &  $+$18$^\circ$09$^\prime$11$^{\prime\prime}$
   &  $\leq$ 78   &  $\leq$ 750 & N \\
L1507A  &  04$^{\rm h}$42$^{\rm m}$38.9$^{\rm s}$
   &  $+$29$^\circ$44$^\prime$19$^{\prime\prime}$
   &  $\leq$ 54 &   $\leq$ 570 & N \\
L1517C  &  04$^{\rm h}$54$^{\rm m}$46.8$^{\rm s}$
   &  $+$30$^\circ$35$^\prime$10$^{\prime\prime}$
   &  $\leq$ 36 &  $\leq$ 300 & Y \\
L1517D  &  04$^{\rm h}$55$^{\rm m}$48.1$^{\rm s}$
   &  $+$30$^\circ$38$^\prime$46$^{\prime\prime}$
   &  $\leq$ 36   &  $\leq$ 770 & N \\
L1523   &  05$^{\rm h}$06$^{\rm m}$17.5$^{\rm s}$
   &  $+$31$^\circ$41$^\prime$58$^{\prime\prime}$
   &  $\leq$ 21 &  $\leq$ 620 & N \\
L1778A  &  15$^{\rm h}$39$^{\rm m}$27.5$^{\rm s}$
   &  $-$07$^\circ$10$^\prime$09$^{\prime\prime}$
   &  $\leq$ 36 &  $\leq$ 350 & N \\
L1778B  &  15$^{\rm h}$40$^{\rm m}$02.5$^{\rm s}$
   &  $-$07$^\circ$07$^\prime$21$^{\prime\prime}$
   &  $\leq$ 30 &  $\leq$  270  & N \\
L134    &  15$^{\rm h}$53$^{\rm m}$31.3$^{\rm s}$
   &  $-$04$^\circ$37$^\prime$03$^{\prime\prime}$
   &  $\leq$ 24  &  $\leq$ 140 & N \\
L134A   &  15$^{\rm h}$53$^{\rm m}$43.9$^{\rm s}$
   &  $-$04$^\circ$35$^\prime$00$^{\prime\prime}$
   &  $\leq$ 42 &  $\leq$ 350 & Y \\
L1721   &  16$^{\rm h}$14$^{\rm m}$28.2$^{\rm s}$
   &  $-$18$^\circ$54$^\prime$44$^{\prime\prime}$
   &  $\leq$ 48 &  $\leq$ 140 & N \\
L1719B  &  16$^{\rm h}$22$^{\rm m}$12.4$^{\rm s}$
   &  $-$19$^\circ$38$^\prime$41$^{\prime\prime}$
   &  $\leq$ 54   &  $\leq$ 160 & N \\
L1709C  &  16$^{\rm h}$33$^{\rm m}$53.4$^{\rm s}$
   &  $-$23$^\circ$41$^\prime$32$^{\prime\prime}$
   &  $\leq$ 30   &  $\leq$ 280 & Y \\
L121    &  16$^{\rm h}$39$^{\rm m}$28.4$^{\rm s}$
   &  $-$14$^\circ$05$^\prime$21$^{\prime\prime}$
   &  $\leq$ 45 &  $\leq$ 530  & N \\
L204B   &  16$^{\rm h}$47$^{\rm m}$33.2$^{\rm s}$
   &  $-$11$^\circ$58$^\prime$47$^{\prime\prime}$
   &  $\leq$ 30   &  $\leq$ 490  & Y \\
L547    &  18$^{\rm h}$51$^{\rm m}$24.7$^{\rm s}$
   &  $-$04$^\circ$16$^\prime$35$^{\prime\prime}$
   &  $\leq$ 87 &  $\leq$ 1200 & N \\
L549    &  19$^{\rm h}$02$^{\rm m}$06.7$^{\rm s}$
   &  $-$04$^\circ$18$^\prime$50$^{\prime\prime}$
   &  $\leq$ 48 &    $\leq$ 1200 & N \\
L581    &  19$^{\rm h}$07$^{\rm m}$26.2$^{\rm s}$
   &  $-$03$^\circ$54$^\prime$49$^{\prime\prime}$
   &  $\leq$ 42 &  $\leq$ 1300 & N  \\
\hline
\end{tabular}
\end{table}

A core was defined as undetected if its
peak 850-$\mu$m flux density was less than 3 times the 1$\sigma$
level of the particular observation.
Table 2 lists the peak and integrated
flux densities of the intermediate cores.
The integrated flux densities were once again
measured in a 150-arcsec diameter circular aperture.
16 of the 52 cores have intermediate brightness under this definition.

Table 3 lists the positions searched and 3$\sigma$ upper limits of the
peak flux densities of the cores we designate as undetected, having peak
850-$\mu$m values less than 3 times the 1$\sigma$ value of the observation.
23 of the 52 cores were undetected.
Note that some of the upper limits are higher than some of the peak
flux densities detected for a small number of the intermediate cores.
This is due to the varying atmospheric conditions during the observations.
It is also possible for the undetected cores that we missed the core peak
if it did not lie within the 2.3-arcmin field of view of SCUBA.

The position centres for L134 and L134A do not coincide exactly
with some listed positions for molecular line observations of
these sources so we list the three sigma upper limits observed in the
locations observed. L1551C was observed in the normal manner, but it was
found afterwards that a bright source (possibly L1551N) was present in the
`off' beam position, making it impossible to map in the beam-switching
mode we were using. Hence we list the three sigma upper limit at the
location observed but note that it may be an underestimate.
There are a small number of cores for which we believe we may have
missed their brightest peaks. These are indicated by asterisks in the three
Tables. In terms of our statistics of numbers of cores in each category, we
estimate that this introduces an error of roughly $\pm$3.

We found that for the bright cores
we could map their 850-$\mu$m structure with
reasonable signal-to-noise ratio but for intermediate cores
we could not map their structure without smoothing to
slightly lower resolution. However, for the 16 intermediate
cores we were still able to make integrated flux density measurements
to obtain their total 850-$\mu$m flux densities. Roughly half of the
intermediate cores were not detected significantly at 450~$\mu$m.
We also note that all of our `bright' detections with SCUBA coincide
with NH$_3$ detections (for the sources in common), whereas four out
of sixteen cores found with SCUBA to be
`intermediate' were not detected in NH$_3$.
In addition, 18 out of 23 of our `faint' cores were not detected in NH$_3$.
This seems to indicate that our SCUBA maps are tracing roughly similar
material to NH$_3$ maps. Since the latter is a volume density tracer,
we believe that our SCUBA mapping (with a finite chop throw) also appears to
be tracing volume density.

Figures 3 \& 4 show the 850- \& 450-$\mu$m maps of the 13 cores
designated as bright. The 450-$\mu$m maps have been
smoothed to the same resolution as the 850-$\mu$m data (14.8 arcsec).
The first thing that is apparent from the maps is that none of the cores
are circular, but rather they show a variety of morphologies. Most of
the cores appear elongated -- some to the extent that one would call them
filamentary.
The cores that we would class as filamentary are L1521D, L183,
L1696A \& L1689SMM. This is based on the fact that these cores show
evidence for a low level ridge or filament in which the core is embedded.

Most cores have only one peak, although some have two or more.
L43 is somewhat of an exception, in that it is the only source in our
sample where there is another source in the field of view as well
as a prestellar core. The more westerly of the two peaks that can be
seen in the map is the classical T Tauri star RNO~91 (Cohen 1980).
L1696A appears to contain two prestellar sources. We here call them
L1696A-N and L1696A-S. L1689B, L63 and B133 all show low level emission
apparently extending beyond at least one edge of the map (e.g. for L1689B
there is extended emission in the east-west direction).

\begin{figure*}
\setlength{\unitlength}{1mm}
\noindent
\begin{picture}(170,220)
\put(-10,220){\includegraphics{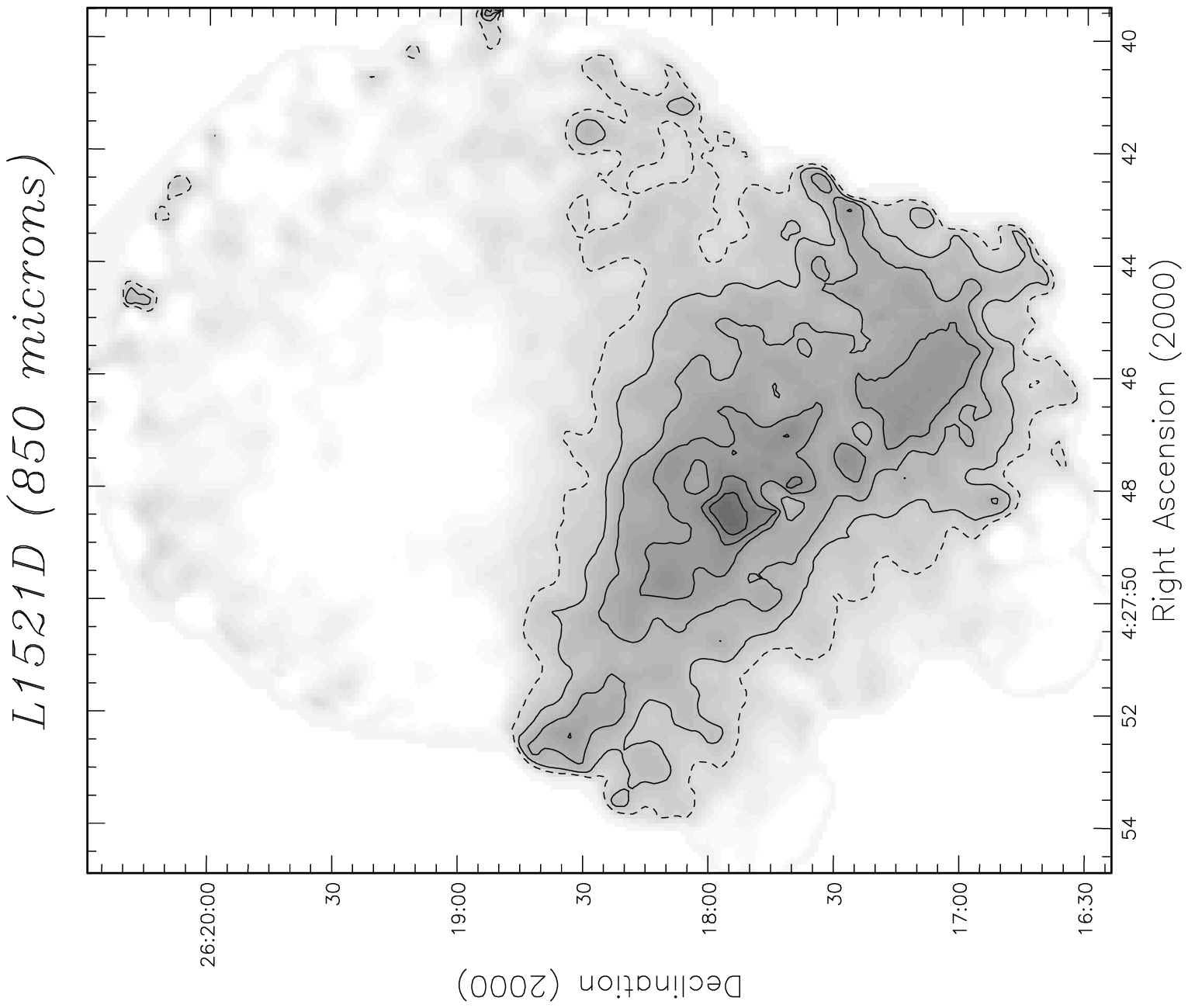}}
\put(50,220){\includegraphics{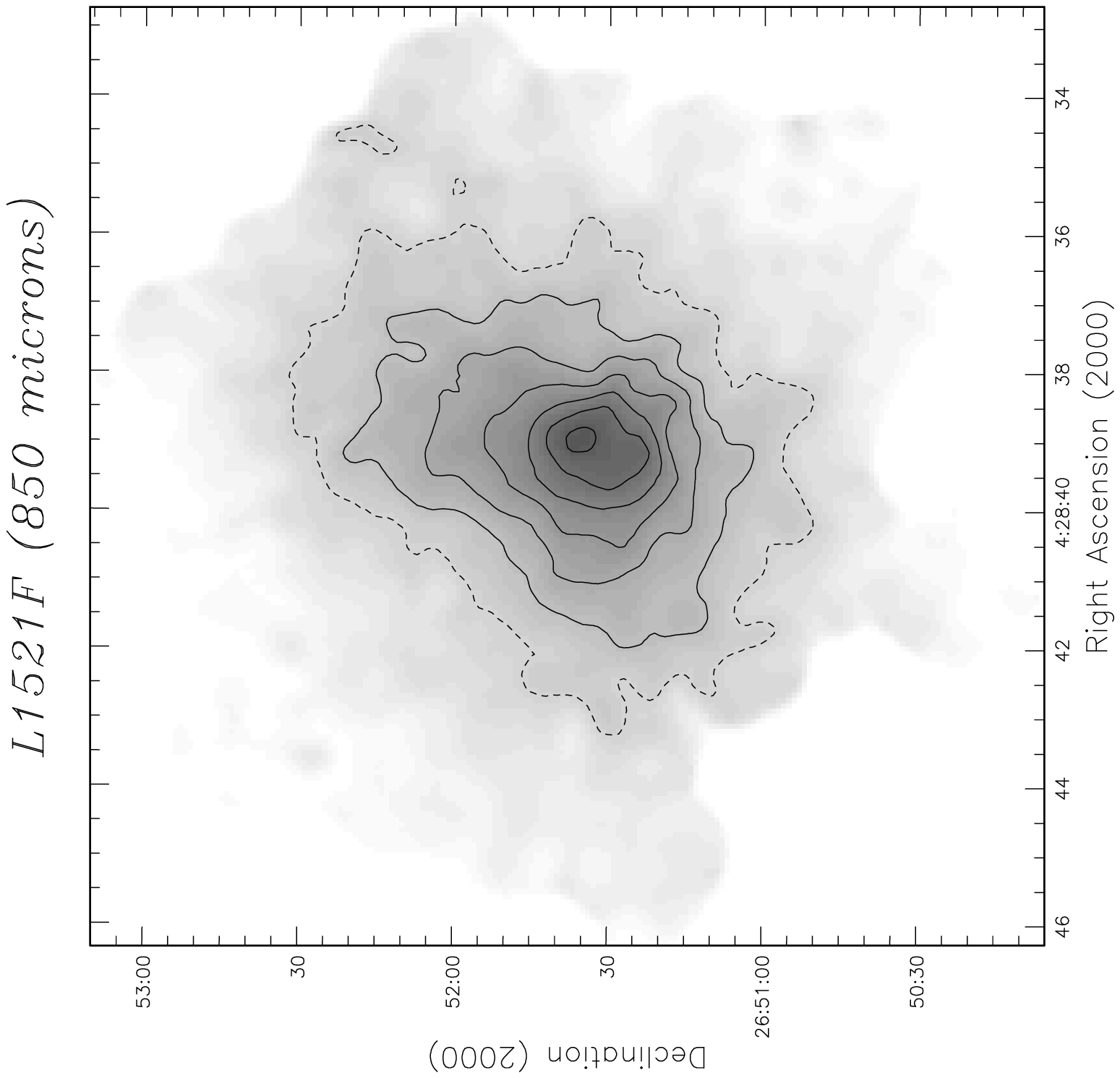}}
\put(110,220){\includegraphics{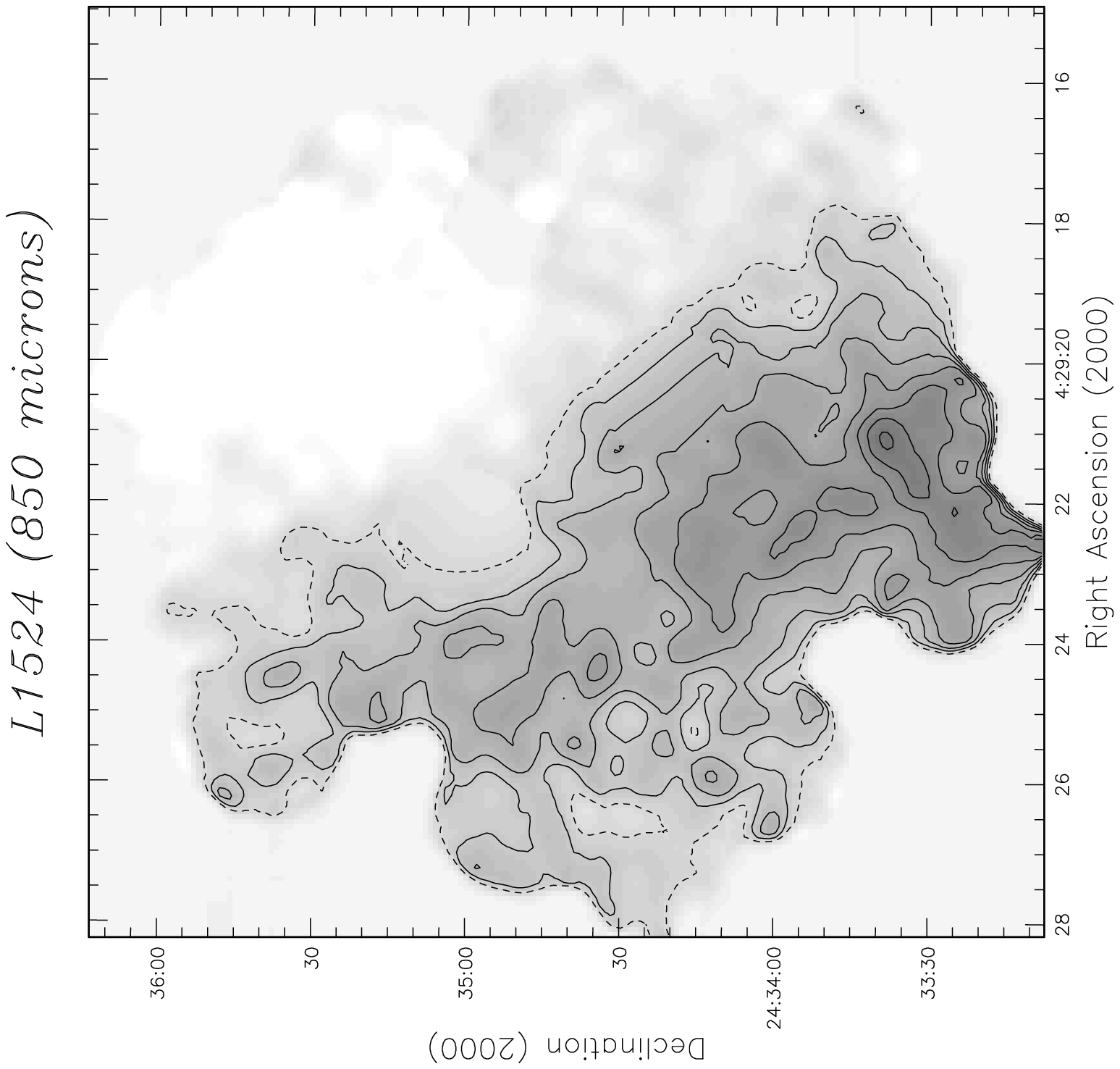}}
\put(-10, 165){\includegraphics{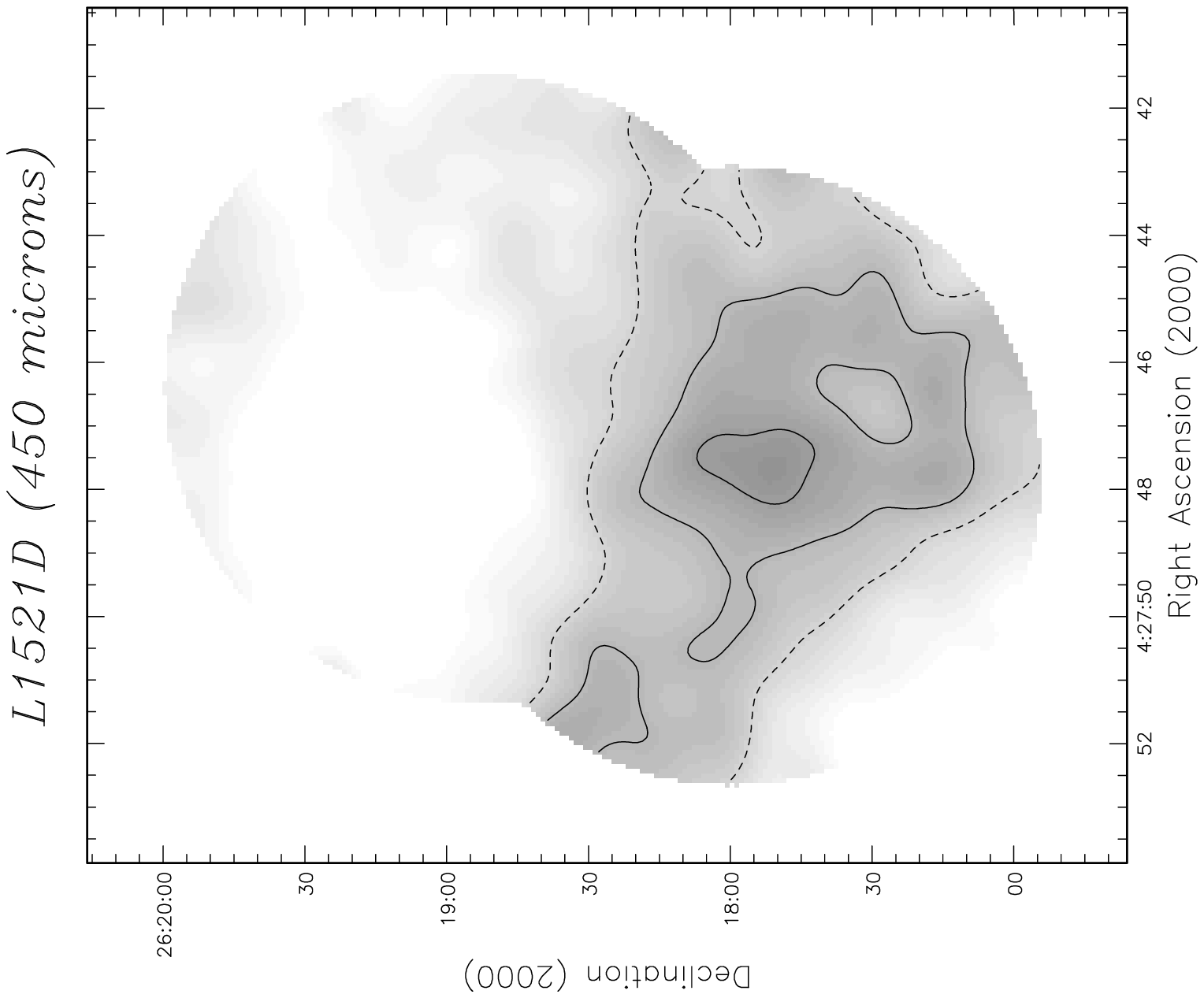}}
\put(50,165){\includegraphics{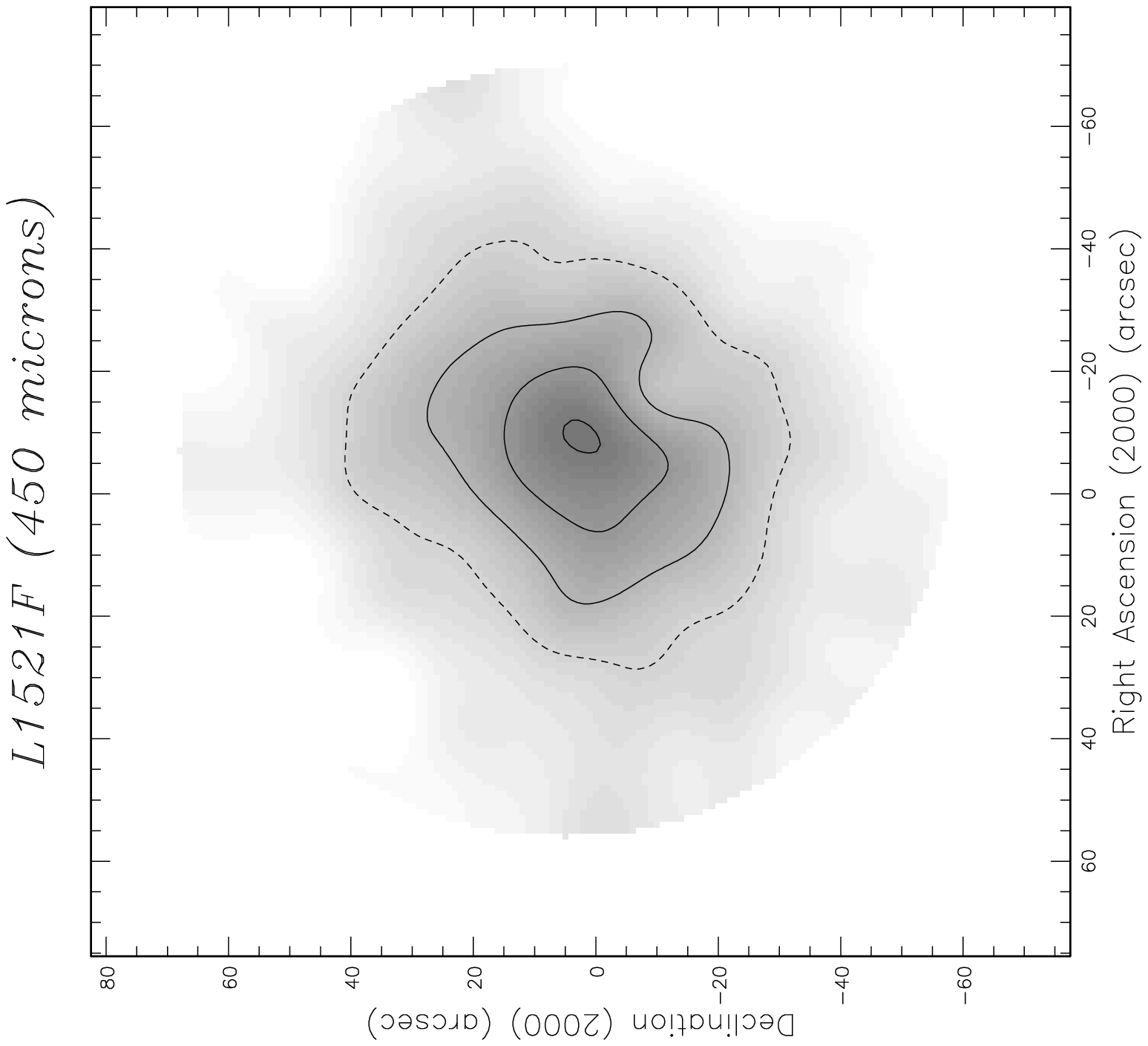}}
\put(110,165){\includegraphics{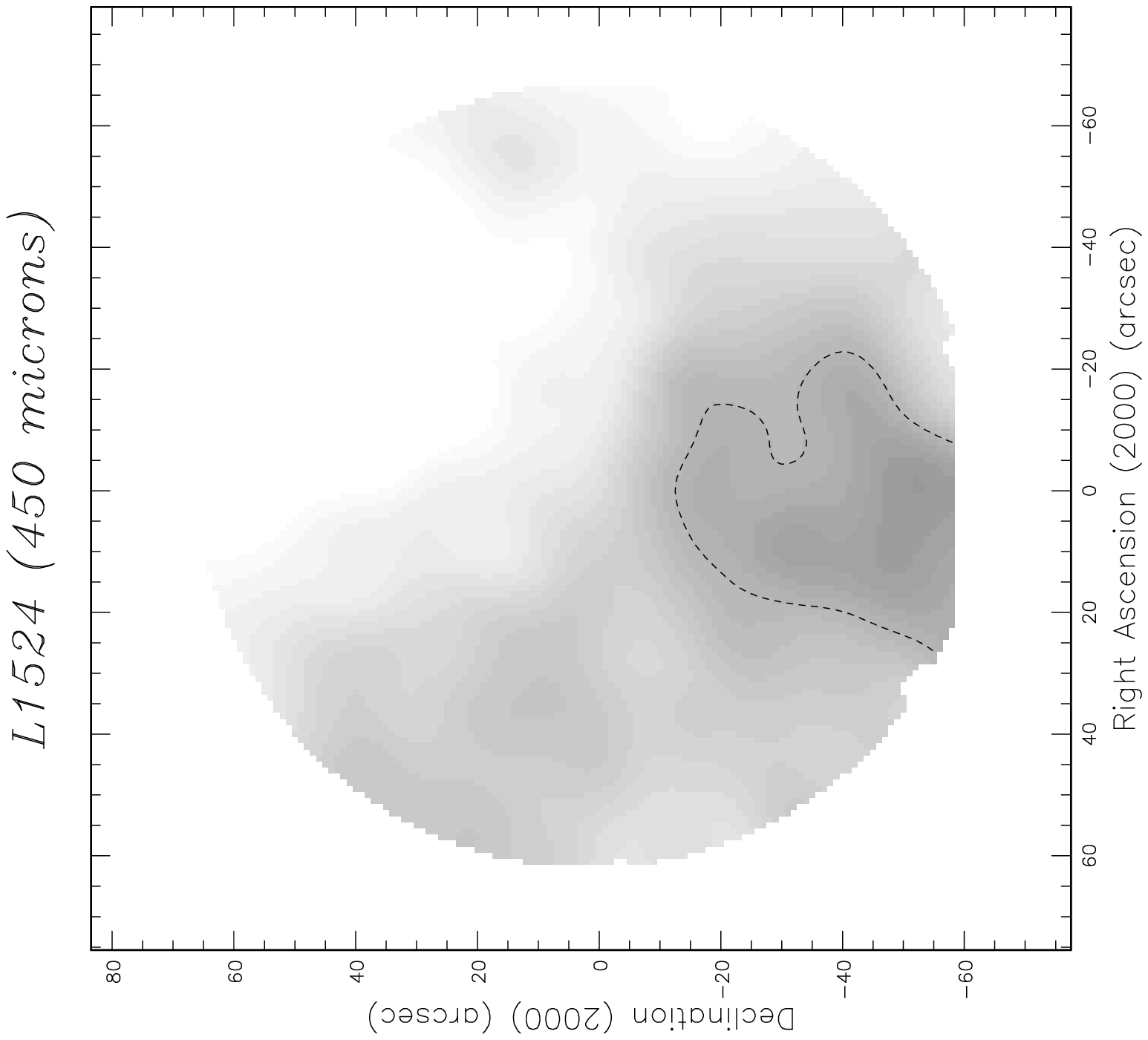}}
\put(-10,110){\includegraphics{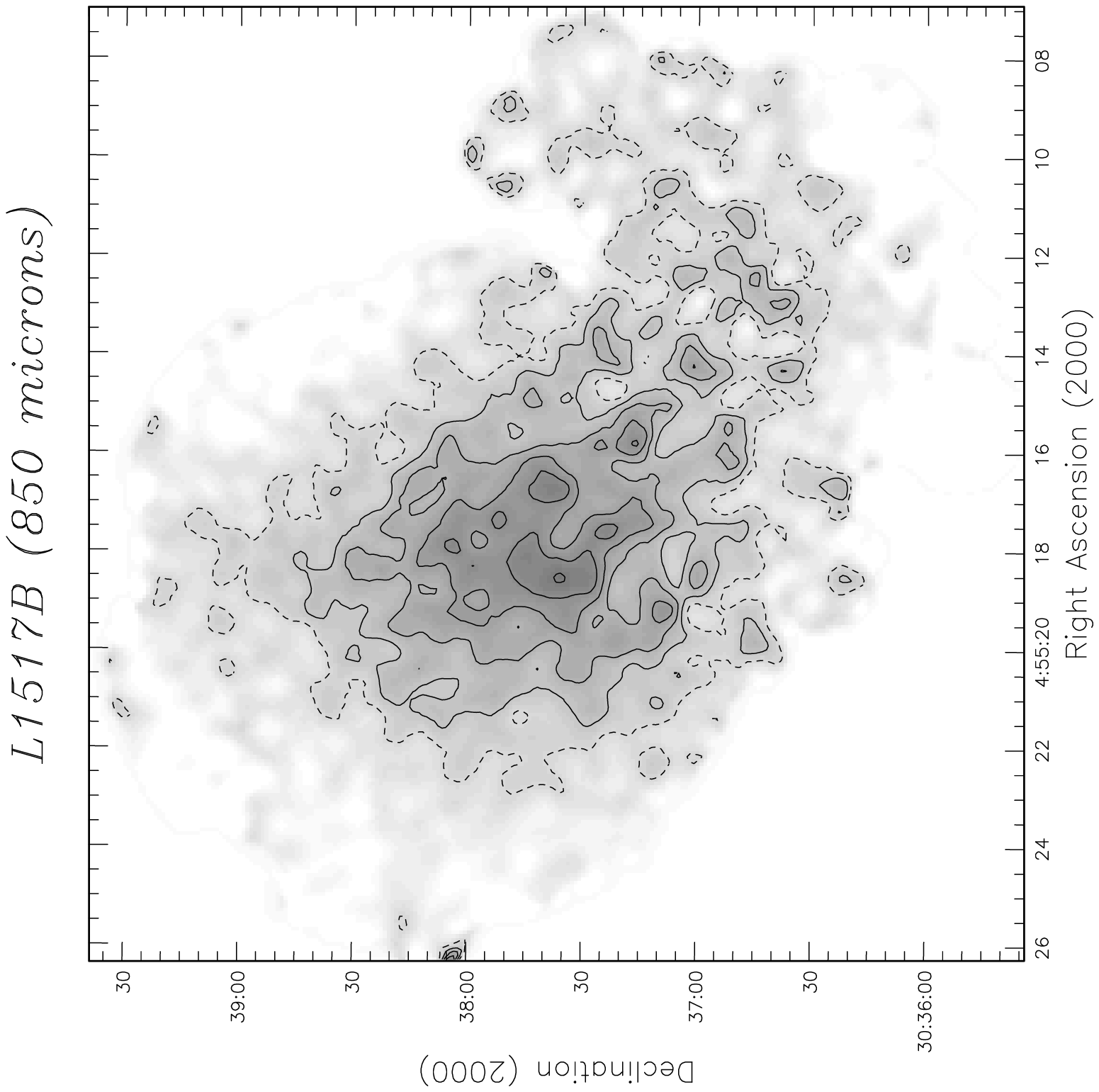}}
\put(50,110){\includegraphics{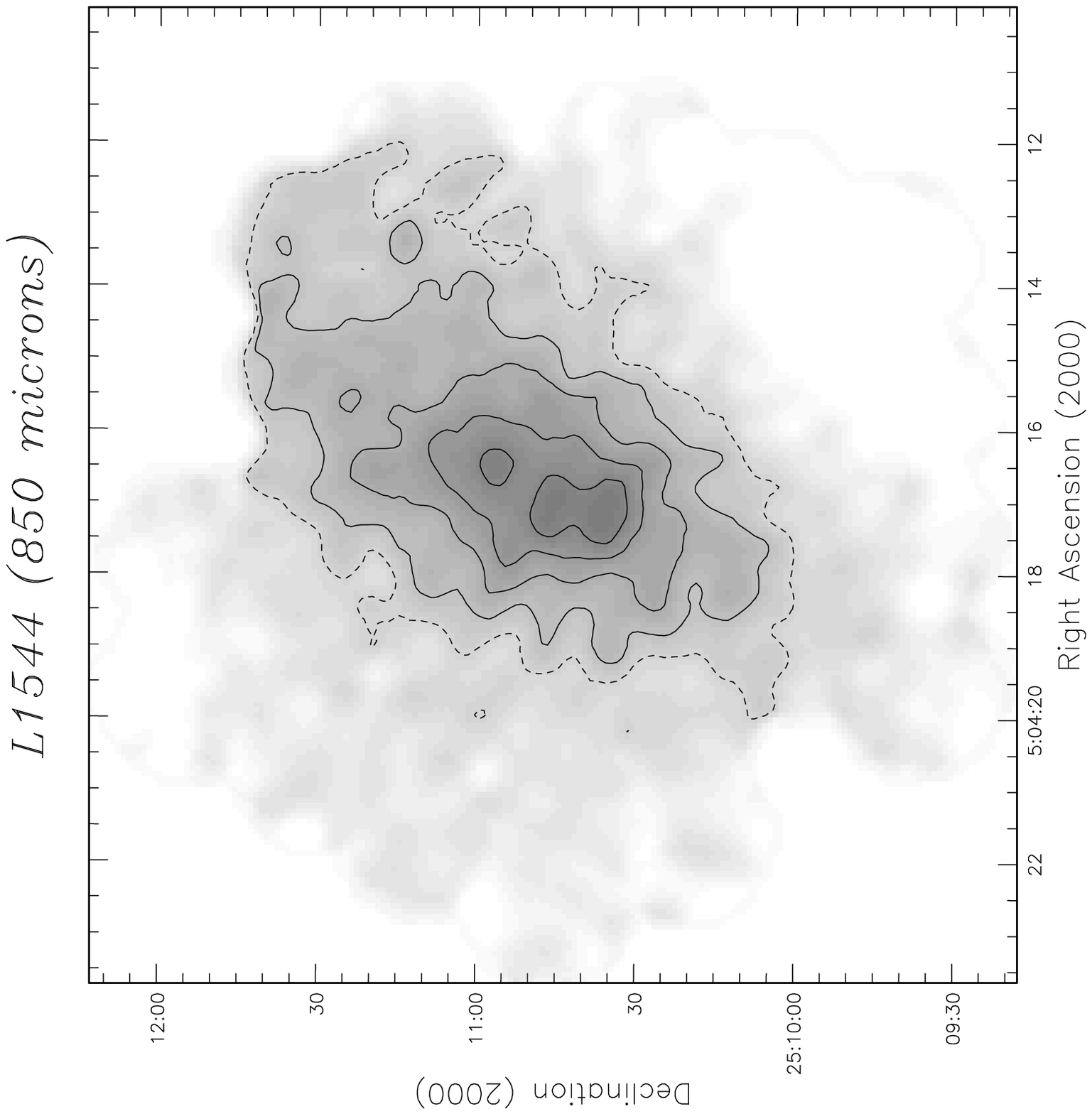}}
\put(110,110){\includegraphics{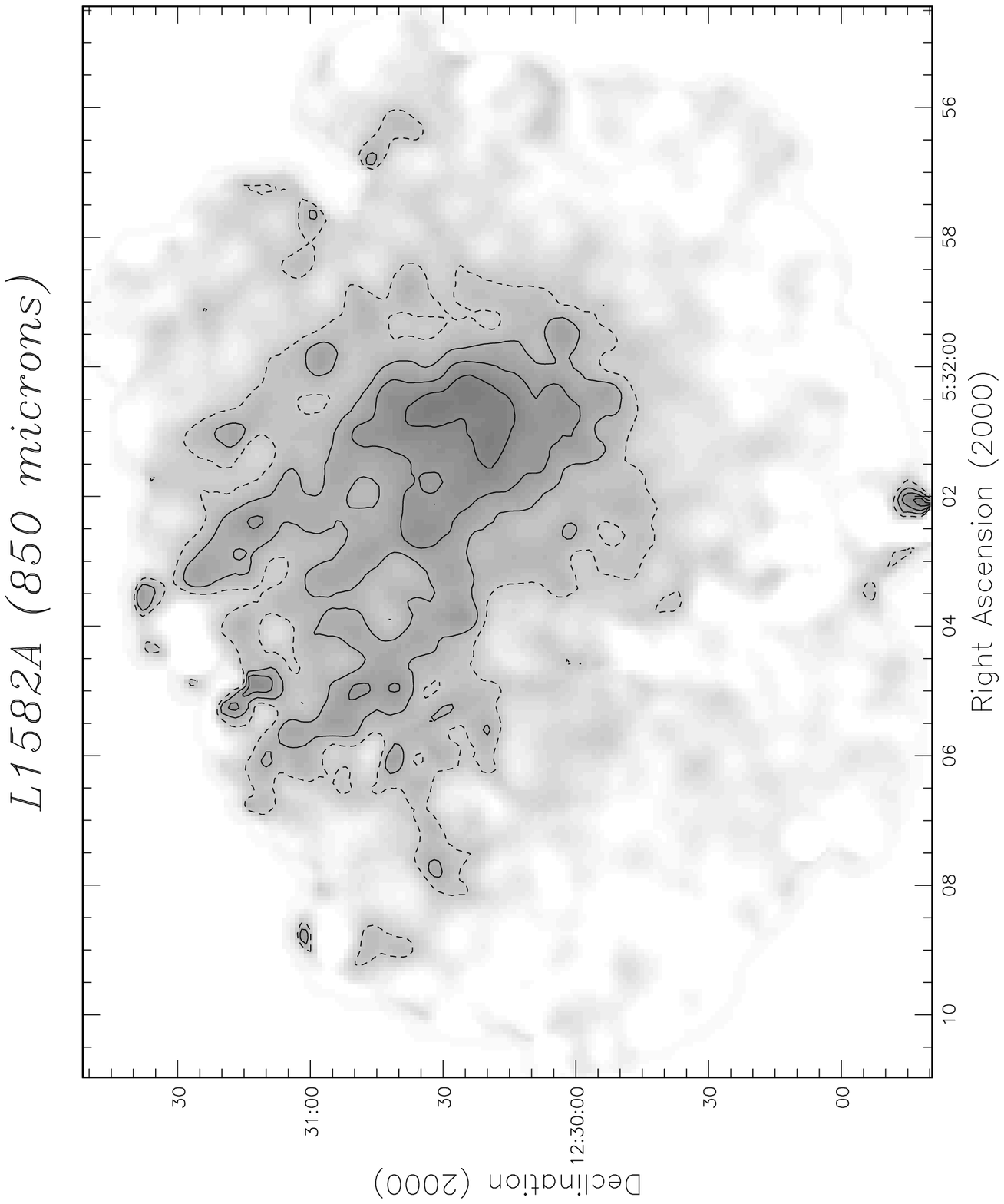}}
\put(-10,55){\includegraphics{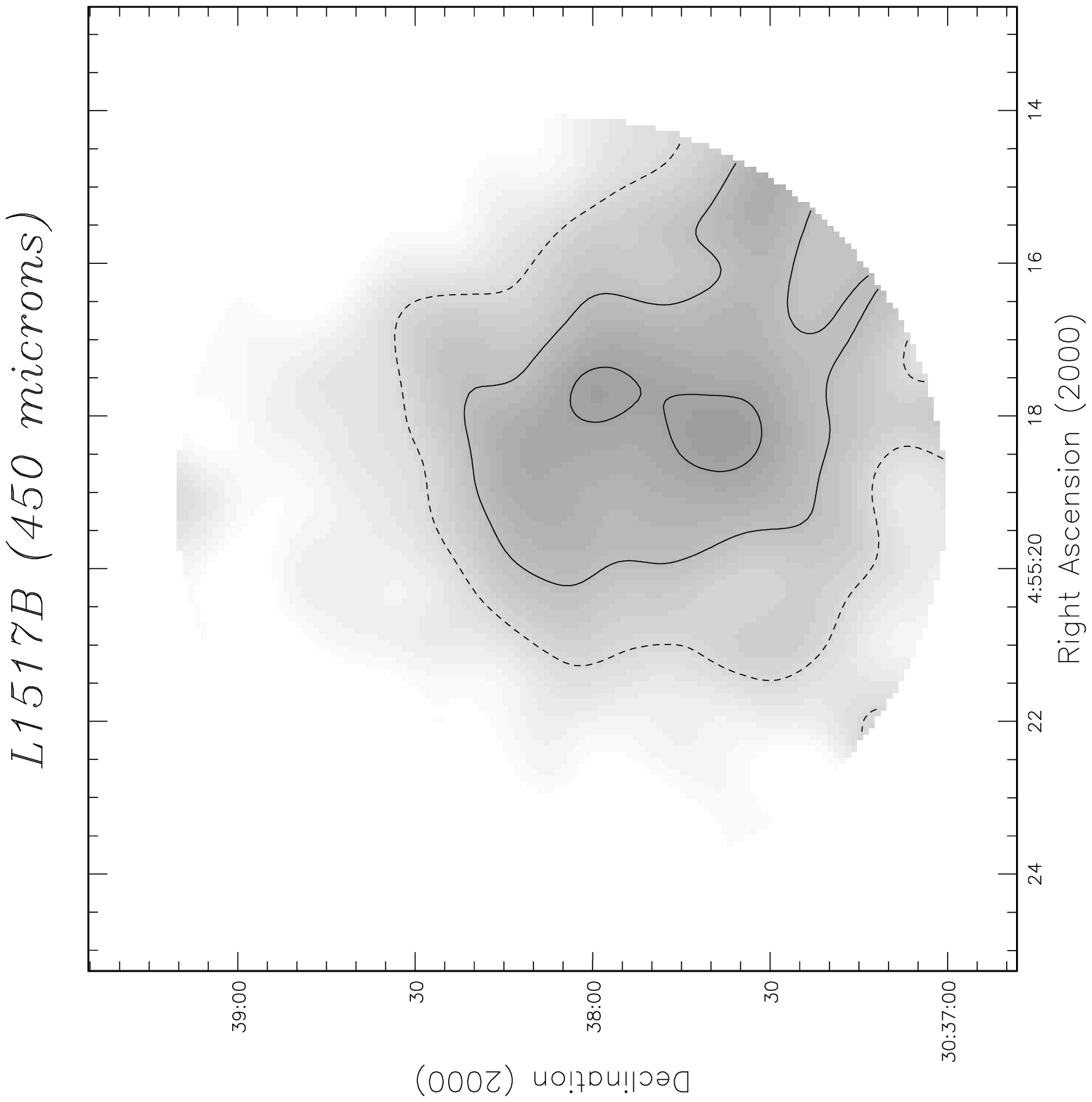}}
\put(50,55){\includegraphics{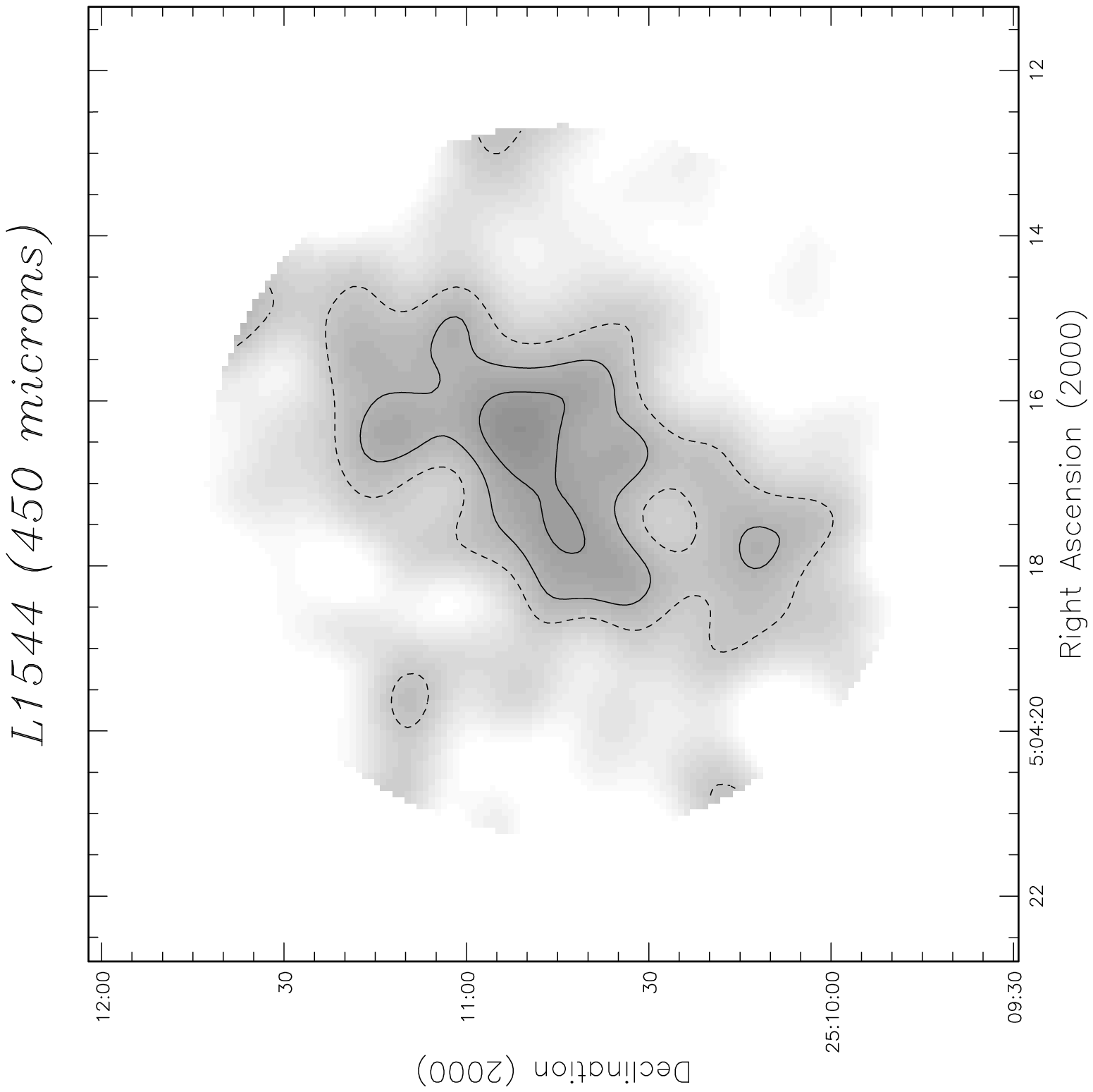}}
\put(110,55){\includegraphics{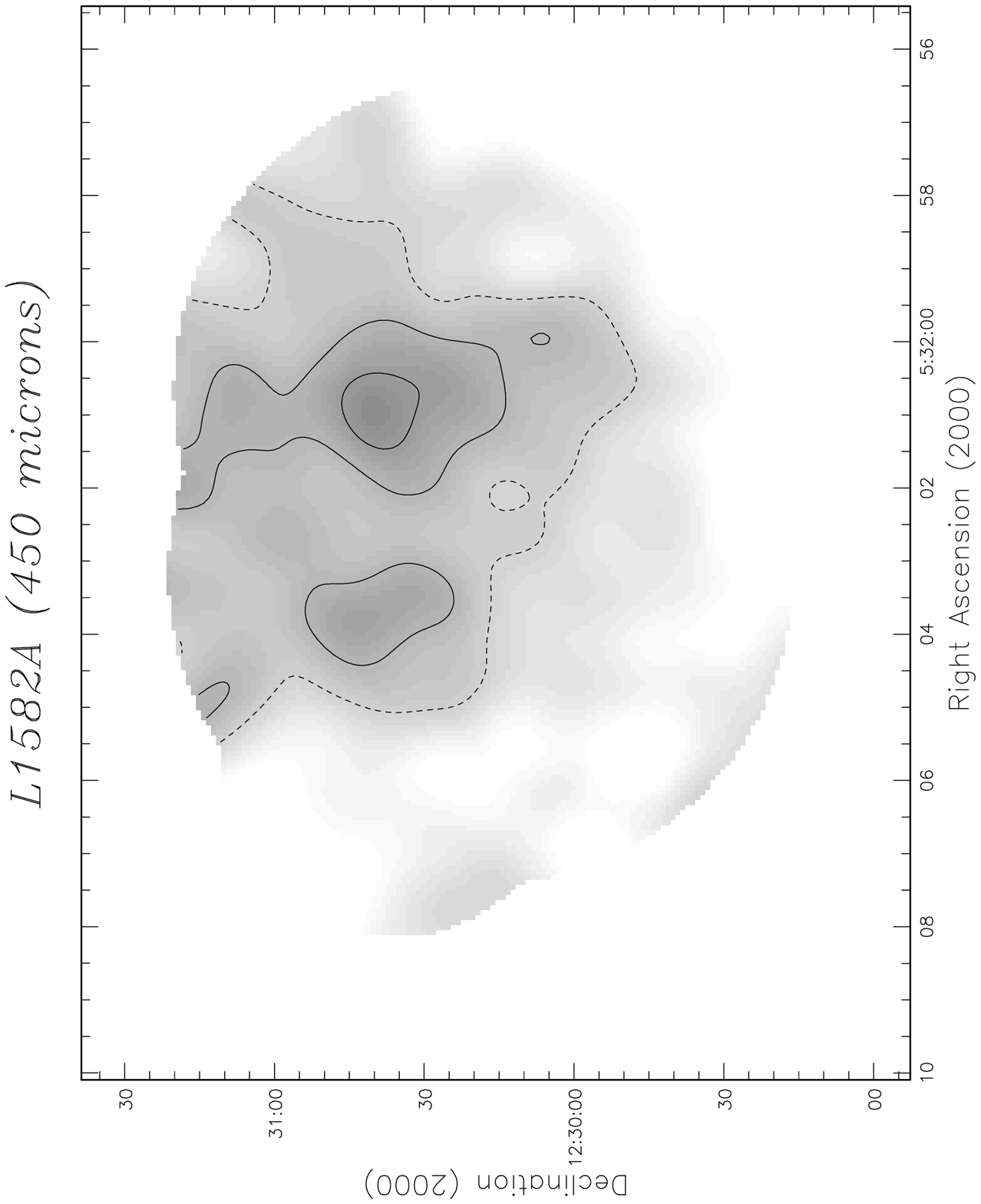}}
\end{picture}
\caption{Greyscale images with contours superposed of 850- and 450-$\mu$m
continuum maps of six of the thirteen cores designated as bright
-- L1521D \& F, L1524, L1517B, L1544 \& L1582A.
The lowest (dashed) contour in each case is at a level of 3$\sigma$.
The solid contours start at a level of 5$\sigma$, and the contour interval is
2$\sigma$. The 1$\sigma$ noise levels for each source at each wavelength are
listed in Table 1. The 450-$\mu$m data have been smoothed to the same
resolution as the 850-$\mu$m data (14.8 arcsec). Note that the peak
of the emission may have been missed in the case of L1524.}
\end{figure*}

\begin{figure*}
\setlength{\unitlength}{1mm}
\noindent
\begin{picture}(170,220)
\put(-20,220){\includegraphics{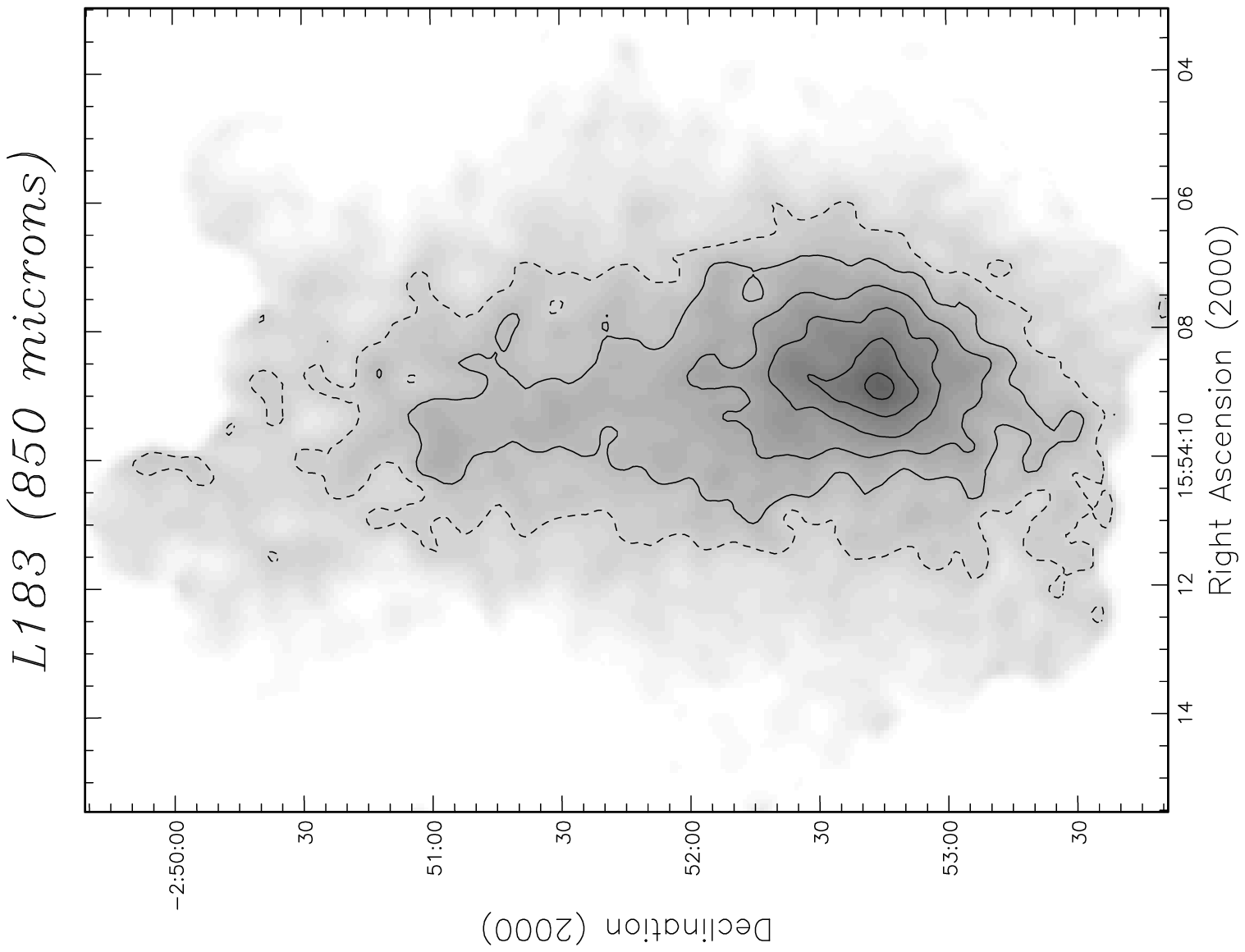}}
\put(20,220){\includegraphics{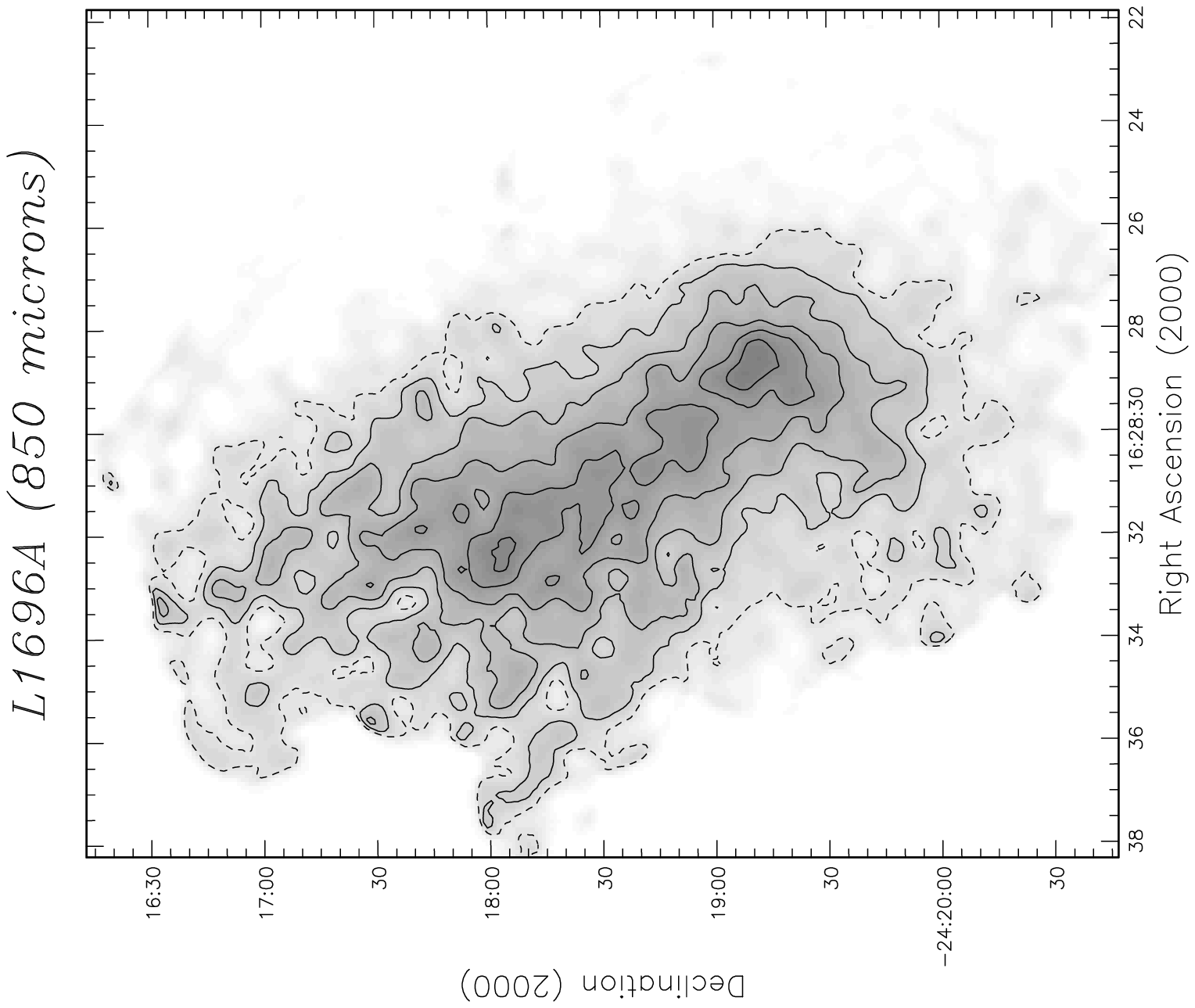}}
\put(70,220){\includegraphics{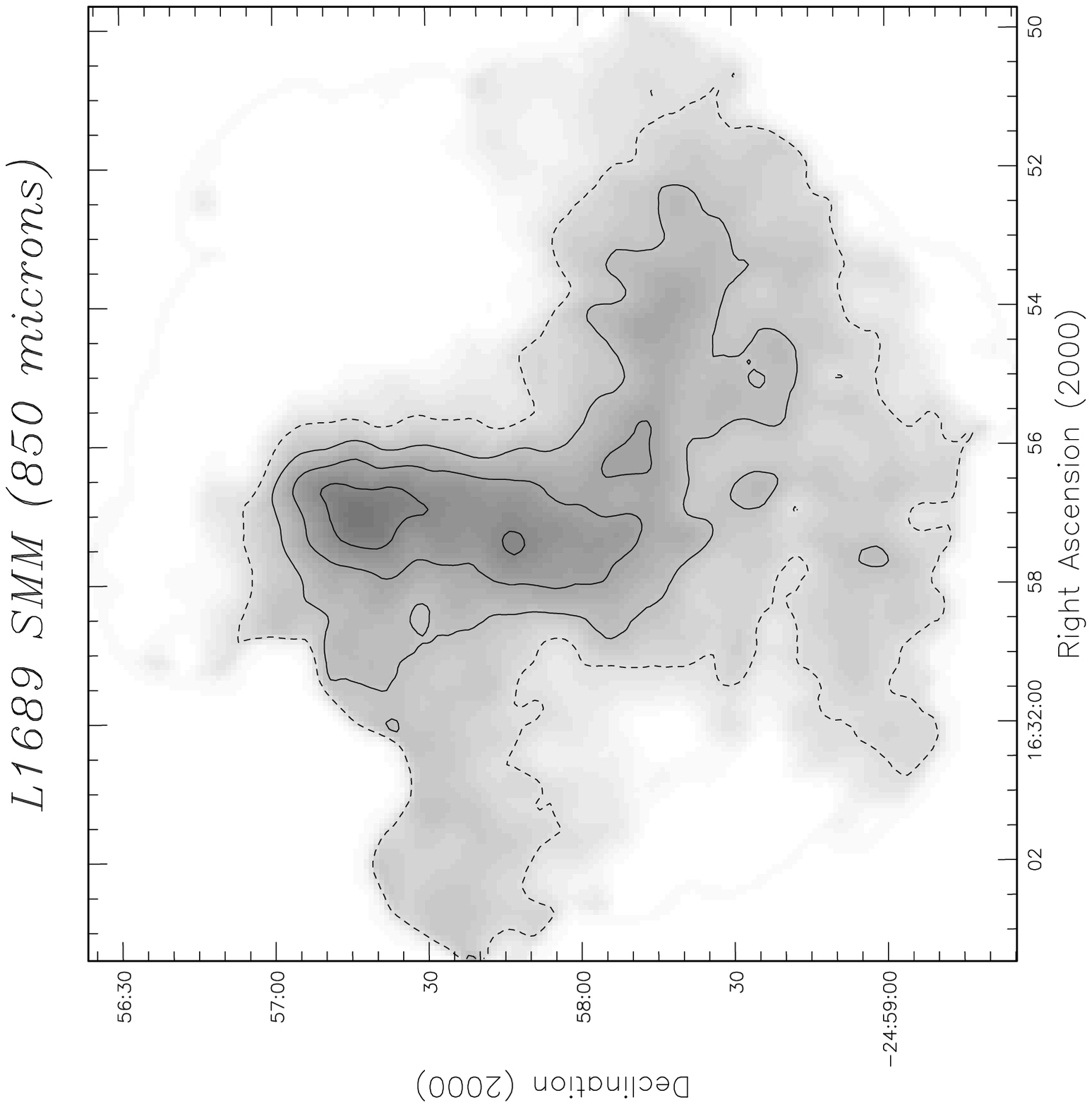}}
\put(120,220){\includegraphics{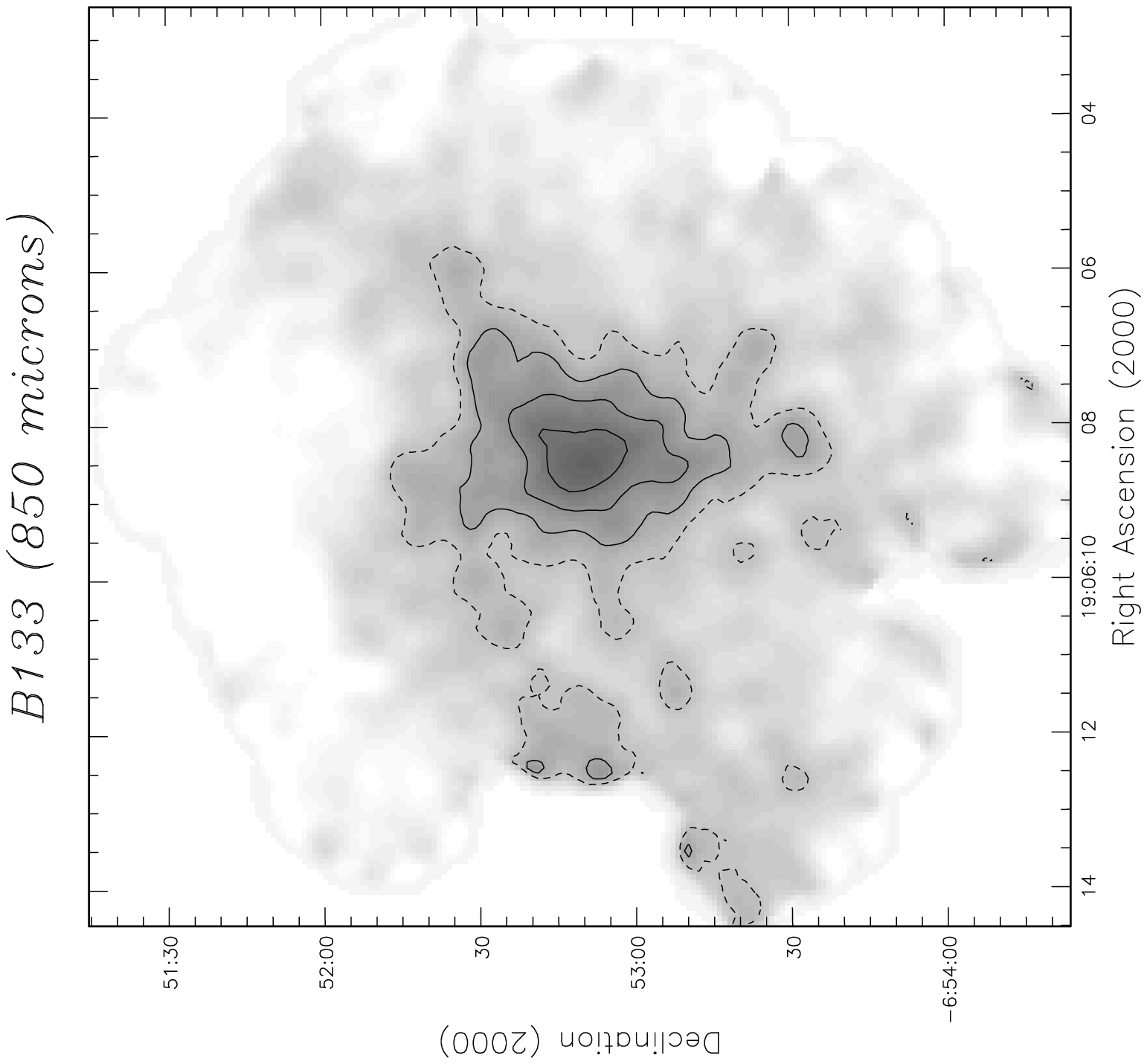}}
\put(-20,165){\includegraphics{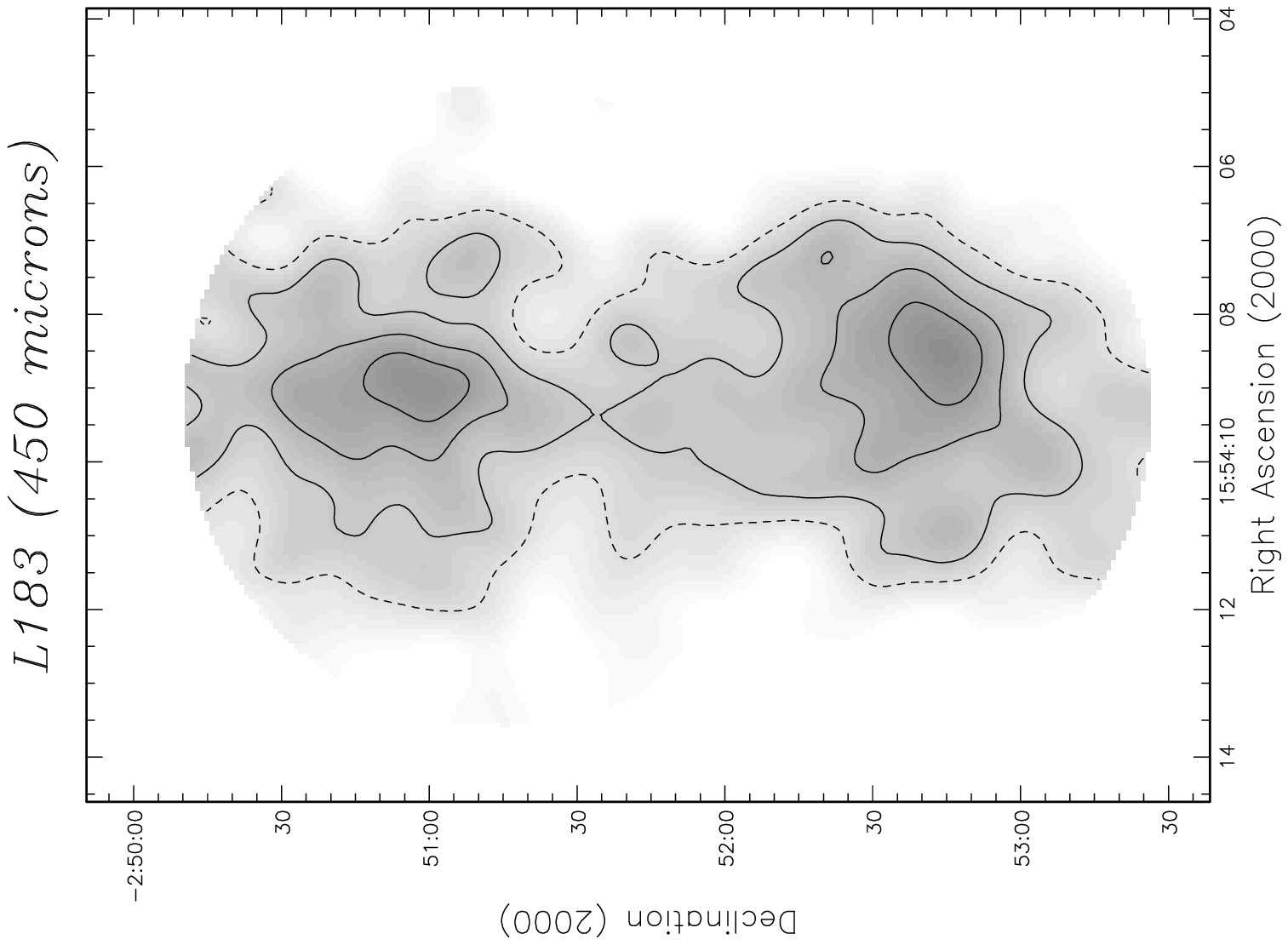}}
\put(20,165){\includegraphics{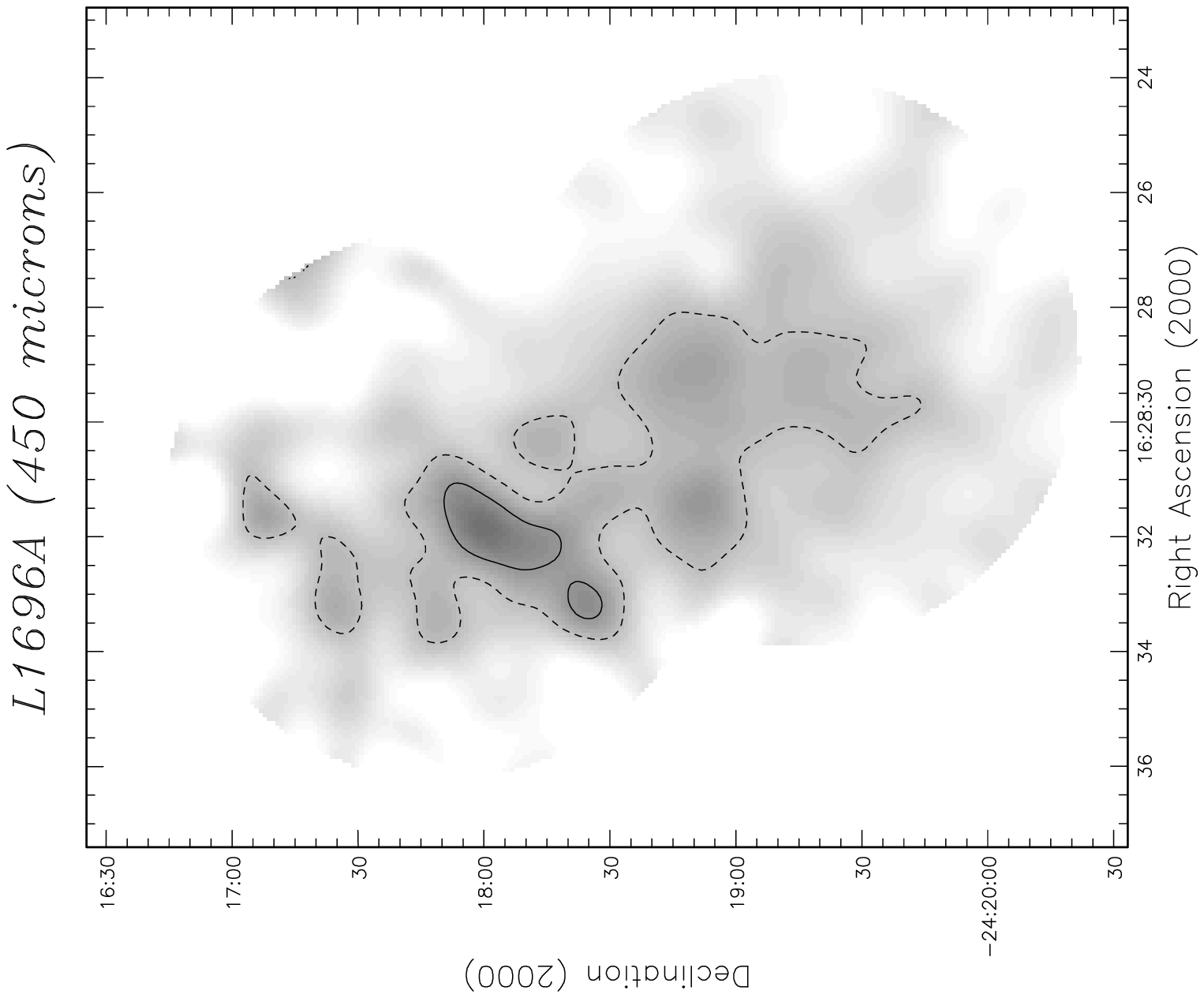}}
\put(70,165){\includegraphics{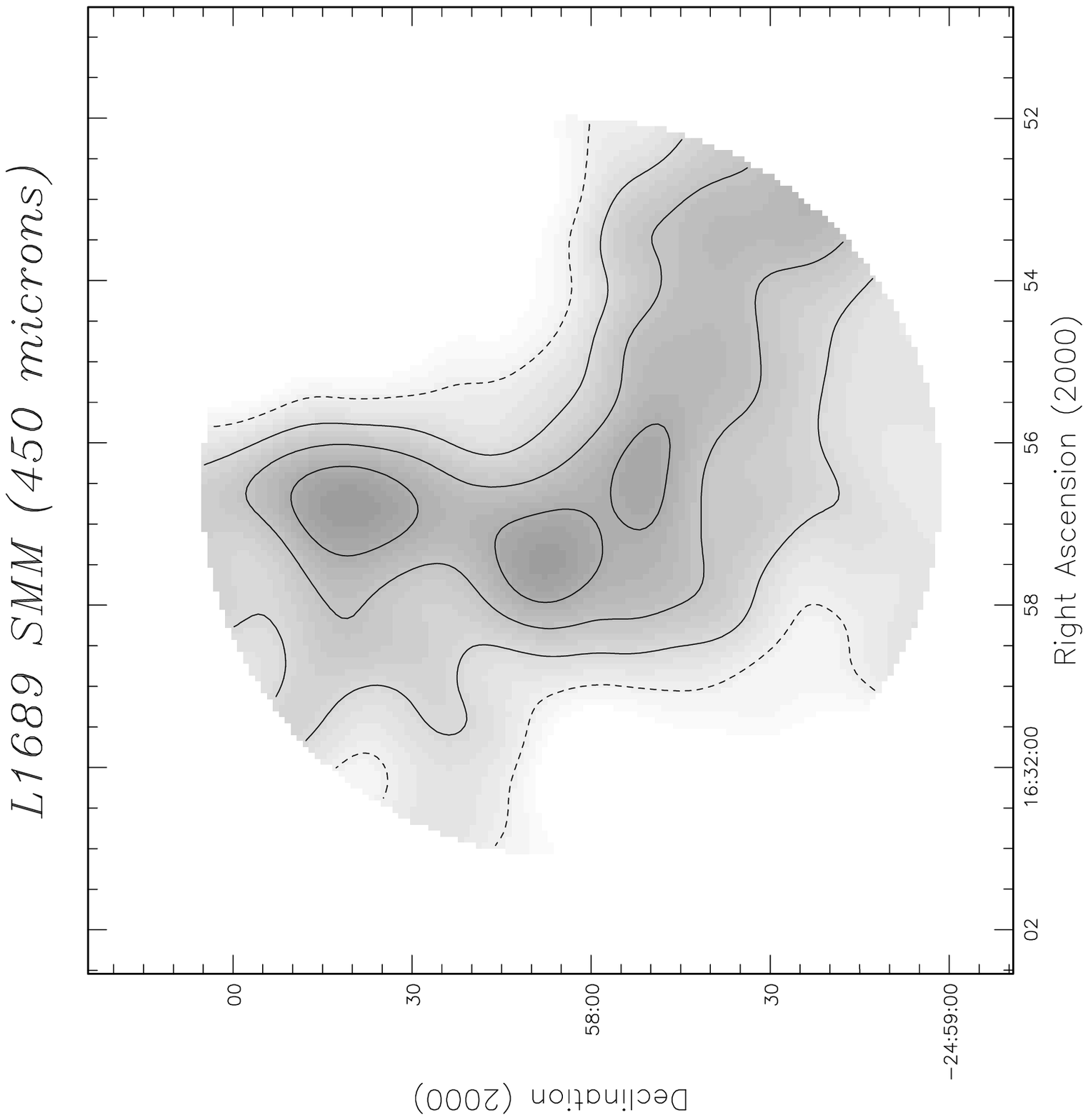}}
\put(120,165){\includegraphics{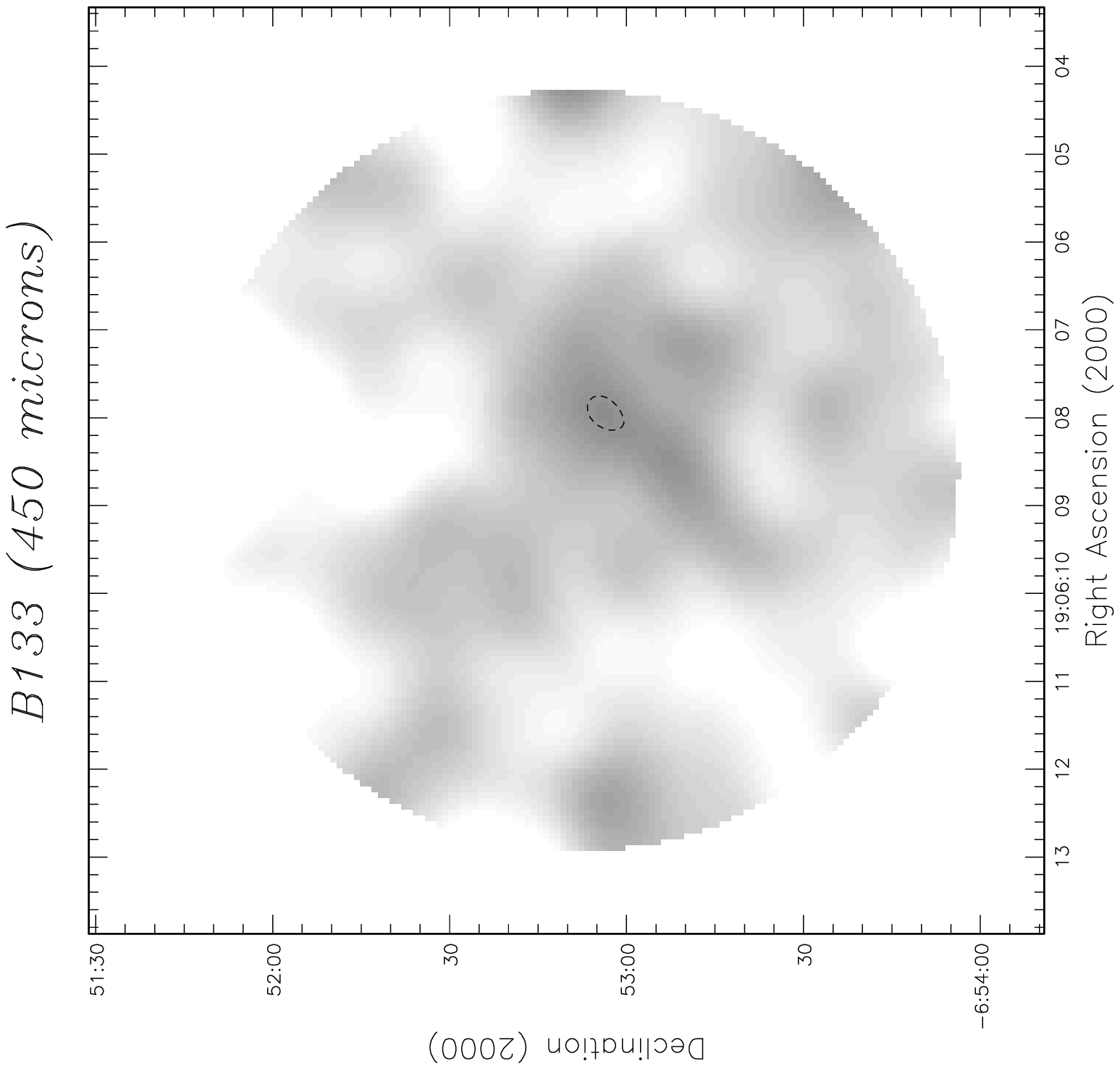}}
\put(-10,110){\includegraphics{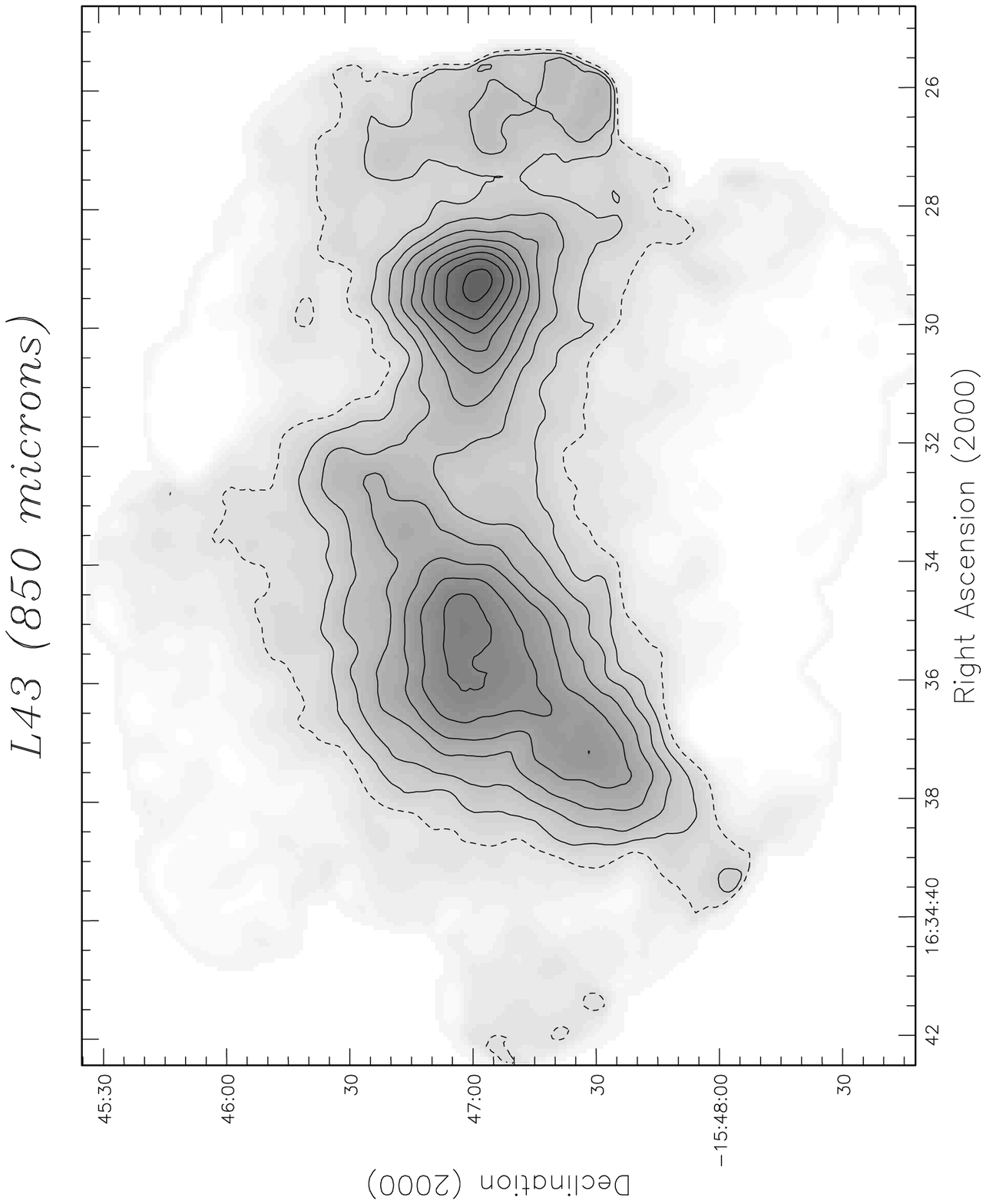}}
\put(50,110){\includegraphics{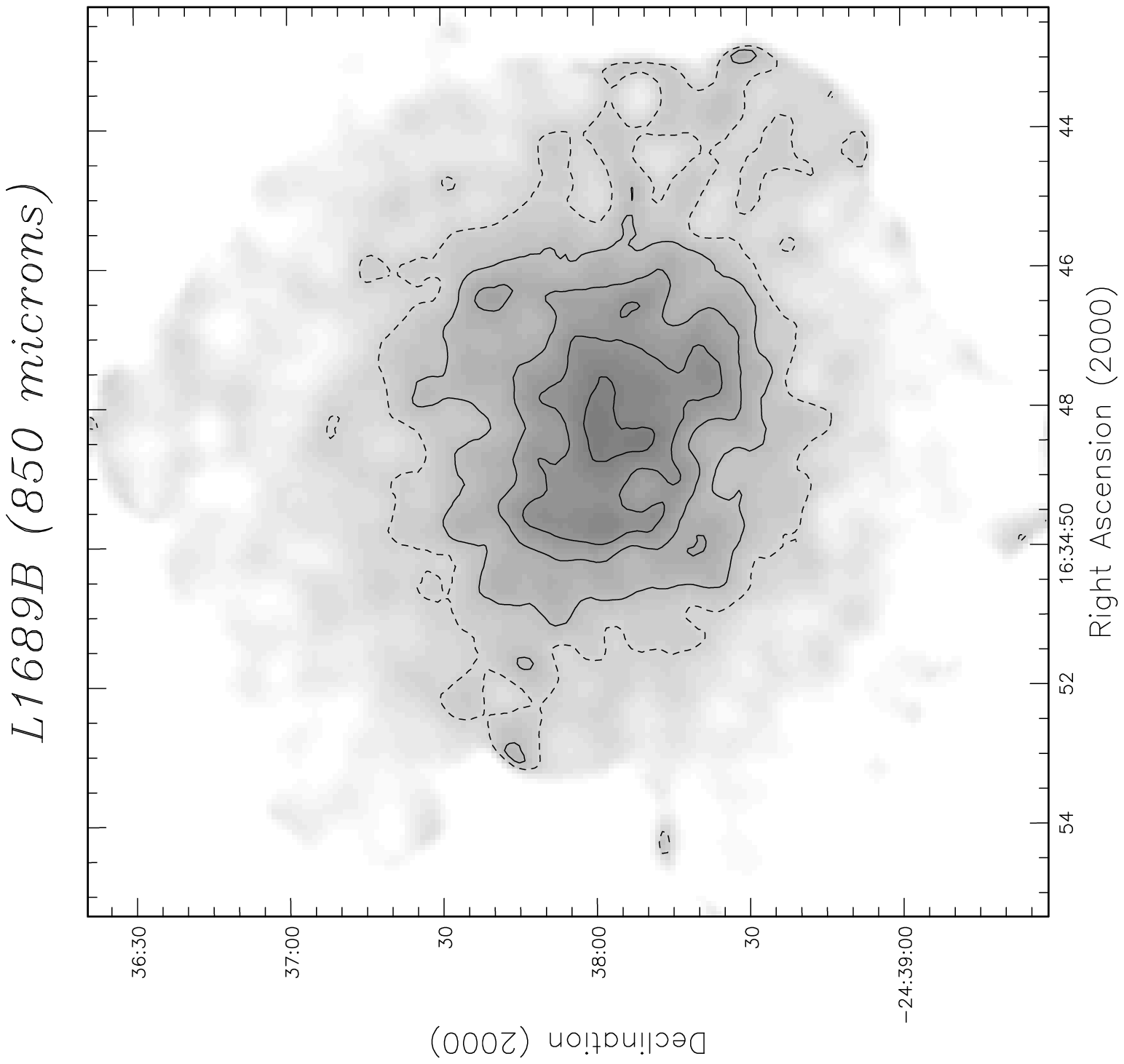}}
\put(110,110){\includegraphics{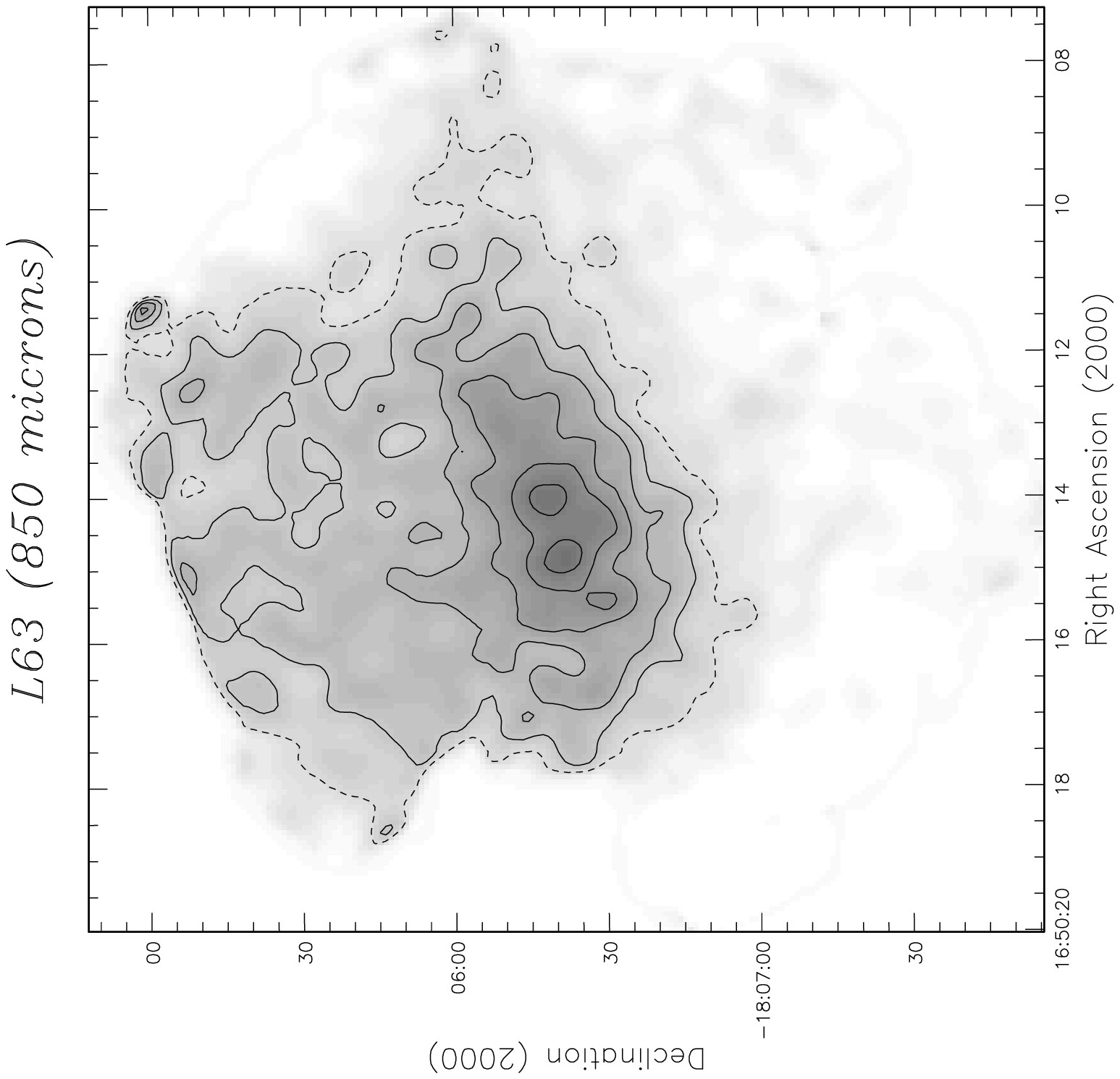}}
\put(-10,55){\includegraphics{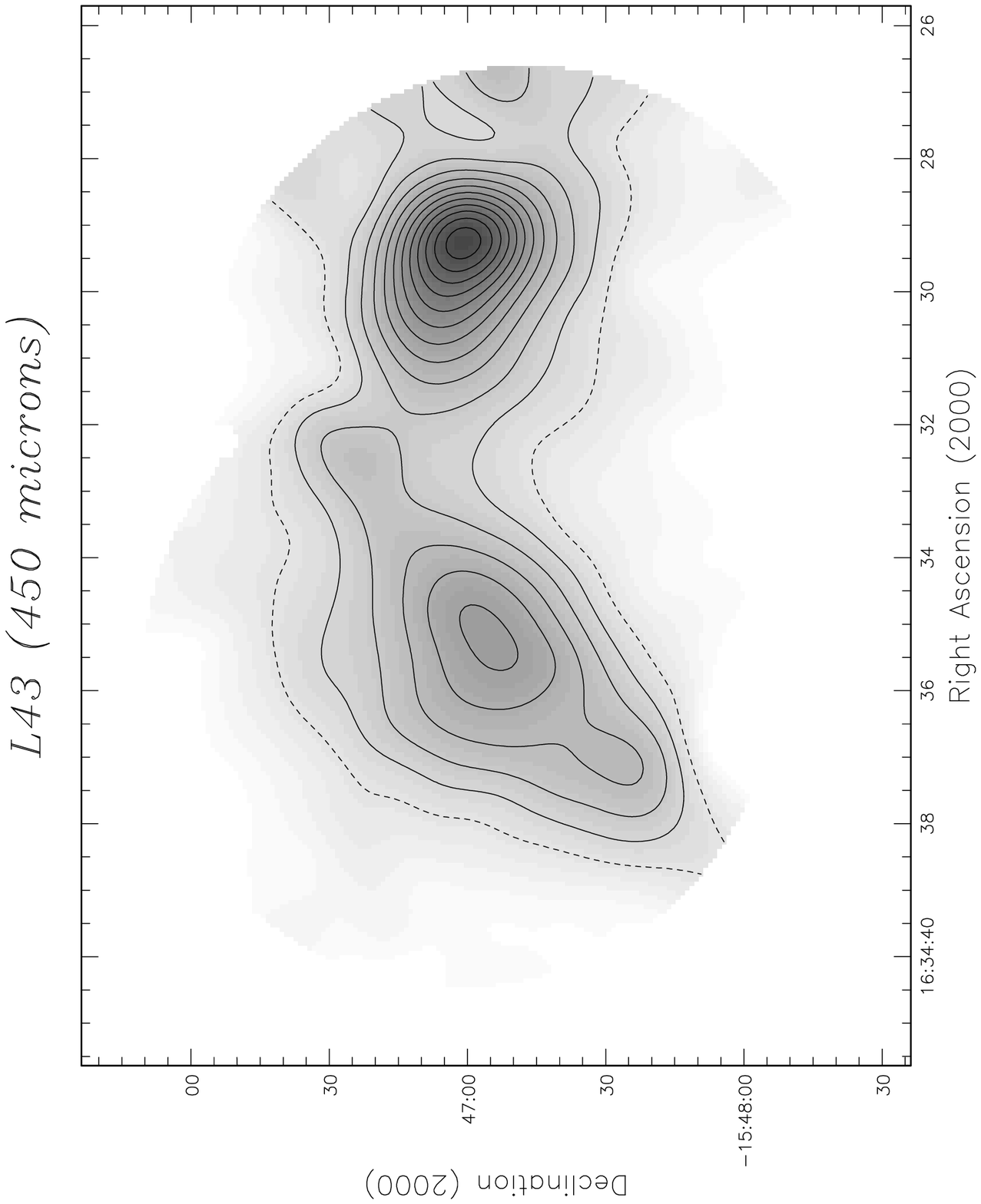}}
\put(50,55){\includegraphics{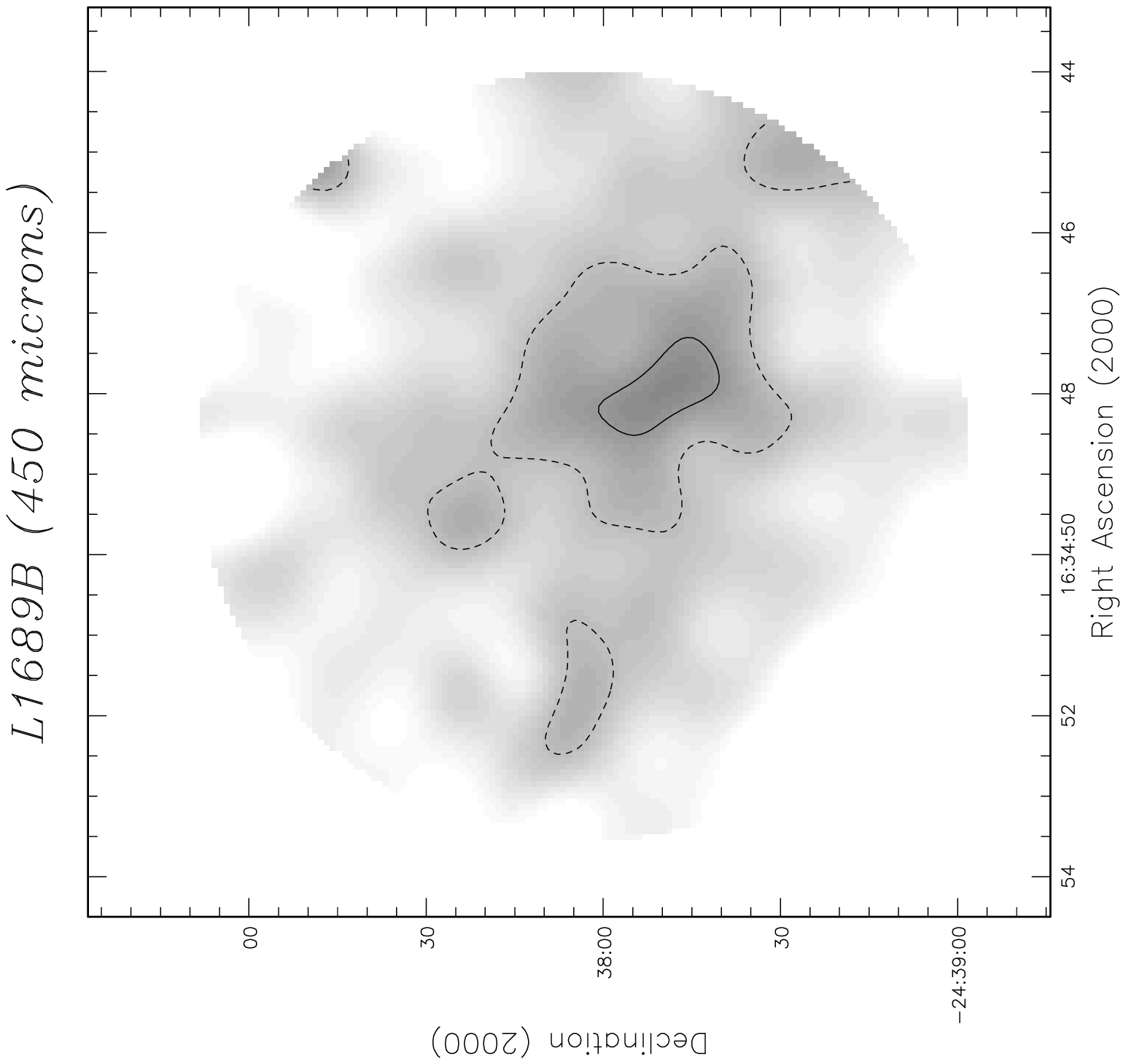}}
\put(110,55){\includegraphics{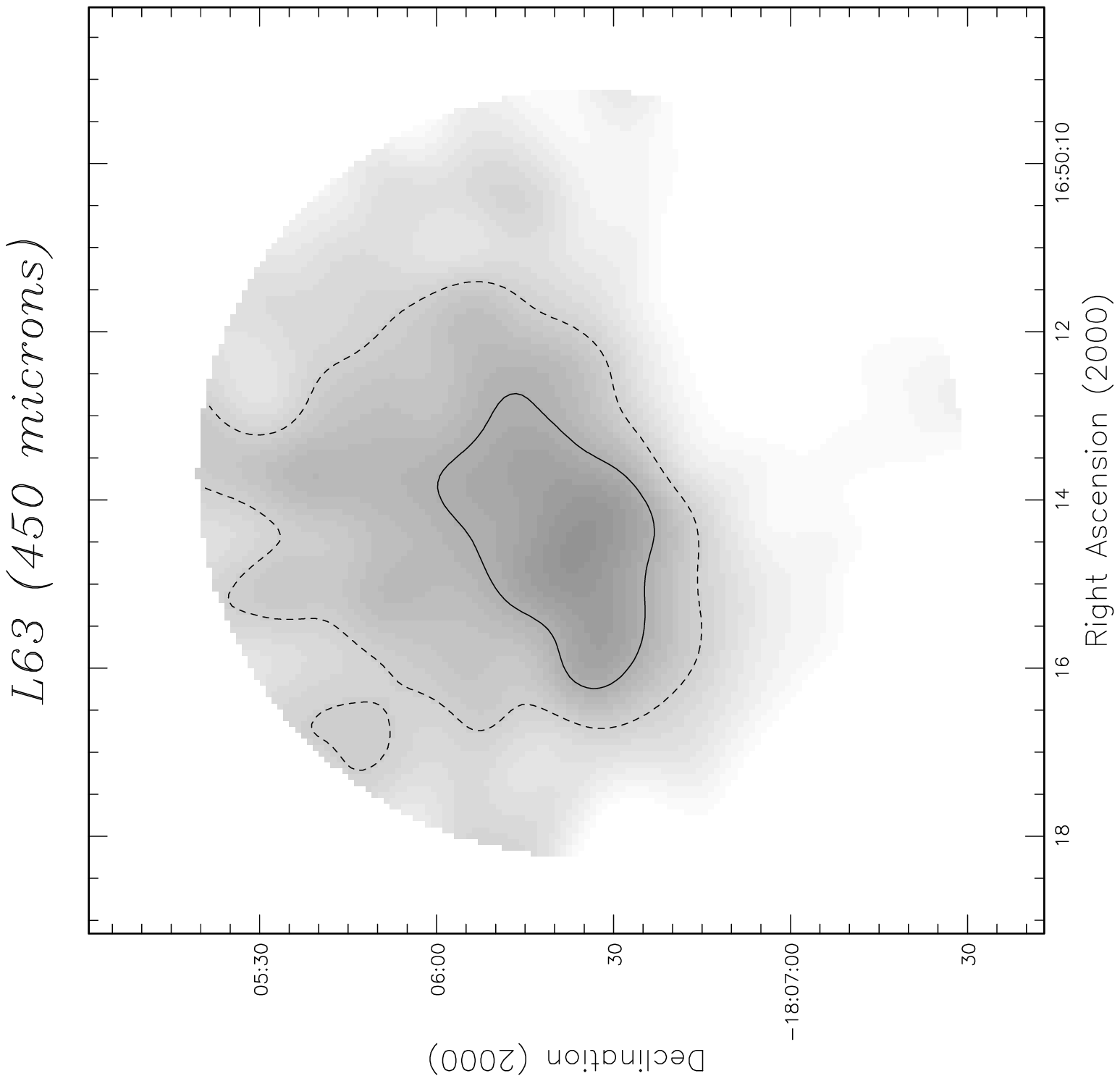}}
\end{picture}
\caption{Greyscale images with contours superposed of 850- and 450-$\mu$m
continuum maps of seven of the thirteen bright cores
-- L183, L1696A, L1689SMM, B133, L43, L1689B \& L63.
Details as in Figure~3.}
\end{figure*}

\begin{figure*}
\setlength{\unitlength}{1mm}
\noindent
\begin{picture}(170,220)
\put(-10,220){\includegraphics{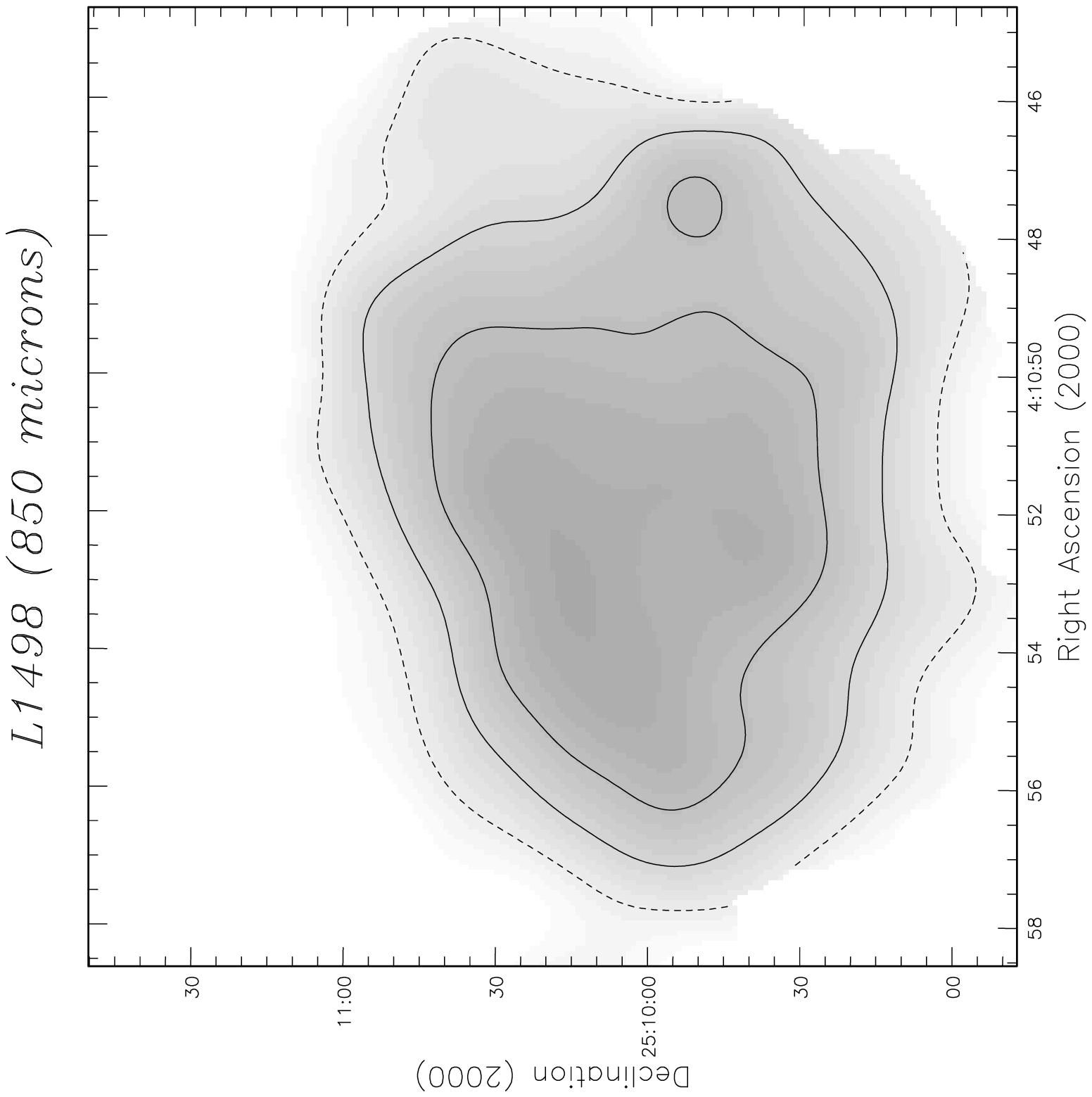}}
\put(50,220){\includegraphics{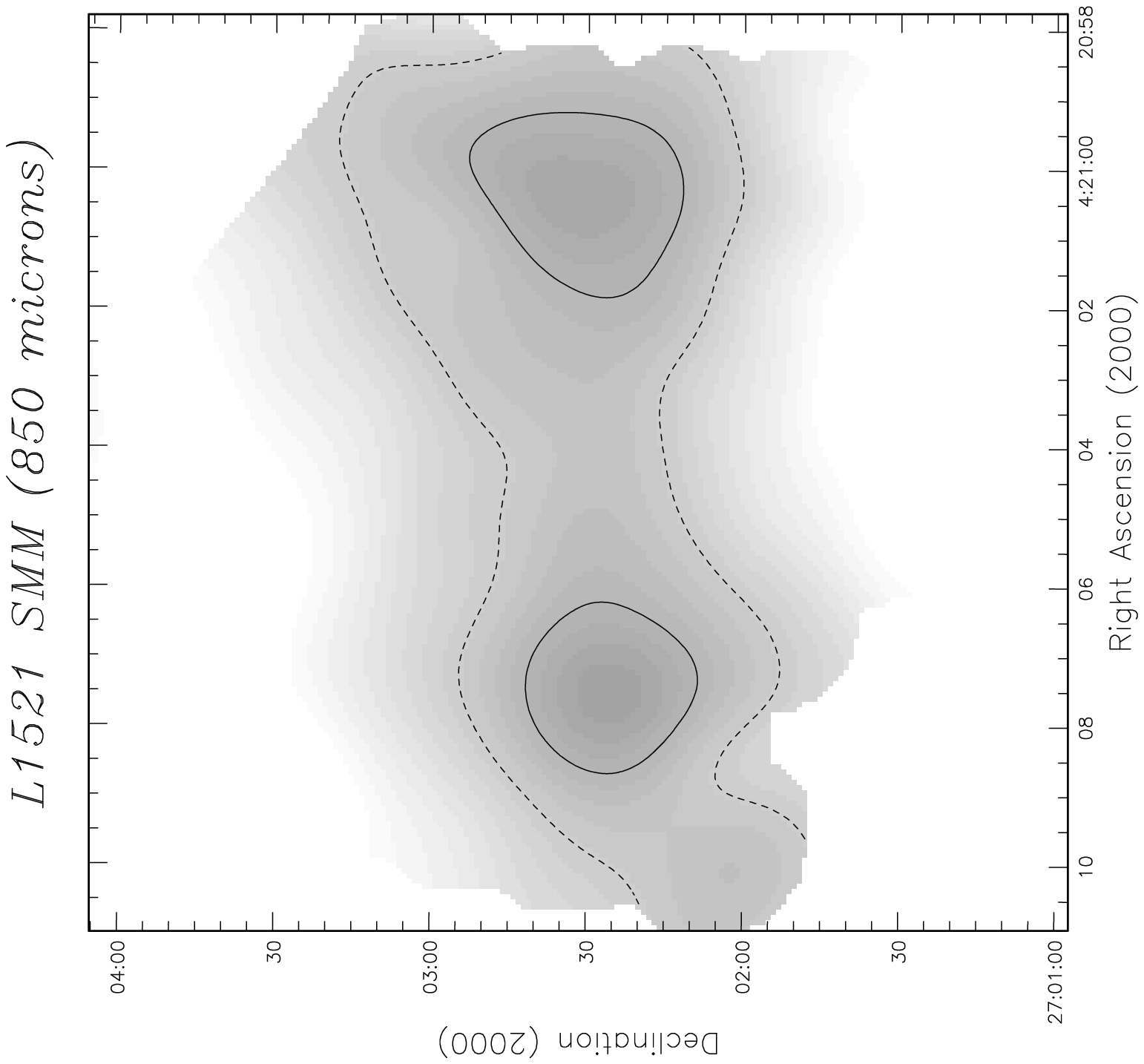}}
\put(110,220){\includegraphics{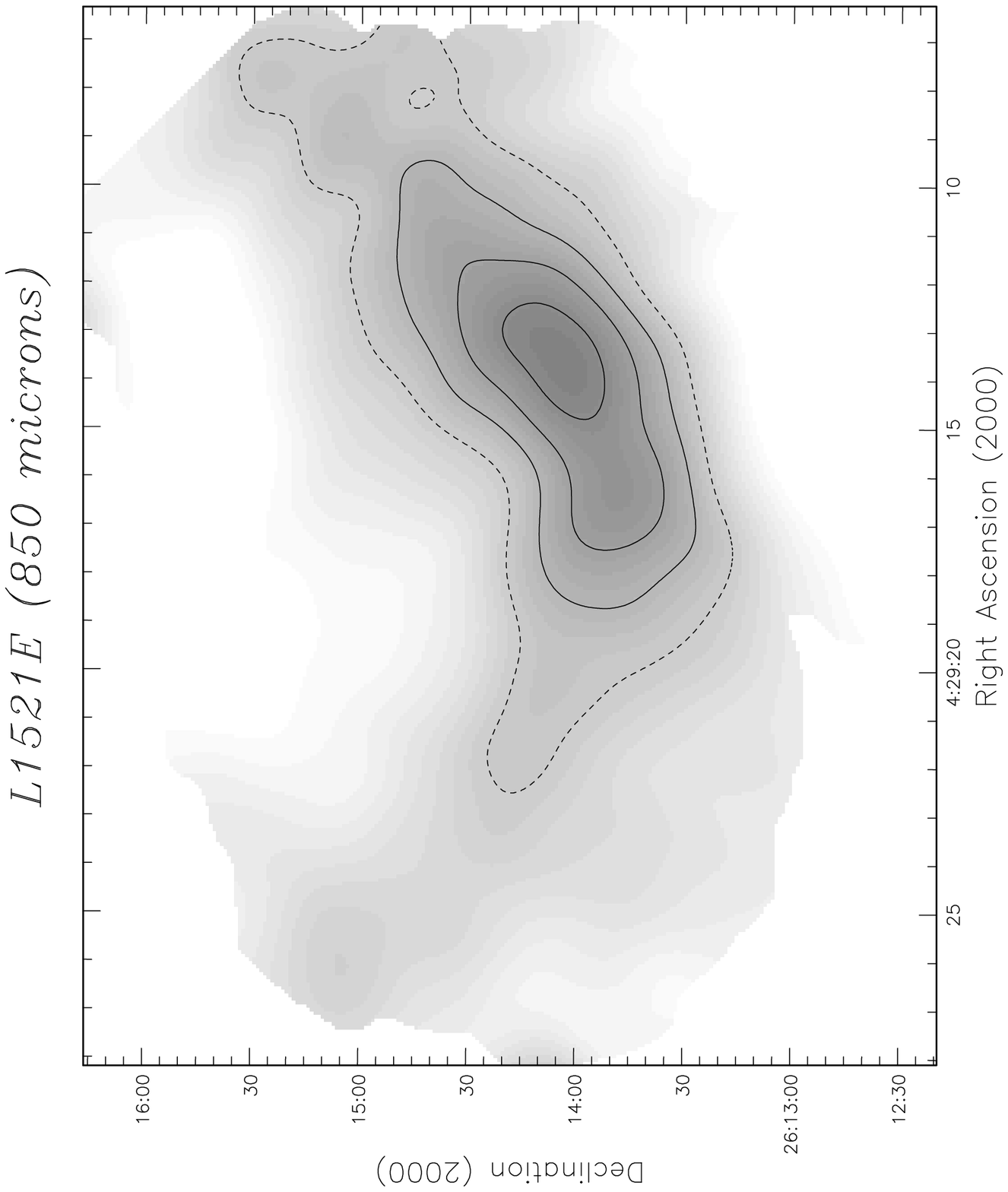}}
\put(-10, 165){\includegraphics{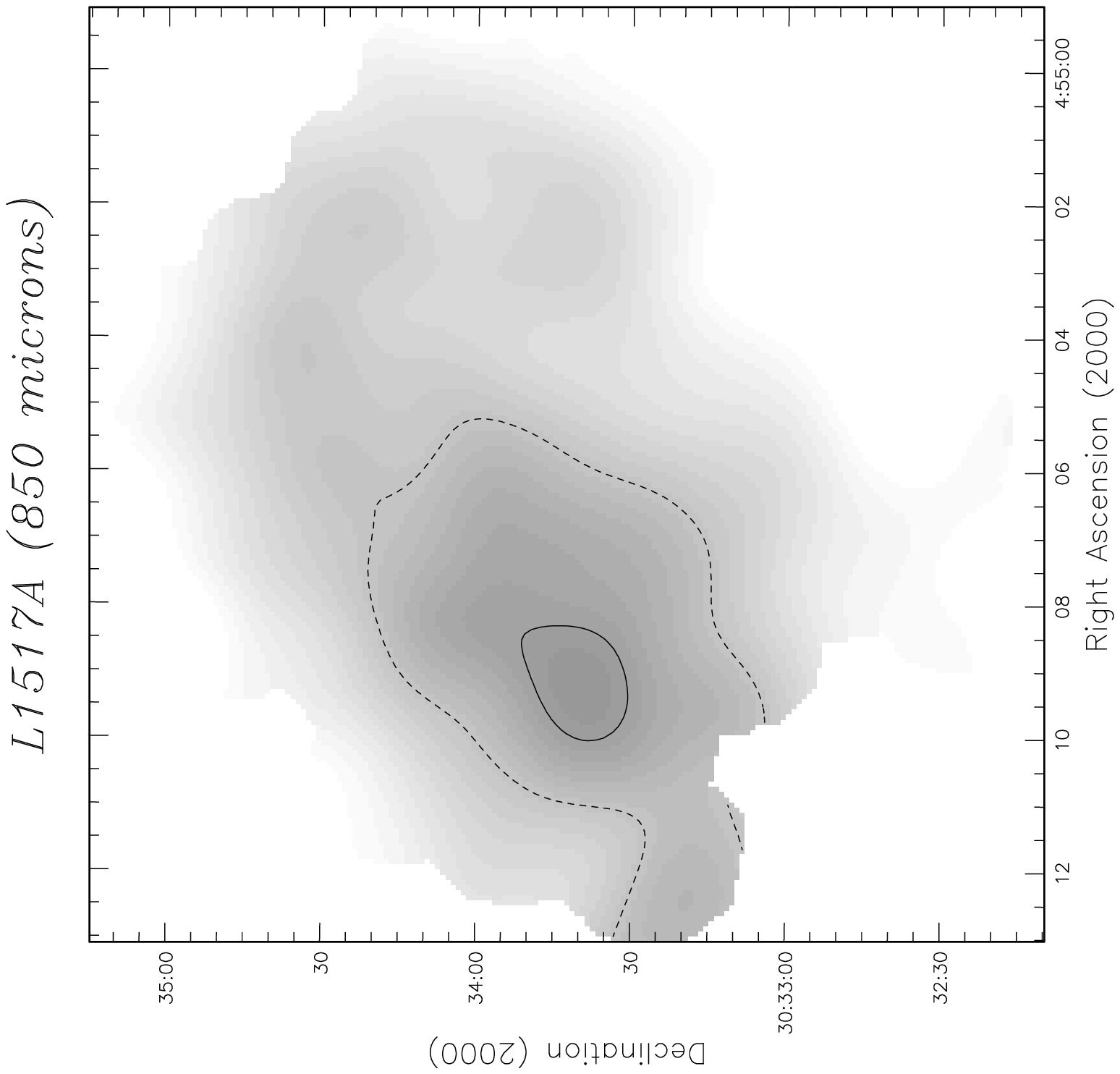}}
\put(50,165){\includegraphics{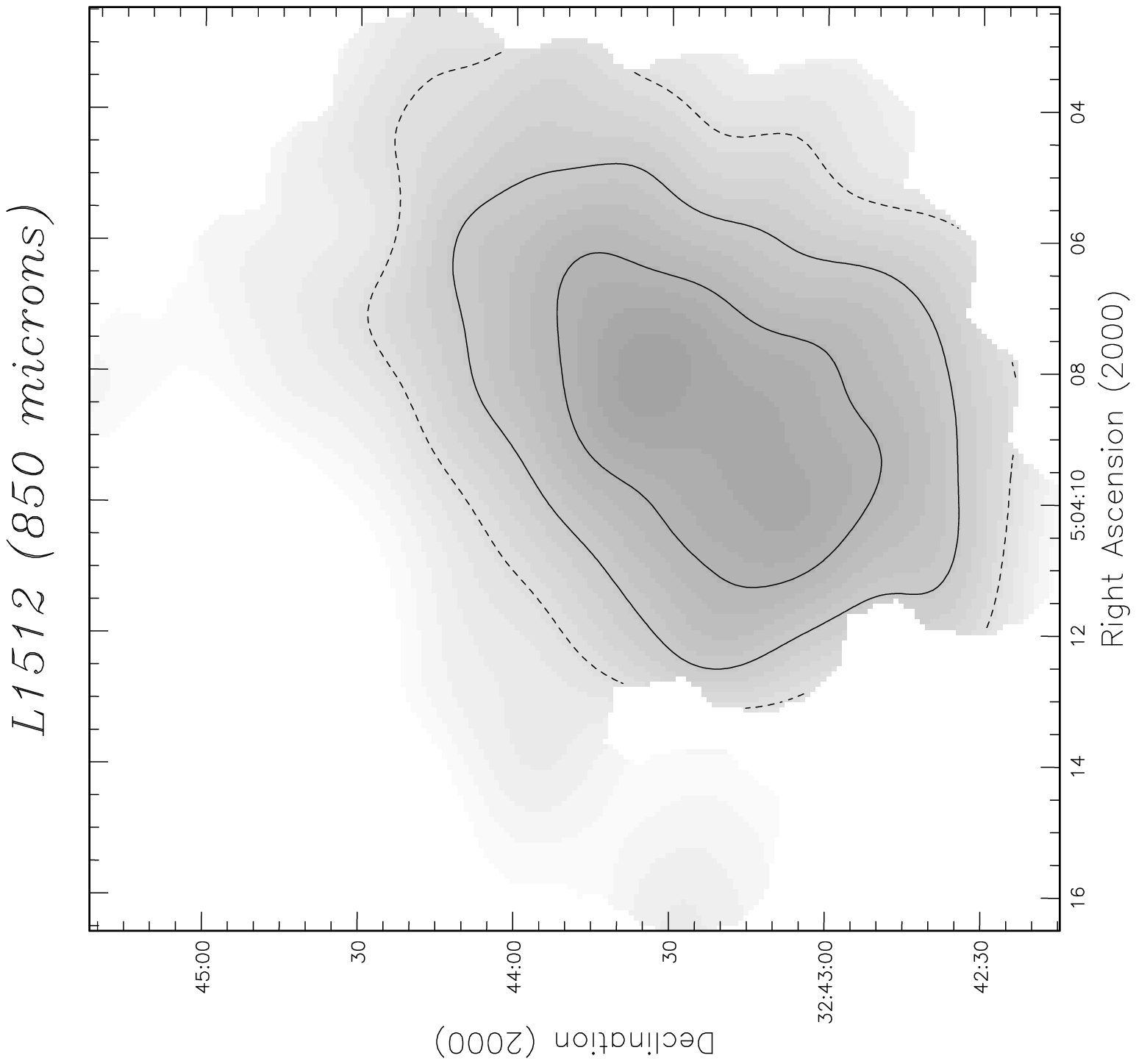}}
\put(110,165){\includegraphics{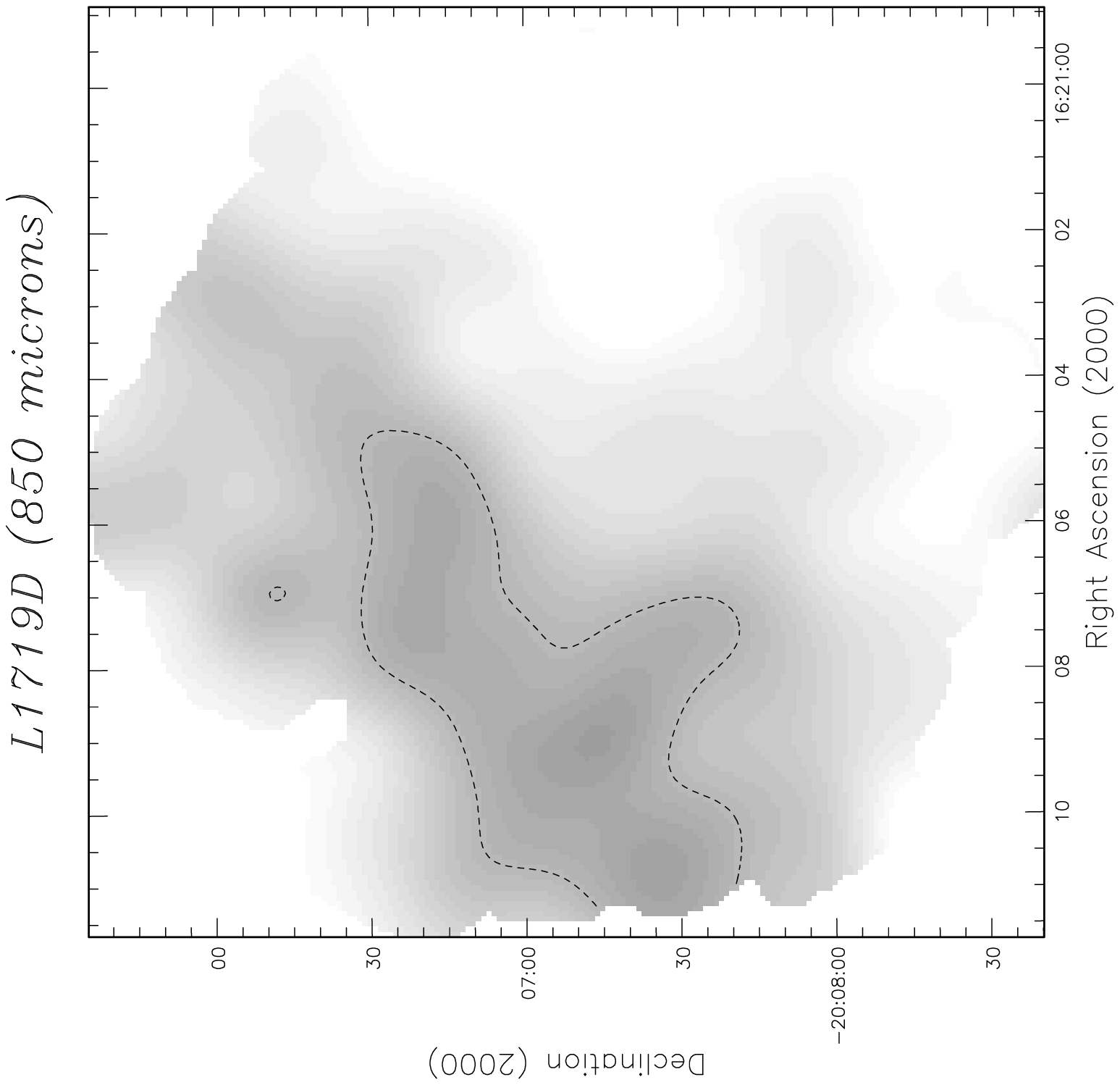}}
\put(-10,113){\includegraphics{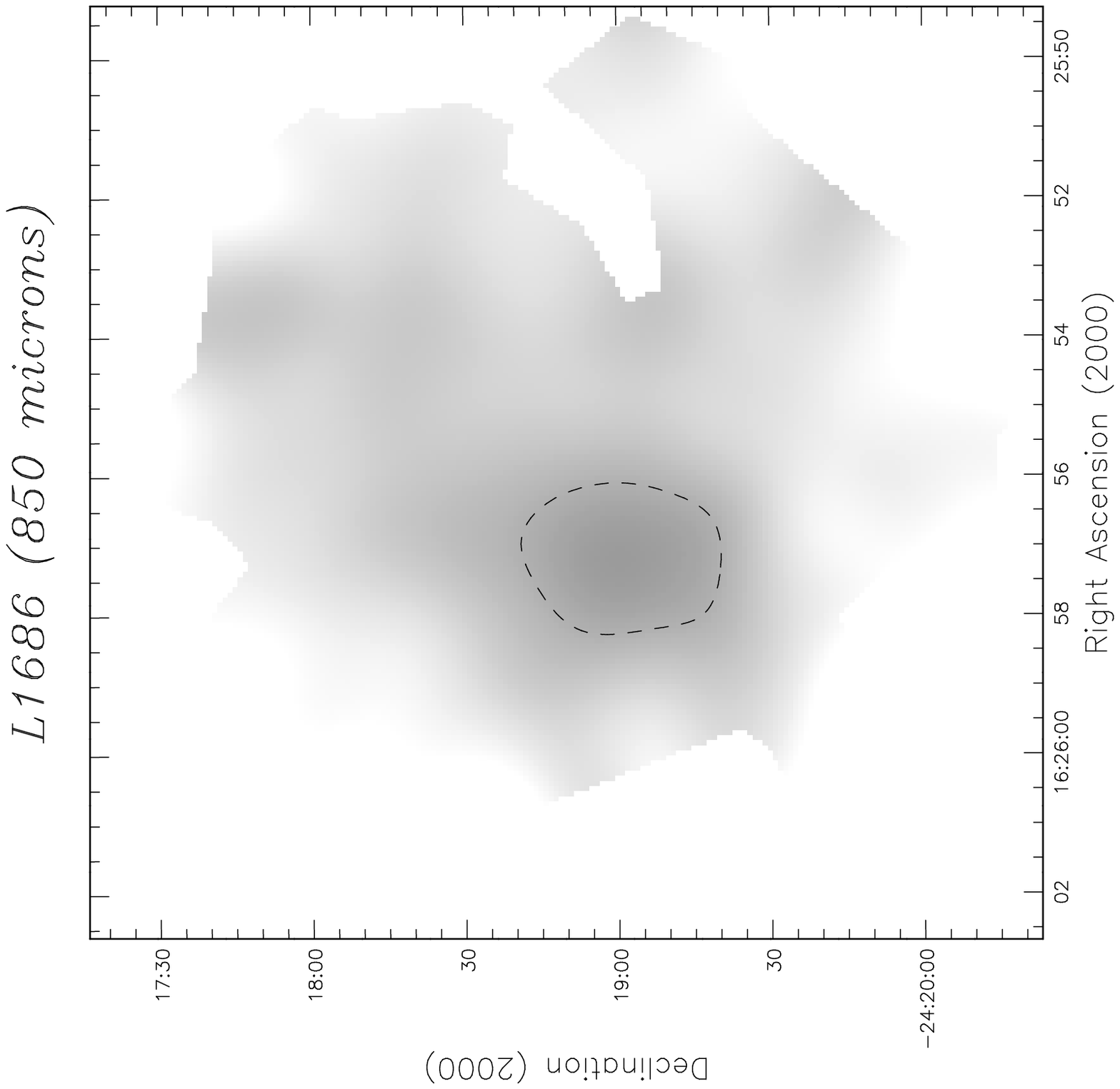}}
\put(50,110){\includegraphics{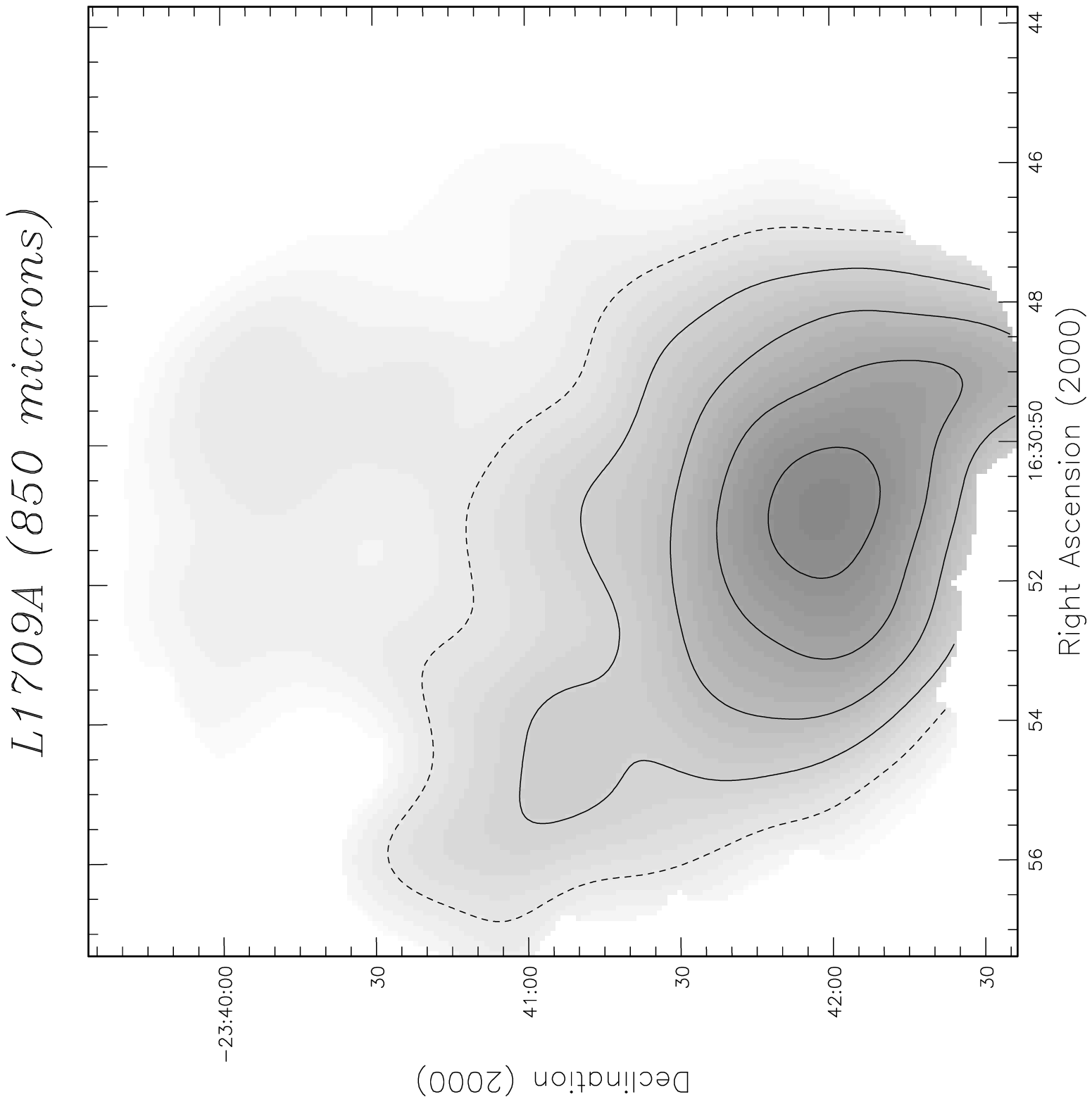}}
\put(110,110){\includegraphics{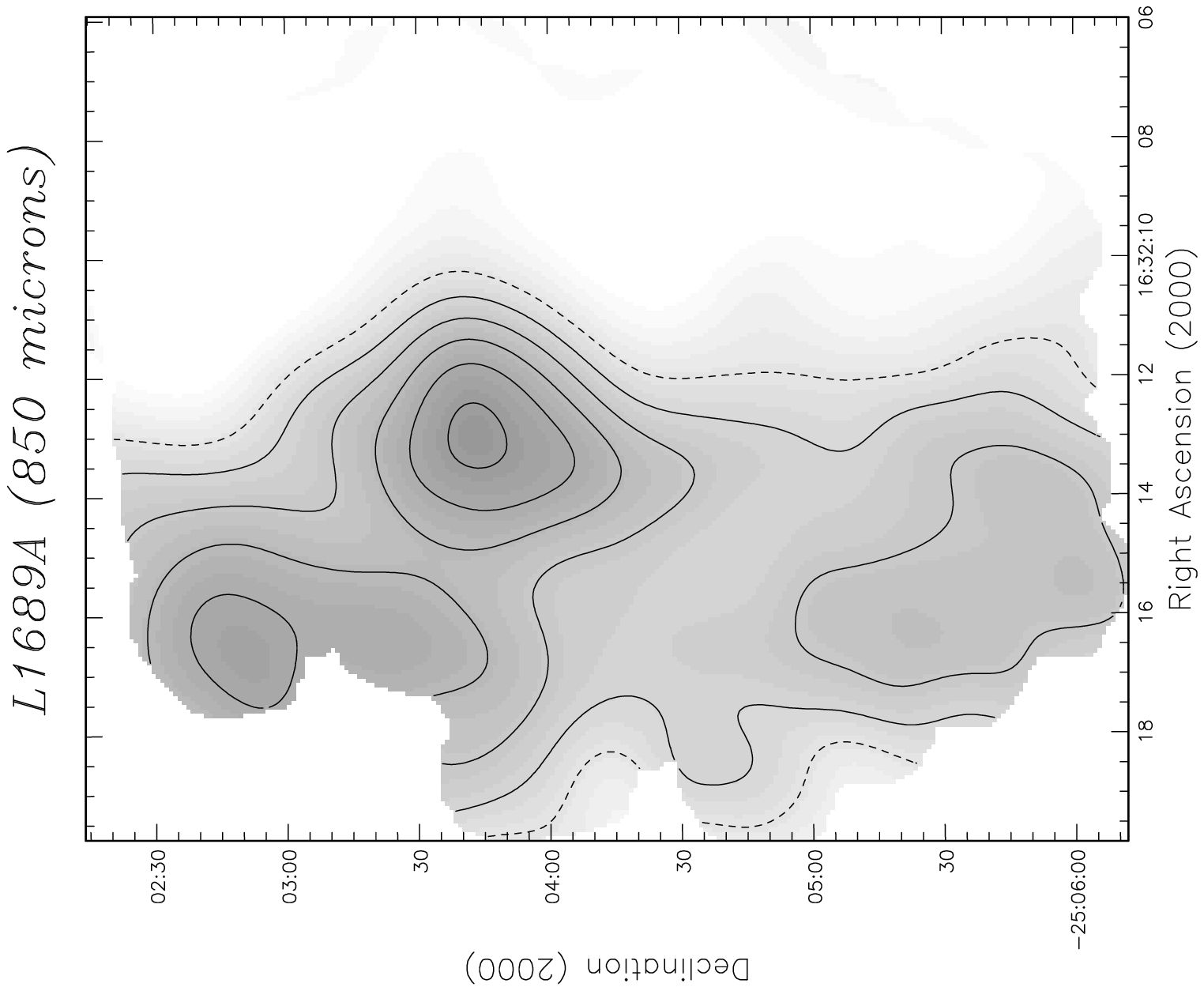}}
\put(-10,55){\includegraphics{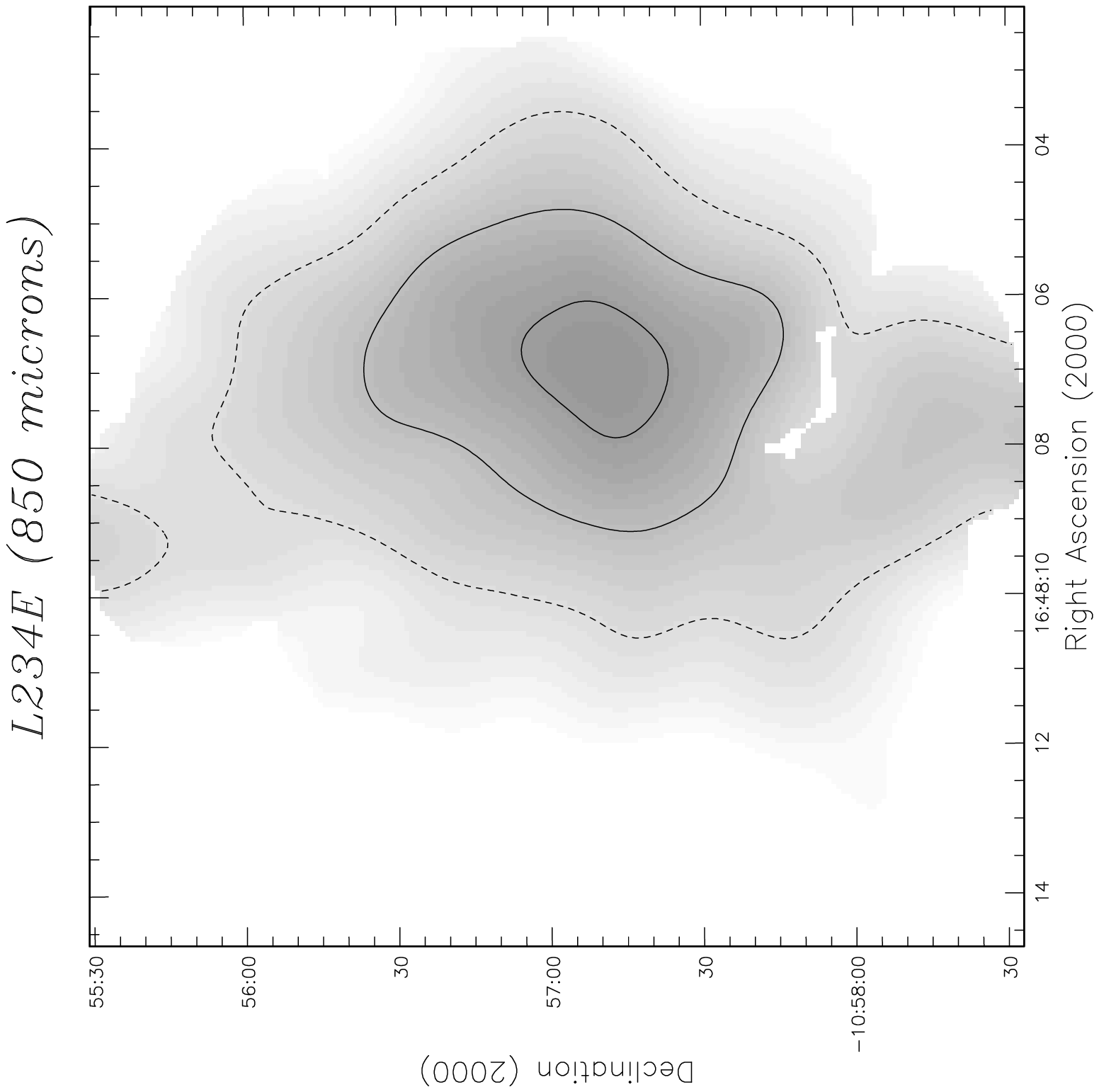}}
\put(50,55){\includegraphics{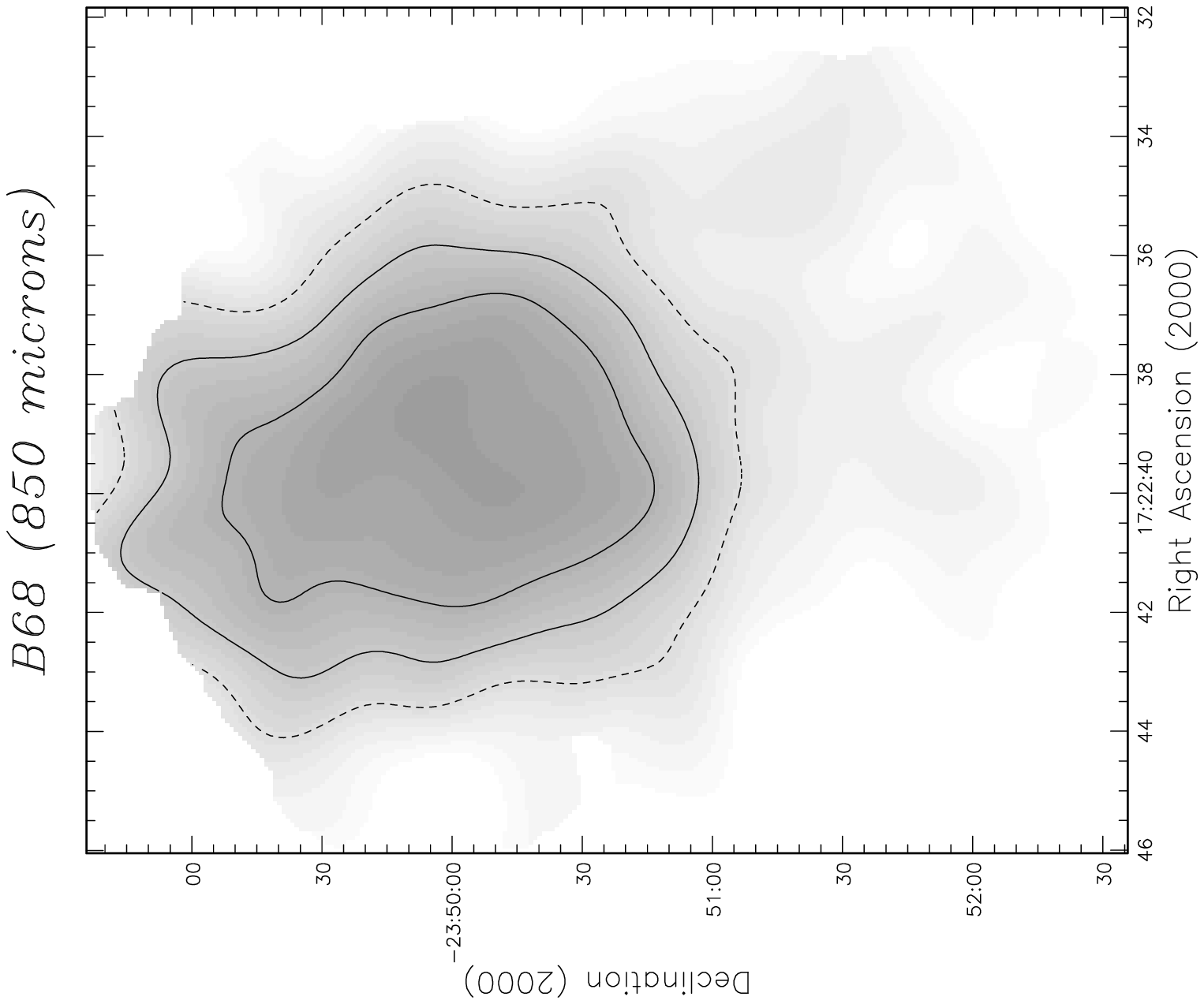}}
\put(110,55){\includegraphics{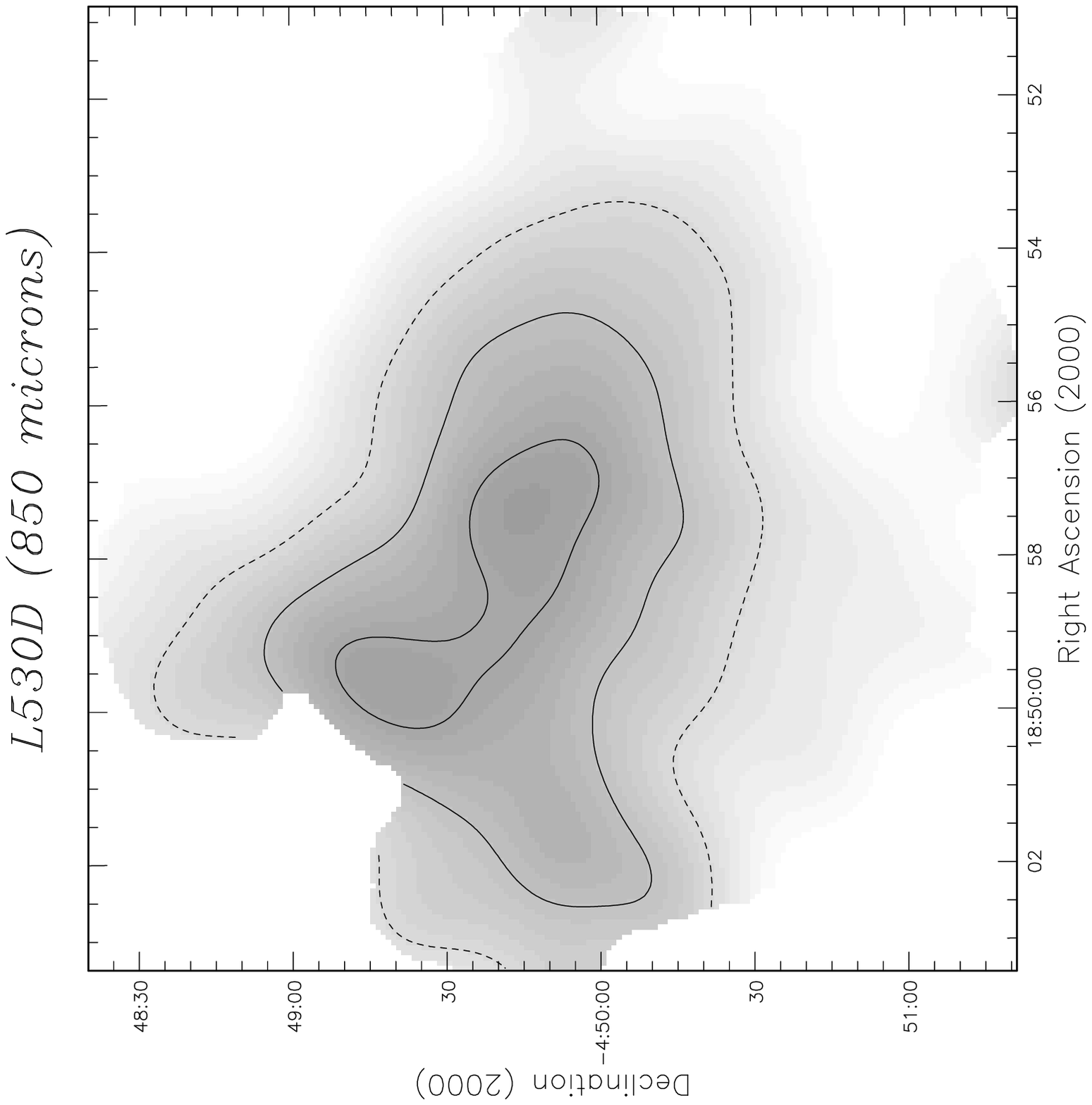}}
\end{picture}
\caption{Greyscale images with contours superposed of 850-$\mu$m
continuum maps of 12 of the 16 cores designated intermediate in brightness
at 850~$\mu$m -- L1498, L1521SMM \& E,
L1517A, L1512, L1719D, L1686, L1709A, L1689A, L234E, B68 \& L530D.
The lowest (dashed) contour in each case is at a level of 3$\sigma$.
The solid contours start at a level of 5$\sigma$, and the contour interval is
2$\sigma$. The 1$\sigma$ noise level for each source is
listed in Table 2. The data have been smoothed to a
FWHM resolution of 25~arcsec.}
\end{figure*}

\begin{figure*}
\setlength{\unitlength}{1mm}
\noindent
\begin{picture}(170,60)
\put(-30,55){\includegraphics{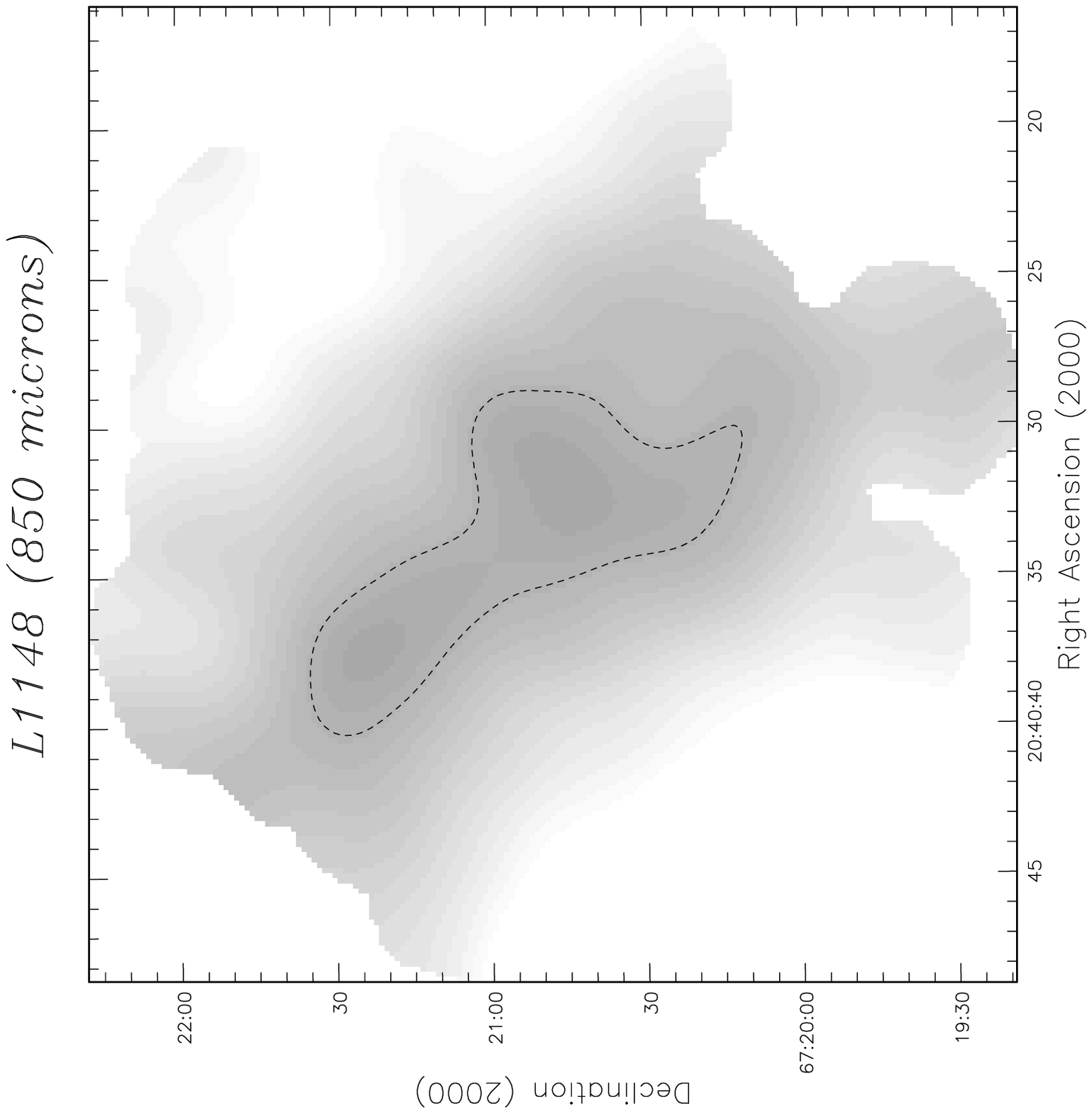}}
\put(20,55){\includegraphics{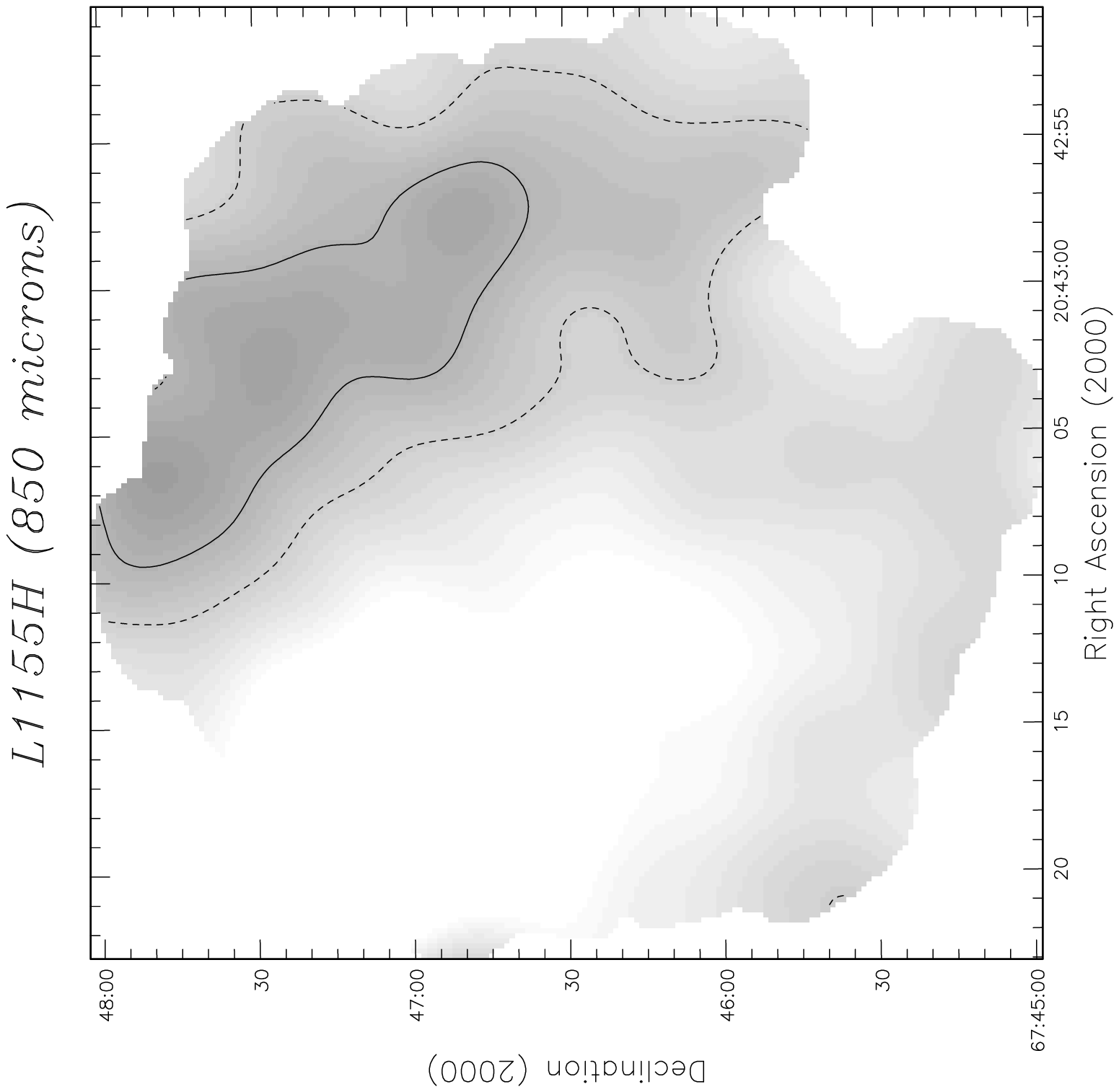}}
\put(70,55){\includegraphics{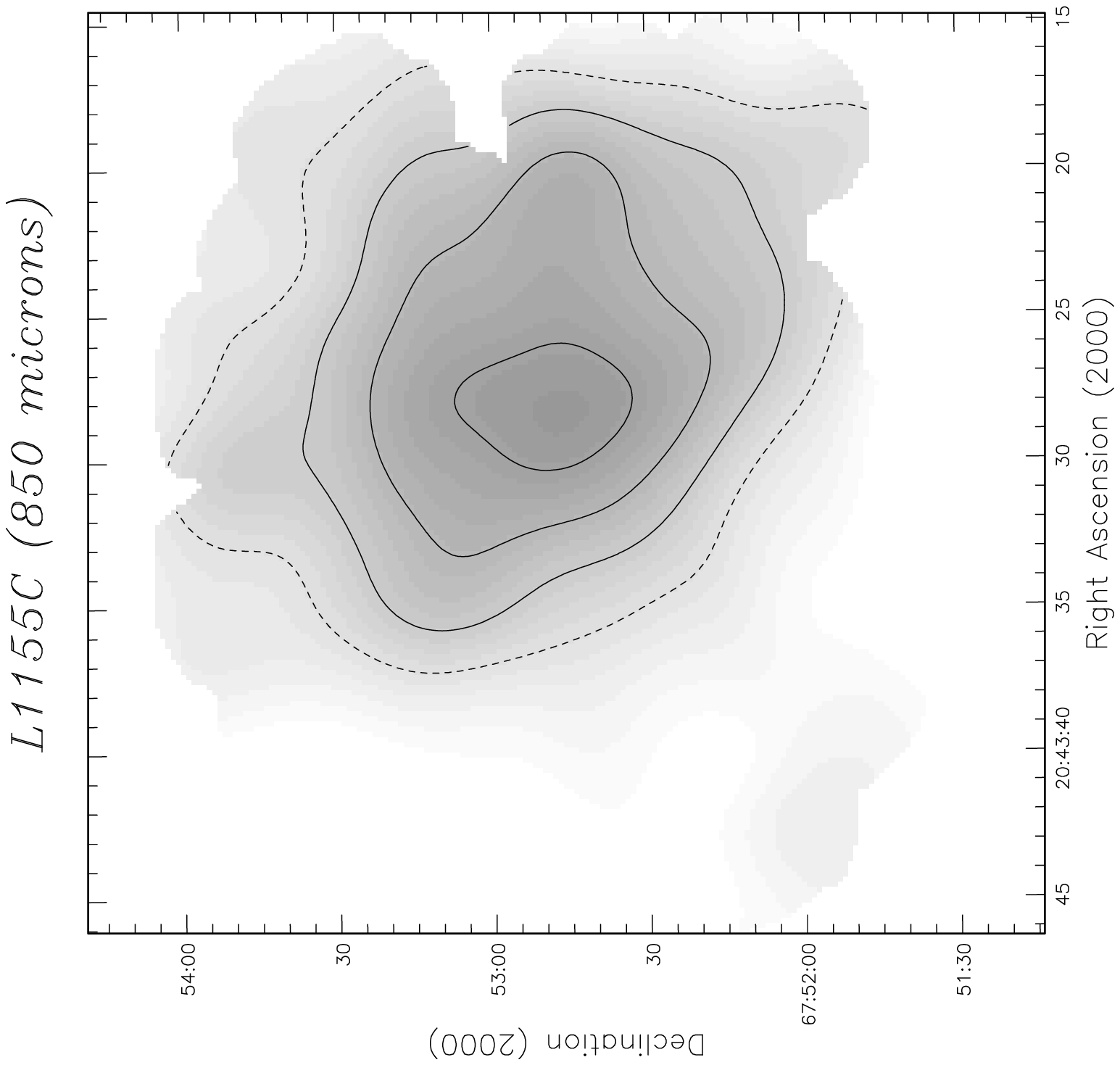}}
\put(120,55){\includegraphics{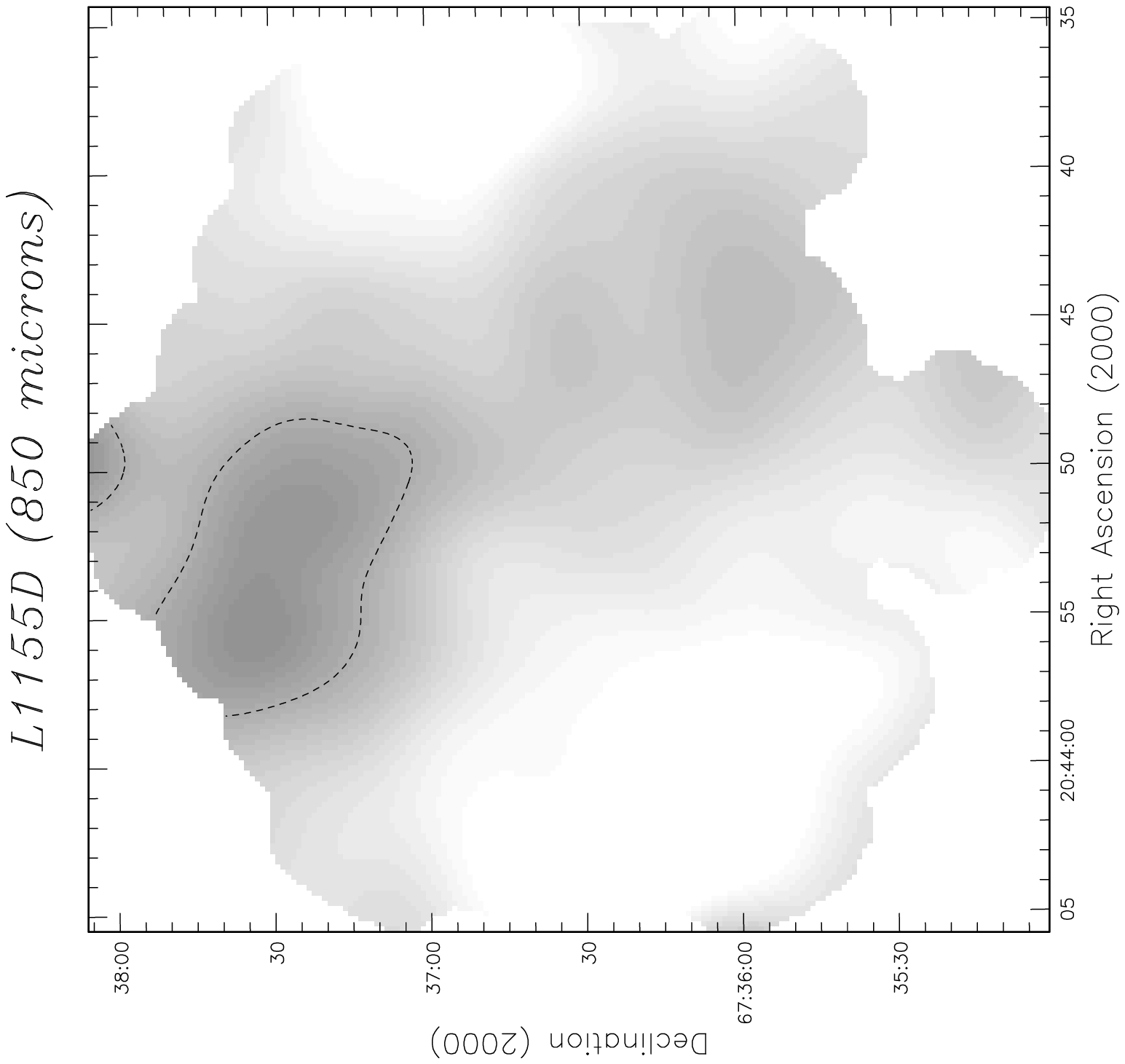}}
\end{picture}
\caption{Greyscale images with contours superposed of 850-$\mu$m
continuum maps of four of the sixteen intermediate cores
-- L1148 and L1155H, C \& D.
Details as in Figure~5.}
\end{figure*}

In nine of the cores (L1521D \& F, L1517B, L1582A, L1544, L1689SMM, L43,
L1689B \& L63) the 850- and 450-$\mu$m maps show a very similar morphology,
clearly indicating that they are tracing the same dust,
albeit with lower signal-to-noise ratio
at 450~$\mu$m. Any slight differences
can be attributed to anomalous refraction effects
(e.g. Zylka et al. 1995).
In two cases (L1524 \& B133) the core does not
appear to have been detected
significantly at 450~$\mu$m despite having relatively high signal-to-noise
ratio at 850~$\mu$m. This may be due
either to varying atmospheric conditions
during the observations, or
to anomalous refraction effects (e.g. Zylka et al. 1995).
Alternatively, in the case of L1524,
the core peak may have been missed.
In general,
if the differences between 850 and 450~$\mu$m
are real, then they must be due to changes in
dust temperature or emissivity across a core.

In the remaining 2 cores (L183 \& L1696A) the 450-$\mu$m map
appears somewhat different to the 850-$\mu$m map.
In both of these latter cases the approximate overall size appears the
same at both wavelengths and the outer contours appear to trace roughly
the same material, but they
appear to disagree somewhat over the peak of the core. L183 has only one
peak at 850~$\mu$m at the southern end of the core, but has apparently
two peaks at 450~$\mu$m -- one coincident with the 850-$\mu$m peak, and the
other at the northern end of the core. For L1696A two peaks are seen at
850~$\mu$m, but only one at 450~$\mu$m. Furthermore, the 450-$\mu$m peak
coincides with the slightly weaker of the two 850-$\mu$m peaks.

Figures 5 \& 6 show the 850-$\mu$m maps of the 16 cores we designated
as `intermediate', due to their somewhat
lower levels of peak emission (although for a few of the cores
the peak may have been missed).
The data in Figures 5 \& 6 have been smoothed to an effective FWHM
resolution of 25 arcsec. Consequently, some of the structure has been
smoothed out of the maps. Nonetheless, in some maps there is some
remaining structure that can be discerned. For example, L1521SMM can still
be seen to have two peaks, and 10 of the cores (L1498, L1521E, L1517A, L1512,
L1719D, L1709A, L1689A, L530D, L1148
\& L1155H) can still be seen to have some elongation.
In all of the designated intermediate sources there was
no structure discernible in the 450-$\mu$m maps after smoothing, even
for the sources detected above the 5$\sigma$ level on peak at 450~$\mu$m.

The majority of the cores shown in Figures 3 \& 4 can be seen
to be non-circular.
Consequently, we fitted an elliptical two-dimensional
Gaussian to each of the
cores. This allowed us to measure the full-width
at half-maximum (FWHM) of the major and minor axis of each core, and to
ascertain the position angle (PA) of the major axis in each case. The
values of FWHM and PA for each core are listed in Table 4, as well as
the ratio of the two FWHM values, which we shall refer to as the aspect
ratio. From our sample of bright cores,
those with double peaks or with a potentially missed peak
(L1524, L1696A and L43) were excluded
from this treatment. Furthermore, L1689SMM was seen to be too `ridge-like'
to determine a one-dimensional radial profile.
For the remaining nine cores listed in Table 4 the weighted
mean aspect ratio is 0.62$\pm$0.15 (c.f. Goodwin et al 2002).

\section{Core properties}

\subsection{Core morphologies and comparison with ISO data}

In Paper V we presented data from the Infrared Space Observatory ISO
(Kessler et al. 1996),
using the long-wavelength photo-polarimeter ISOPHOT (Lemke et al. 1996),
of 18 prestellar cores that were
a subset of the 52 cores discussed in this paper. We can therefore
compare the far-infrared observations in Paper V
with the submillimetre observations presented here.
All of the 18 cores observed with ISOPHOT were detected at
both 170 \& 200~$\mu$m, but most were undetected at 90~$\mu$m,
indicating that they are cold (Paper V).

We note that of the 18 cores observed with ISOPHOT, 8 have been
classified in this paper as bright cores in terms of their SCUBA
850-$\mu$m flux densities -- L1517B, L1544, L1582A, L183, L1696A,
L1689B, L63 \& B133. 7 of the 18 cores appear in the intermediate
brightness category -- L1498, L1517A, L1512, L1709A, L1689A, B68 \& L1155C.
The remaining three cores were undetected by SCUBA -- L1517C, L1709C \&
L204B. Consequently, the ISOPHOT subsample represents a cross-section
of the cores. Note that the typical flux densities of the cores at
200~$\mu$m are significantly greater than their flux densities at
850~$\mu$m simply due to the spectral shape of cores at the typical
temperatures observed (see below).

\begin{figure*}
\setlength{\unitlength}{1mm}
\noindent
\begin{picture}(170,220)
\put(-20,230){\includegraphics{fig7a.ps}}
\put(80,230){\includegraphics{fig7b.ps}}
\put(-20,160){\includegraphics{fig7c.ps}}
\put(80,160){\includegraphics{fig7d.ps}}
\put(-20,90){\includegraphics{fig7e.ps}}
\put(80,90){\includegraphics{fig7f.ps}}
\end{picture}
\caption{Composite images of SCUBA 850-$\mu$m maps (bold white contours)
and ISOPHOT 200-$\mu$m maps (narrower grey contours)
superposed on POSS optical (grey-scale) images
for 6 regions -- L1498, L1517, L1512, L1544, L1582A \& L183.
The dashed contours designate the edge of the ISO mapped regions.
Each field is 16 arcmin square except for L1517.}
\end{figure*}

\begin{figure*}
\setlength{\unitlength}{1mm}
\noindent
\begin{picture}(170,220)
\put(-20,230){\includegraphics{fig8a.ps}}
\put(80,230){\includegraphics{fig8b.ps}}
\put(-20,160){\includegraphics{fig8c.ps}}
\put(80,160){\includegraphics{fig8d.ps}}
\put(-20,90){\includegraphics{fig8e.ps}}
\put(80,90){\includegraphics{fig8f.ps}}
\end{picture}
\caption{Composite images of SCUBA 850-$\mu$m maps (white bold contours)
and ISOPHOT 200-$\mu$m maps (grey narrower contours)
superposed on POSS optical (grey-scale) images
for 6 cores -- L1696A, L1155C, L1689A \& B, B68 \& B133.
The dashed contours indicate the edge of the SCUBA mapped regions.
Each field is 16 arcmin square.}
\end{figure*}

Of the three non-detections, L1517C was seen to be the weakest core
at 200~$\mu$m, implying that the SCUBA sensitivity limit may not have
been sufficient to detect it. However, L1709C and L204B were not among
the faintest cores at 200~$\mu$m, possibly indicating that we may have
missed their peaks with SCUBA. ISOPHOT had a much lower resolution
at 200~$\mu$m ($\sim$90~arcsec) than SCUBA at 850~$\mu$m ($\sim$14~arcsec),
allowing us to make much larger maps of typically 10--20 arcmin
with ISOPHOT (see Paper V).

For the sources that we have here classified as
bright or intermediate cores we can compare the morphologies of the SCUBA
850-$\mu$m maps with the ISOPHOT 200-$\mu$m maps. Figures 7 \& 8
show the SCUBA and ISOPHOT contour maps for these sources
superposed on POSS optical grey-scale images of the same regions.
We exclude L1709A from further consideration in this paper as
its peak may have been missed. We also exclude L63 from this analysis
as its 200-$\mu$m map shows that its far-IR peak may have been missed
(see Paper V).

Study of Figures 7 \& 8 shows that
in each case we see a strong coincidence of peak 200-$\mu$m emission
with peak 850-$\mu$m emission and peak optical obscuration.
In most cases the morphology of the far-infrared emission also
closely traces the optical obscuration. For example, for 10 of the cores
in Figures 7 \& 8, the peak 200-$\mu$m emission and 850-$\mu$m emission
overlie one another precisely. Furthermore, the elongation
direction seen in the 200-$\mu$m emission in L1582A exactly matches the
elongation in optical obscuration in the same region. Similarly, the
850-$\mu$m emission in L1696A lines up with the edge of the optical
obscuration. However, in this case it is not quite so clear where the
200-$\mu$m peak lies, as the source seems quite extended. Hence we also
exclude this from further analysis (we note parenthetically that this
core is the closest to the cluster-forming core of $\rho$ Oph A and
may not be typical of the rest of the sample).

There appears to be a slight exception in the case of L183, where the
200-$\mu$m peak appears to lie slightly to the north of the 850-$\mu$m
peak. We note that this is also the source for which the 450-$\mu$m
data peaked to the north of the 850-$\mu$m data (see Figure 4). It is
possible that there is a north-south temperature gradient across this
region (Juvela et al 2002), which may account for an apparent shift
of the peak with wavelength.

\begin{table*}
\caption{Fitted source properties for the sub-set of bright cores
(above the line), together with a subset of the intermediate cores (below
the line), for which
we have ISOPHOT data, and hence can determine the SED and dust temperature
(also including L1521D \& F, whose
temperatures are taken from Codella et al. 1997).
Columns 1, 2 \& 3 list the source name,
its distance and its temperature found from fitting its SED.
Typical error-bars on the temperatures are $\pm$2~K.
Columns 4, 5 and 6 list the full-width at half-maximum (FWHM) of
the major and minor axes, the position angle of the major axis
(measured north through east) and
the ratio of the two axes, as determined by a two-dimensional
Gaussian fit to the 850-$\mu$m data. The beamwidth has been
deconvolved from the FWHM quoted, and we estimate the error on
the core FWHMs to be equivalent to half a beamwidth ($\sim$7 arcsec)
for the bright cores and roughly double this ($\sim$14 arcsec)
for the intermediate cores.
The error on the position angle is $\sim\pm$5$^{\circ}$ and
$\sim\pm$10$^{\circ}$ for bright and intermediate cores respectively.
Columns 7 and 8 list the
values found for the central column density N$(H_2)_{C}$ and maximum
extinction A$_V$ from the 850-$\mu$m flux densites.
Typical errors on these values are $\pm$20--30\%.
Columns 9 and 11 give the mean central
($\overline{n}_{c}$) and FWHM ($\overline{n}_{F}$)
volume number densities.
Error-bars on these values may be as high as 50\%.
The FWHM mass (M$_{F}$) and the total mass in a 150-arcsec
aperture (M$_{150}$),
calculated from the FWHM submillimetre flux density
and the total flux density in a 150-arcsec aperture respectively,
are listed in columns 10 \& 12 using the method explained in the text.
Error-bars are in the region of $\pm$30\%.
Column 13 lists the virial mass (M$_{vir}$)
calculated from the NH$_3$ linewidth
(Benson \& Myers 1989) in a region of radius equal to the FWHM.
Error-bars here are also $\sim\pm$30\%.}
\begin{tabular}{lcccccccccccc}
\hline
Source  & D & T & FWHM & Aspect & P.A. & N$(H_2)_{C}$ & A$_{V}$ &
$\overline{n}_{c}$ & M$_{F}$ & $\overline{n}_{F}$ & M$_{150}$  & M$_{vir}$\\
& (pc) & (K)  & (pc) & Ratio & ($^\circ$) & (cm$^{-2}$) & (mag)
& (cm$^{-3}$) & (M$_{\odot}$) & (cm$^{-3}$) & (M$_{\odot}$) & (M$_{\odot}$)\\
\hline
L1521D  & 140 & 10 & 0.048$\times$0.021  & 0.44 & 42  &
 5$\times 10^{22}$  & 53  & 5$\times 10^{5}$ & 0.5    &
3$\times 10^{5}$ & 2.0 & 2.6 \\
L1521F  & 140 & 9  & 0.019$\times$0.012  & 0.66 & 153 &
 1$\times 10^{23}$ & 130 & 3$\times 10^{6}$ & 0.4     &
2$\times 10^{6}$ & 2.6 & 0.9 \\
L1517B  & 140 & 10 & 0.028$\times$0.019  & 0.68 & 21  &
 4$\times 10^{22}$  & 41  & 5$\times 10^{5}$ & 0.3    &
4$\times 10^{5}$ & 1.7 & 1.3 \\
L1544   & 140 & 9 & 0.057$\times$0.029  & 0.50 & 148  &
 1$\times 10^{23}$  & 102  & 7$\times 10^{5}$ & 1.6   &
6$\times 10^{5}$ & 3.1 & 2.4 \\
L1582A & 400  & 15 & 0.054$\times$0.026  & 0.47 & 18  &
 2$\times 10^{22}$  & 18   & 1$\times 10^{5}$ & 0.4   &
1$\times 10^{5}$ & 4.8 & 2.3 \\
L183    & 110 & 10 & 0.020$\times$0.012  & 0.59 & 179 &
 8$\times 10^{22}$  & 85 & 1$\times 10^{6}$ & 0.3     &
1$\times 10^{6}$ & 1.8 & 0.9 \\
L1689B  & 130 & 11 & 0.027$\times$0.022  & 0.82 & 68  &
 5$\times 10^{22}$  & 49  & 6$\times 10^{5}$ & 0.4    &
5$\times 10^{5}$ & 1.4 & 1.5 \\
L63     & 130 & 11 & 0.024$\times$0.011  & 0.48 & 111 &
 5$\times 10^{22}$  & 50  & 9$\times 10^{5}$ & 0.2    &
8$\times 10^{5}$ & 1.6 & 1.0 \\
B133    & 200 & 13 & 0.048$\times$0.037  & 0.78 & 175 &
 2$\times 10^{22}$  & 25   & 2$\times 10^{5}$ & 0.4   &
1$\times 10^{5}$ & 1.4 & 4.5 \\
\hline
L1498   & 140 & 10 & 0.103$\times$0.052  & 0.50 &  100   &
 3$\times$10$^{22}$  & 30   & 1$\times 10^{5}$ & 0.5   &
3$\times 10^{3}$ & 1.5 & 2.6 \\
L1512   & 140 & 12 & 0.080$\times$0.034  & 0.43 &  135   &
 2$\times 10^{22}$  & 20   & 1$\times 10^{5}$ & 0.3   &
3$\times 10^{3}$ & 0.6 & 2.5 \\
B68     & 130 & 13 & 0.074$\times$0.047  & 0.63 &   23  &
 1$\times 10^{22}$  & 11   & 5$\times 10^{4}$ & 0.2   &
2$\times 10^{3}$ & 0.4 & 2.1 \\
\hline
\end{tabular}
\end{table*}

\subsection{Core Temperatures}

The similarity in morphology between the ISOPHOT and SCUBA data
for each core seen in Figures 7 \& 8 lead us to
ascertain that for most of the cores
it is the same dust that we
are seeing in emission at both 200 \& 850~$\mu$m.
The coincidence of emission at both of these wavelengths
with optical obscuration show that this dust is
also absorbing at shorter wavelengths. Therefore,
for these sources we can
combine the integrated flux measurements presented in Tables 1 \& 2
with the far-infrared data to produce
spectral energy distributions (SEDs) of the sources.
We have previously used the measurements integrated
in a 150-arcsec diameter aperture in
Tables 1 \& 2 of this paper and table 5 of
Paper V and plotted
the resultant spectral energy distributions in Paper V.
We also incorporated 1.3-mm data from Paper III.

The solid curves on each of the plots in Paper V are
modified black-body curves, often known as grey-body curves.
The monochromatic flux density,
$F_{\nu}$, of a grey-body, at frequency $\nu$,
radiated into solid angle $\Omega$ is given by:

\begin{equation}
F_{\nu}=\Omega f B_{\nu}(T) ( 1 - e^{-(\nu/\nu_{c})^\beta}) ,
\end{equation}

\noindent
where $B_{\nu}(T)$ is the Planck function, $\nu_{c}$ is the frequency
at which the optical depth is unity, $\Omega$ is the solid angle of the
aperture, $f$ is the filling factor of the source within the
aperture and $\beta$ is the dust emissivity index.
The temperatures of the grey-body fits found in this way are
listed in Table 4 for reference. The error-bars on the temperatures are
typically $\pm$2K. This could introduce uncertainties in the mass
estimates (see below).

\subsection{Core Masses}

Submillimetre continuum emission is optically thin, and hence
it is a direct tracer of the mass content of molecular cloud cores.
For a spherical isothermal dust source
at distance $d$, the total (dust~$+$~gas) mass,
$M(r<R)$, contained within a radius $R$ from the centre,
is related to the submillimetre flux density
$S_{850\mu m}(\theta)$ integrated over a circle of
projected angular radius $\theta = R/d$ by the equation:

\begin{equation}
M(r<R) =  [S_{850\mu m}(\theta)\, d^2]/[ \kappa_{850}\,
B_{850}(T)],
\end{equation}

\noindent
where $\kappa_{850}$ is the dust opacity per unit mass column density
at $\lambda$ = 850$\mu$m
and $B_{850}(T)$ is the Planck function at
the same wavelength, for a dust temperature $T$.
The dust temperatures are taken from Paper V and Table 4,
except for L1521D \& F, which are rotational line temperatures
taken from Codella et al (1997), and which therefore
assume thermal equilibrium between the gas and dust for these sources.
For the dust opacity, we follow the method we adopted in
Paper II, and use $\kappa_{850}$ = 0.01 cm$^2$g$^{-1}$ (see Papers II \& III
and AWB93 for detailed justifications both of this value of $\kappa_{850}$
in particular and this method of obtaining masses in general). The
uncertainties in the masses due to a combination of uncertainties in
$\kappa$ and in measuring the temperature could be as high as a
factor of a few.

Table~4 lists the masses, column densities and volume densities
for all of the cores within various radii.
The column density follows directly from the mass.
The peak
volume density is calculated from the peak column density,
assuming that the extent of the column is the geometric mean of the
other two core axis FWHMs.
The FWHM volume density is calculated from the FWHM mass
and the volume enclosed by a tri-axial ellipsoid,
again assuming that the line of sight FWHM
is the geometric mean of the other two FWHMs.
Table~4 also lists the total mass of each core within a 150-arcsec diameter
aperture (M$_{150}$). In addition we calculate the virial mass
(M$_{vir}$) for each
core, using the NH$_3$ linewidth (Benson \& Myers 1989).
The measured masses of the bright cores all appear to be
of a similar order to their virial masses.

There are two cores in common (L1544 and L1689B) between Table 4 and the
sample of Bacmann et al. (2000). They calculated the central
volume densities of cores based on their mid-IR absorption seen in
ISOCAM data. In both of these cases they found a central volume
density of 6 $\times$ 10$^5$ cm$^{-3}$. We find values of 5 and 7
$\times$ 10$^5$ cm$^{-3}$ respectively, in very good agreement.
In fact all of the volume densities
of the bright cores lie in the range of
$\sim$10$^5$--10$^6$ cm$^{-3}$.
The intermediate cores all have significantly larger FWHM, and
generally lower column densities, which together entail significantly
lower volume densities, all around a few 10$^3$ cm$^{-3}$.
Therefore we deduce that bright cores are significantly more
centrally condensed than intermediate cores.

\subsection{Core radial profiles}

The radial density profiles of the cores can be
measured in the case of our `bright' cores and
used for comparison with
theoretical predictions to attempt to
assess the stability and potential evolution
of the cores. This is most easily assessed for spherically symmetric
cores. Despite the fact that the majority of the cores are
non-circular we still make radial profiles for the bright cores,
with the exception of those with double peaks or with a potentially missed
peak (L1524, L1696A and L43). Furthermore, L1689SMM and L183 are
seen to be too `ridge-like' to determine a one-dimensional radial profile.

We generate azimuthally averaged radial profiles for the remaining 8 cores
by assuming that our observed cores are ellipsoids (as defined
by the parameters in Table 4) which are aligned with the plane of
the sky. We then remove instrumental effects that might cause us to
miscalculate the profiles. Chief amongst these effects is that caused
by `chopping'.

The JCMT uses a chopping secondary mirror to remove the bright, slow, varying
component of the submillimetre sky (see above). This
introduces a characteristic differential pattern in the data. Every point
in the output map is equal to the true flux density
at that point minus the mean of the
true flux density at two points offset by $\pm$120 arcsec
in a direction given by the position angle of the chop throw.
The two positions arise due to the effect of `nodding' the telescope between
`left' and `right' beams. The two `off-beam' positions serve
to improve the sky removal.

If the true sky flux density distribution is $S_{sky}(x,y)$,
and the off-beams are at offsets
$(+a,+b)$ and $(-a,-b)$, then the flux density distribution of
the output map, $S_{map}(x,y)$, will be given by:

\[S_{map}(x,y) = S_{sky}(x,y) - 0.5*[ S_{sky}(x+a,y+b) + \]

\begin{equation}
\hspace*{4.6cm}  S_{sky}(x-a,y-b)] .
\end{equation}

\noindent
For a source that is significantly smaller than the chop throw:

\begin{equation}
S_{sky}(x+a,y+b) = S_{sky}(x-a,y-b) = 0 ,
\end{equation}

\noindent
so:

\begin{equation}
S_{map}(x,y) = S_{sky}(x,y)
\end{equation}

\noindent
and:

\[ S_{map}(x+a,y+b) = S_{map}(x-a,y-b) \]

\begin{equation}
\hspace*{2.65cm} = - 0.5 \times S_{sky}(x,y).
\end{equation}

\noindent
However, for more extended sources, chopping
causes one to underestimate the true extent of the flux
density distribution parallel to the chop throw,
whilst having relatively little effect on the extent of
the flux density distribution orthogonal to the chop throw.
This has important
implications for fitting flux density profiles to
extended sources, and must be taken into account.

To replicate the effect of chopping on a given core,
we generate a synthetic core map
by taking a known, circularly symmetric column
density distribution and compressing it along a minor axis to
obtain the core's
observed axis ratio. The direction of the minor axis is
chosen so as to match the observed position angle of the elliptical core.
The synthetic map is then convolved with a Gaussian beam
with the same FWHM as
the average FWHM of the measured telescope beam during our observations.
A simulated chop that has the same position angle
and chop throw as the actual chop is
applied to the synthetic map by using the
standard SURF software routine `add-dbm'
(Jenness \& Lightfoot 1998).

A number of synthetic maps are produced with different input
parameters. The synthetic maps and the real maps are then analysed in
exactly the same fashion and the profiles are compared (see
section 5.5 below).
When a real core profile matches one of its synthetic core profiles
then the parameters of the real core can be deduced from the synthetic
core, thus effectively removing the instrumental effects of the beam
and the chopping.
The maps of L1582A and L1517B have multiple pointing centres and multiple
chop position angles. In these cases
a single mean chop position angle
was used in each case to generate the synthetic maps.

The profiles are generated on both real and synthetic maps
by binning the data into elliptical annuli (with the aspect ratios given in
Table 4) of radial width equal to the
half-width at half maximum (HWHM) of the 850-$\mu$m beamsize (7.4 arcsec).
The mean value in each annulus and the
standard deviation of the points inside that annulus are then calculated.

These values are plotted against the mean annulus semi-major axis
for each of the cores and are shown in Figure 9 on a logarithmic scale,
normalised to the peak value in each case. Note that any
departure of the geometry of a core from its mean ellipticity would serve
to increase the scatter of the pixel values within a given annulus. This
would cause an associated increase in the standard deviation measured
for that annulus. As such the one sigma bars shown in Figure 9 represent
a robust determination of the aggregate errors on these radial flux
density profiles.

Each of the profiles in Figure 9 shows the now familiar profile for a
prestellar core with a flat inner region and a gradient that steepens
towards the edge (c.f. Papers I--III).
The submillimetre continuum flux density within a beam is approximately
proportional to the mass of dust emitting within the beam area and the
temperature of the emitting dust (assuming that the dust is optically thin).
The dust mass can also be related to the total mass using simple
assumptions about the dust-to-gas ratio (see above).

We know that our cores are roughly isothermal (see section 1 above), so we
can convert our measured flux density profiles into column density profiles.
This allows us to use the radial flux density
profiles to determine the internal volume density structure of the core
by making simple assumptions about the distribution along the line of sight
relative to the other two dimensions.
We have on previous occasions (e.g. Papers I--III)
used two power laws to model the radial profiles obtained for
prestellar cores. In this paper we take the analysis a stage further.

\begin{figure*}
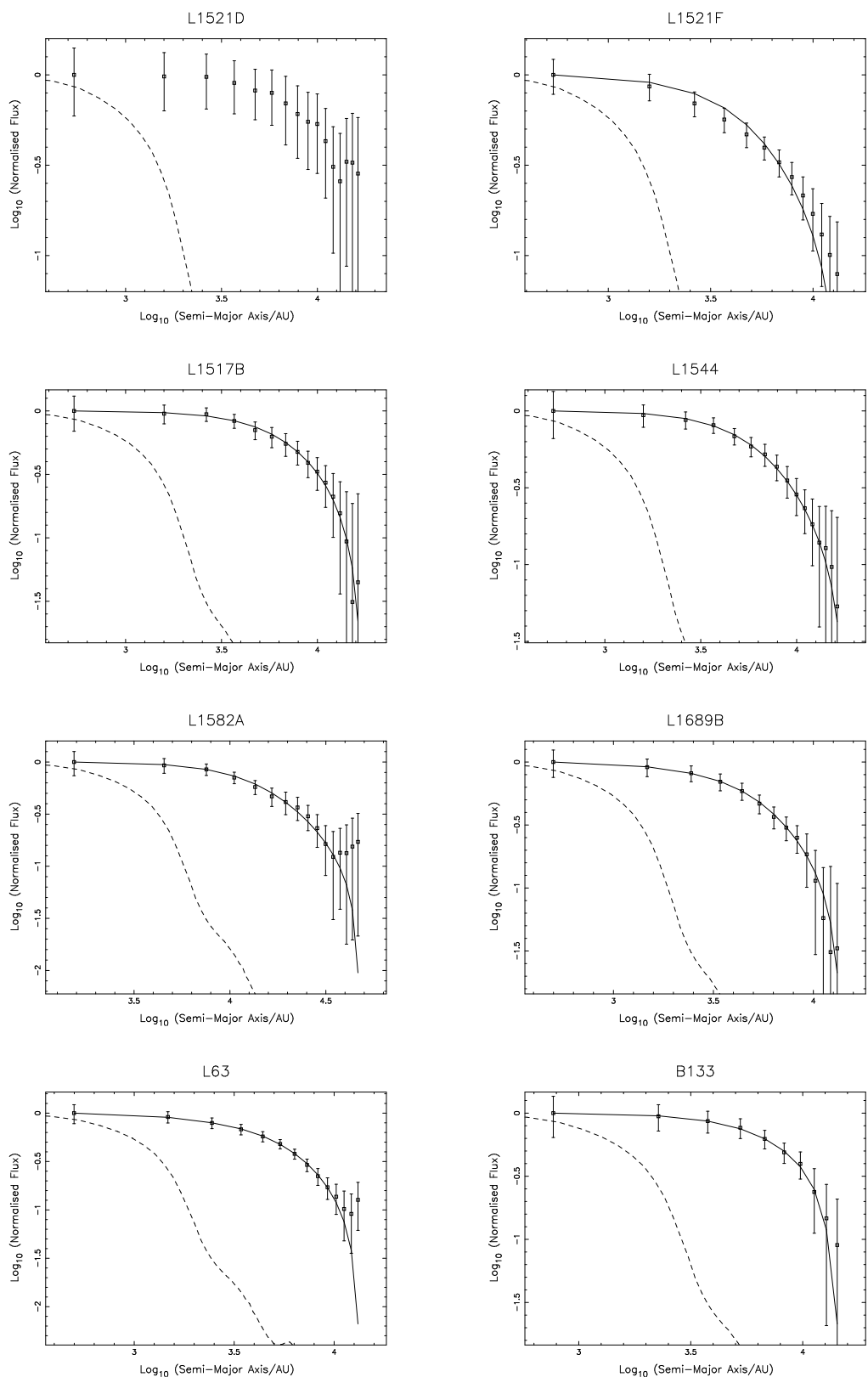

\setlength{\unitlength}{1mm}
\noindent
\begin{picture}(170,220)
\put(10,215){\includegraphics{fig9a.ps}}
\put(85,215){\includegraphics{fig9b.ps}}
\put(10,160){\includegraphics{fig9c.ps}}
\put(85,160){\includegraphics{fig9d.ps}}
\put(10,105){\includegraphics{fig9e.ps}}
\put(85,105){\includegraphics{fig9f.ps}}
\put(10,50){\includegraphics{fig9g.ps}}
\put(85,50){\includegraphics{fig9h.ps}}
\end{picture}
\caption{850-$\mu$m azimuthal elliptically-averaged, normalised radial
flux density profiles (reading left-to-right from top)
of: L1521D, L1521F, L1517B, L1544, L1582A, L1689B, L63 and B133.
Data points are shown on a logarithmic scale
at half beam spacings with 1$\sigma$ error bars.
The solid curves are the best fit Bonnor-Ebert profiles
(no unique fit could be obtained for L1521D),
while the dashed curves are the 850-$\mu$m beam profiles.}
\end{figure*}

\begin{table}
\caption{Fitted source properties for the bright cores in Table 4 (apart from
L1521D and L183, which appear too elongated or filamentary to be approximated
as circularly symmetric).
Column 1 lists the source name. Columns 2,
3 \& 4 list the values found for $R_{flat}$, $\xi_{max}$ and $R_{edge}$
from Bonnor-Ebert fits to the radial density profiles (see text for details).
See text for discussion of the
fitting methodology and associated errors.
The critical value of $\xi_{max}$ for a stable Bonnor-Ebert sphere (6.5) is
significantly exceeded in 6 out of 7 cores.
The effective temperature required by the Bonnor-Ebert fit,
T$_{{\rm eff}}$, is listed in column 5. The values of T$_{{\rm eff}}$ are in
all cases significantly higher than our measured values from the SED
fitting. The need for such high effective temperature values implies
that the cores are not Bonnor-Ebert spheres.}
\begin{center}
\begin{tabular}{lcrcc}
\hline
Source & $R_{flat}$ & $\xi_{max}$  & $R_{edge}$ & T$_{{\rm eff}}$ \\
       & (AU) &                    &    (AU) & (K) \\
\hline
L1521F & 3400  & $\geq$7.5  & $\geq$13000  &  52 \\
L1517B & 5800  & $\geq$10   & $\geq$29000  &  35 \\
L1544  & 4800  &     10     &  24000       &  71 \\
L1582A & 11000  &      9     &  50000       &  33 \\
L1689B & 3800  & $\geq$14   & $\geq$27000  &  28 \\
L63    & 3600  & $\geq$15   & $\geq$27000  &  26 \\
B133   & 6800  &        4   &  13000       &  24 \\
\hline
\end{tabular}
\end{center}
\end{table}

\subsection{Comparison with Bonnor-Ebert spheres}

If we assume that the cores are supported
against gravity by hydrostatic pressure
and we neglect magnetic and turbulent
contributions, then for a spherically
symmetry sphere of gas, Poisson's
equation can be written as:

\begin{equation}
\frac{1}{r^{2}}\frac{\partial}{\partial\,r} \left( \frac{r^{2}}{\rho}
\frac{\partial\,P}{\partial\,r} \right) +4\pi\,G\rho = 0 ,
\label{eq:poisson}
\end{equation}

\noindent
where $r$ is the radius, $\rho(r)$ is the density
and $P(r)$ is the pressure at radius $r$. Given that we find the cores
to be roughly isothermal, we can use an equation of state of the form:

\begin{equation}
P=\rho\frac{kT}{m}+\frac{a}{3}T^{4} .
\label{eq:state}
\end{equation}

\noindent
Then equation \ref{eq:poisson} reduces to:

\begin{equation}
 \frac{1}{\xi^{2}}\frac{d}{d\,\xi} \left( \xi^{2} \frac{d\,\psi}{d\,\xi}
\right) - e^{-\psi} = 0 ,
\end{equation}

\noindent
where we have introduced
the usual dimensionless parameters
$\xi = (2r/R_{flat})$, and
$\psi = - \ln(\rho/\rho_{c})$. In these expressions
$\rho_{c}$ is the central volume density at $r=0$, and
$R_{flat} = (kT/\pi\,mG\rho_{c})^{1/2}$, where
$T$ is the temperature, $m$ is the
mean particle mass, $k$ is Boltzmann's constant and $G$ is the gravitational
constant (Chandrasekhar 1939).
We note parenthetically that
the parameter $R_{flat}$ is two thirds of the King Length
(e.g. Binney \& Tremaine 1987),
which is analogous to the Jeans length, but is defined using the central
density of a sphere rather than its mean density.

To compare this theoretical result with the data in Figure 9
we can transform the function $\psi(\xi)$
into $\rho(r)$ and project this into column density
$\Sigma(b)$ by integrating along the line of site ($b$ is the radial
impact parameter).
The form of the solution is governed by two parameters, $R_{flat}$ and
$R_{edge}$. $R_{flat}$ can be thought of as being roughly the point on
the radial profile where the central `flat' region ends and the profile
becomes steeper (c.f. Paper II).

The model definition of
$R_{edge}$ is the radius at which $\Sigma(b) = 0$. It is
also known as the truncation radius.
In practice, it was taken to be the radius at which the core
emission either becomes indistinguishable from the background
cloud emission, or becomes steeper in volume density profile
than r$^{-3}$ (c.f. Bacmann et al. 2000).
The family of solutions for $\xi_{max}=(2R_{edge}/R_{flat})$
are known as Bonnor-Ebert spheres
(Ebert 1955; Bonnor 1956).
The critical value of $\xi_{max}$ is 6.5. Any core with a
value greater than this is unstable against collapse.
Others have used a similar form of profile to infer the
stability of dense cores (e.g. Alves, Lada \& Lada 2001).

The two radii $R_{flat}$ and $R_{edge}$
can be fitted to a data-set, such as those
shown in Figure 9, using a $\chi^2$ minimisation routine.
This was achieved by minimising the variance
between an observed data-set and a grid of synthetic profiles
(see section 5.4 above)
parameterised by $R_{flat}$ and $\xi_{max}$.
Table 5 lists the values of $R_{flat}$, $R_{edge}$ and $\xi_{max}$
found for seven of the eight cores illustrated in Figure 9.
We could not obtain a constrained fit for L1521D,
so we exclude it from further analysis.
The seven cores that we fit typically
have $R_{flat}\sim$3000--6000~AU.

The errors on the resultant fit can be estimated from a contour
map of the $\chi^2$ variation in the $R_{flat}$--$\xi_{max}$ plane.
$R_{flat}$ for our data is found to be generally
well-defined with a relative 1-$\sigma$ error-bar of around 10--15\%.
However, $\xi_{max}$ is in general somewhat
less well constrained for our data, because chopping
creates a degeneracy between Bonnor-Ebert
solutions that have large values of $R_{edge}$.
Hence the error-bars on the values in Table 5 are quite large --
typically up to 50\%. This is because the drop
in signal-to-noise towards the edge of our maps
and the finite extent of the maps combines with the tendency of chopping
to filter out information on large spatial scales.
It is found to be impossible to generate
reasonable fit solutions for a
value of $R_{edge}$ greater than approximately twice
the chop throw, because our data contain no
information on these size scales (the chop throw corresponds to radii
of $\sim$15,000 -- 40,000 AU, depending on the distance to the source).

All of the above profile fitting is based on the normalised flux density
profiles in Figure 9. However, the absolute values of the central
column densities quoted in Table 5 can be used
to calculate the effective temperature, $T_{{\rm eff}}$,
of the Bonnor-Ebert fit.
Table 5 lists the effective temperature, $T_{{\rm eff}}$, required
by the Bonnor-Ebert fitting for each core. These Bonnor-Ebert effective
temperatures can be compared with the temperatures calculated from the
observed spectral energy distribution of each core listed in Table 4.
In all cases the Bonnor-Ebert effective temperature is a factor of
2 or more greater than the observed core temperature.
This tells us that, at least for these bright cores in
our sample, a critical Bonnor-Ebert sphere is not
consistent with the radial density profiles of the cores.

The parameter $\xi_{max}$ describes the degree of central condensation,
with cores with larger flat central regions being more
stable against gravitational collapse.
Remembering that the critical value of $\xi_{max}$ for a stable solution
is 6.5, we see that six out of seven of our cores significantly
exceed this value.

\begin{figure}
\setlength{\unitlength}{1mm}
\noindent
\begin{picture}(70,45)
\put(10,50){\includegraphics{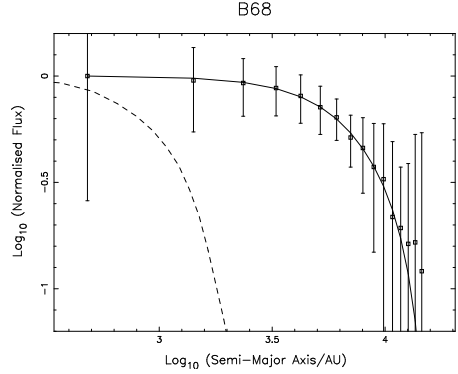}}
\end{picture}
\caption{850-$\mu$m azimuthal elliptically-averaged, normalised radial
flux density profile of B68.
Data points are shown on a logarithmic scale
at half beam spacings with 1$\sigma$ error bars.
The solid curve is the normalised fit to a critical Bonnor-Ebert profile
found by Alves et al. (2001) for their infra-red absorption
data. The two data-sets appear to be consistent.}
\end{figure}

Our conclusion is in agreement with the findings of Evans et al (2001),
who studied a smaller number of pre-stellar cores and found that they
also could not be fitted as critical Bonnor-Ebert spheres. This is
interesting, because their fitting analysis was somewhat different
from ours, and yet they reach ultimately the same conclusion. Of the
cores in common between their small sample and ours there are some
differences in fitting parameters. For example, they allow the temperature
to decrease towards the centre and consequently find smaller values
of $R_{flat}$ than those reported here assuming isothermal cores.
As already mentioned in section 1 above, our findings from ISO (Paper V)
suggest that the central core regions are nearly isothermal.

Evans et al. (2001) also do not treat the
chopping in two dimensions, as we do. Furthermore, for one of the
cores, L1689B, they use a value for the inter-stellar radiation
field (ISRF) that is too small for the environs of Ophiuchus in which
L1689B is situated (c.f. Andr\'e et al. 2003).
Nevertheless they arrive at the same
conclusion as we do that the cores are not critical Bonnor-Ebert spheres.
Hence this conclusion is probably reasonably robust
and independent of fitting model.

Other work has also shown that a purely thermal Bonnor-Ebert equilibrium
model can be ruled out in some cases because
the central temperature predicted by the model is much higher than
the observed temperature (e.g. Andr\'e et al. 2003; Harvey et al., 2003b).
We can also compare our result to the previous work of Alves et al. (2001)
mentioned above. They studied one of our sample, B68, and found that
it was consistent with a critical Bonnor-Ebert sphere. We have classified
B68 as an intermediate core, and it can be seen from Table 2 and
Figure 5 that it is one of the fainter members of that group.

Figure 10 shows the radial flux density profile for our 850-$\mu$m data.
We calculated the normalised
Bonnor-Ebert profile for B68 based on the parameters
quoted by Alves et al. (2001). This is shown as the solid line in
Figure 10. It can be seen that the shape of the flux density profile
that we observed for B68 is consistent with a Bonnor-Ebert profile
with the value of $\xi_{max}$ = 6.9 found by Alves et al.
We may conjecture, therefore, that cores might pass through
a phase (during part of the intermediate core phase) in
which they resemble the structure of a critical
Bonnor-Ebert sphere while their mean central
densities are still relatively low --
only a few 10$^4$cm$^{-3}$ (see discussion in section 6.1 below).
Then by the time they reach mean central densities of a few
10$^5$cm$^{-3}$, typical of the
bright cores, they have passed through a critically stable
equilibrium, and
are either collapsing or else have some other means of support,
even though the shape of their radial density profiles still
resembles that of a super-critical Bonnor-Ebert sphere.

The interpretation of density structure in terms of Bonnor-Ebert
profiles is interesting, although there are other caveats to
studies of this nature.
Some recent work has shown that even highly non-equilibrium cores can
demonstrate a Bonnor-Ebert form of profile
(e.g. Ballesteros-Paredes et al. 2003).
Furthermore, it can be difficult to differentiate between forms of profile
based on a two-dimensional representation such as an astronomical image.

Recent work by Harvey et al. (2003a) has shown that for some data-sets
the profiles can often be fitted by many different forms of profile.
This would tend to suggest that interpreting a particular
profile in terms of a pressure-balanced equilibrium may be over-interpreting
a limited amount of data.
Nonetheless, even if there may be ambiguity over exactly how to
model the density profiles, some forms of profile can be ruled out.
For instance both the singular isothermal sphere and the logotropic
non-isothermal sphere models can be ruled out (c.f. Bacmann et al., 2000).

Taken at face value, our result says that the majority of our
bright cores are not critical, Bonnor-Ebert, pressure-supported
spheres.
This could either be indicating that the cores are already collapsing,
or that there is some additional form of support that is operating,
such as the support of a magnetic field or turbulence.
These possibilities are discussed in the next section.

\section{Discussion}

\subsection{Categorising cores}

We have observed 52 starless cores, including 50 positions from the
catalogues of Myers and co-workers (e.g. Myers et al. 1983; Benson \&
Myers 1989). For the remainder of this paper we will ignore the two
additional positions that we discovered (L1689SMM \& L1521SMM) and
merely discuss the remaining 50 cores. Of these 50, we detected 27
(54\%) and failed to detect 23. Of the 27 detections, we classified
12 as bright and 15 as intermediate.

There is a clear difference between the bright and intermediate cores.
Bright cores are more massive and submillimetre bright
than intermediate cores. They also have higher central densities.
It is a necessity of star formation that to form
a star a cloud must evolve from an extended, low-density state to a much
higher-density, more centrally-condensed core with eventually an embedded
protostar. Therefore it is reasonable to conjecture that bright cores
may be more evolved than intermediate cores, and hence closer to forming
a protostar. We can use this hypothesis to infer evolutionary timescales
for each category of core.

We saw in Tables 1--3 that there is a clear correlation between our SCUBA
detections and the NH$_3$ detections of Benson \& Myers (1989). All of our
bright cores have NH$_3$ detections, and all except 3 of our intermediate
cores have NH$_3$ detections (ignoring the two additional sources that we
added to the Benson \& Myers list). Similarly, only 4 of our non-detections
were detected in NH$_3$. Therefore, we can say that a detection in NH$_3$
is roughly equivalent to a SCUBA detection.
However, we note that almost all of the positions we surveyed
(both detections and non-detections) have CO detections
(Myers et al. 1983).

The amplitude of flux density fluctuations to which
we are sensitive, given our detection sensitivity limit,
corresponds to fluctuations in column density
of $\sim$10$^{22}$cm$^{-2}$.
Furthermore, our observing method is only sensitive to
structures on scales smaller than roughly twice our chop throw,
because the method of chopping that we use effectively subtracts away
the larger-scale emission from extended clouds, on scales of $\sim$240
arcsec or greater. Thus our observing method is most sensitive to
structures on smaller scales than this. Consequently, our data are
sensitive to fluctuations on scales that correspond to roughly
$\leq$0.1~pc at the typical distances of our sources.
Hence our data are picking up the emission from dense
molecular cloud cores on these size-scales.

The typical central
volume densities of our bright cores are in the range of
$\sim$2 $\times$ 10$^5$ -- 2 $\times$ 10$^6$ cm$^{-3}$ (Table 4).
The intermediate cores
(Table 2) have flux densities that translate into
typical volume densities of $\sim$5 $\times$ 10$^4$ -- 2 $\times$ 10$^5$
cm$^{-3}$.
For our third group of non-detected cores we can only place an upper
limit on their column densities. This limit is roughly $\sim$10$^{22}$
cm$^{-2}$. However, if they were of similar size to the detected cores,
this would translate into an upper limit volume density of approximately
$\leq$5 $\times$ 10$^4$ cm$^{-3}$ (c.f. Kirk 2003).
We do not see any statistically significant
differences between the cores in Taurus and those in Ophiuchus.
However, this is perhaps not surprising, as none of our Ophiuchus
cores lie in the cluster-forming region of $\rho$ Oph (see section 2).
Hence all of our cores are relatively isolated,
whatever region they lie in.

\subsection{Core lifetimes}

The numbers of cores detected can be used to determine typical statistical
time-scales for particular evolutionary stages. This method was first
employed for these cores by Beichman et al. (1986), who found that there
were roughly equal numbers of their sources with and without embedded
protostars. Those without embedded sources were labelled starless cores.
Extrapolating from the typical T Tauri star lifetime to the lifetime of
cores with embedded sources, they estimated the starless core
lifetime to be roughly a few times 10$^6$ years (c.f. Paper I).

This was subsequently refined by Lee \& Myers (1999), who used a larger
sample, to be $\sim$0.3--1.6 $\times$ 10$^6$ years.
This value is based upon an estimated range in lifetimes for Class I sources
of $\sim$1--5 $\times$ 10$^{5}$ years. Within this range the best estimate
for the Class I lifetime is $\sim$2 $\pm$ 1 $\times$ 10$^5$ years
(e.g. Greene et al., 1994; Kenyon \& Hartmann 1995). Consequently,
this corresponds to a starless core lifetime of $\sim$6 $\pm$ 3
$\times$ 10$^5$ years.

We have detected 27 out of 50 starless cores, so we arrive at a timescale
for cores detected in the submillimetre of
$\sim$3 $\times$ 10$^5$ years. We note that
this is only an approximate value,
as is the following. We have previously called
the submillimetre detected cores prestellar cores (c.f. Papers I--V).
We also note that this lifetime we derive
is consistent with that which we have
found in previous studies (e.g. Paper I). We shall refer to this
time-scale as $t_{submm}$.

The minimum central volume density of a submillimetre detected core is
$\sim$5 $\times$ 10$^4$ cm$^{-3}$. At this density the free-fall time
$t_{ff}$ is
$\sim$10$^5$ years. Hence the statistical timescale we find
for prestellar cores is roughly three times the free-fall time
at the typical minimum density of a submillimetre detected
prestellar core -- $t_{submm} \sim 3 t_{ff}$.

We can make a similar calculation for the cores we classified as `bright'.
These constitute 12 out of the total of 27 prestellar cores, or 44\%.
Hence we derive a statistical time-scale for the bright cores, $t_{bright}$,
of $\sim$1.5 $\times$ 10$^5$ years.
The minimum central volume density of a bright core is
$\sim$2 $\times$ 10$^5$ cm$^{-3}$.
At this density the free-fall timescale is
$\sim$7 $\times$ 10$^4$ years.
Once again the observed timescale is longer than
the free-fall time
at the typical minimum density of a bright core, this time
twice as long  -- $t_{bright} \sim 2 t_{ff}$.

One additional caveat to our lifetime calculations is that if some
of the cores do not go on to form stars (perhaps because they are
dispersed before they can collapse) then our lifetimes will be
over-estimated. We believe this is unlikely because the cores are
so centrally condensed (and mainly exceed their virial mass -- see Table 4)
that they cannot easily self-disperse, and the probability of a
significant number being over-run by externally-induced shocks on a
timescale before they collapse is very low, given their relatively
isolated locations.

Thus it appears that prestellar cores may live a few times
longer than their free-fall collapse times.
Hence we see quite clearly that prestellar cores cannot generally
all be in free-fall collapse, and there must be some mechanism
responsible for retarding the collapse.
We now consider what that mechanism might be. Most
star formation models appear to fall into one (or both) of two `camps'
when it comes to the mechanism they use for retarding collapse --
turbulence, and magnetic fields. We will treat each in turn.

\subsection{Comparison with turbulent models}

One possibility is that the cores could be supported by turbulence.
A recent model has produced
simulated cores that mimic many of the observed
properties of actual
cores, such as their densities, temperatures and radial profiles
(Ballesteros-Peredes et al. 2003). In this model the cores live for
approximately a free-fall time.

Taken at face value, this is a factor of $\sim$2--3 times shorter than
our estimated timescale
for submillimetre cores above (see also Andr\'e et al., 2004).
Hence, this model that appears to be able to mimic some of
the observed properties of prestellar cores,
may not be able to reproduce the lifetimes of cores.
We also note that this model does not reproduce the narrow linewidths
observed in these cores. The observed linewidths have a mean value
of 0.27~kms$^{-1}$ (Benson \& Myers 1989), whereas the model predicts
linewidths of $\sim$1.5~kms$^{-1}$ (Ballesteros-Peredes et al. 2003).

We have noted elsewhere
(c.f. Ward-Thompson et al. 2000) that turbulent models which include
magnetic fields can only reproduce polarisation maps of prestellar cores
if they are in the regime where the turbulent and magnetic energies
are roughly equal (e.g. Ostriker, Gammie \& Stone 1999;
Ostriker, Stone \& Gammie 2001; Crutcher et al. 2004).

However, these same models can only produce the observed core
radial density profiles described above (see also Bacmann et al.,
2000) if they are in the regime where the magnetic field
dominates over the turbulence, in apparent contradiction with the
polarisation results (Ward-Thompson et al., 2000).
Consequently, turbulent models may be able to reproduce
some of the observations,
but they cannot match all of the data
simultaneously.

\subsection{Comparison with magnetic models}

Another possibility is that magnetic fields
alone are responsible for retarding
collapse, and the cores are undergoing ambipolar diffusion.
This is the process by which the neutrals drift past
the ions, which are supported by the magnetic field,
and precipitate the collapse process (e.g. Mouschovias 1991).

We have previously compared our observations to ambipolar diffusion models
in Papers II and III. However, they were based on either individual cores
or small number statistics. Now we are able to make a detailed comparison
based on our large-scale survey. As a specific illustration,
we begin with a comparison of the models
of Ciolek \& Mouschovias (1994 -- hereafter CM94).

We have derived an approximate timescale, t$_{submm}$, for our
complete sample of detected cores of
$\sim$3 $\times$ 10$^5$ years.
This corresponds to the timescale for a core to
evolve from our minimum detection limit of $\sim$5 $\times$ 10$^4$ cm$^{-3}$
to the point at which it contains a protostar.
The timescale predicted by CM94 `model A', for a core to evolve
from this density to forming a protostar is $\sim$3 $\times$ 10$^6$ years,
or a factor of $\sim$10 too long.

Taking the bright cores only, we have derived a timescale of $\sim$1.5
$\times$ 10$^5$ years. This corresponds to the timescale for a core to
evolve from a lower density limit of $\sim$2 $\times$ 10$^5$ cm$^{-3}$
to the point at which it contains a protostar.
The timescale predicted by CM94 `model A', for a core to evolve
from this density to forming a protostar is $\sim$3 $\times$ 10$^5$ years.
This is in better agreement, although it still disagrees by a factor
of $\sim$2.
We see similar results for other models of CM94. We have previously reported
similar findings (c.f. Paper III).

However, we have other constraints on the magnetic models. Observations of
magnetic fields in molecular clouds have shown that most clouds have a
ratio of mass to magnetic flux that is within a factor of
$\sim$2 of the critical ratio whereby the two are in balance
(e.g. Crutcher 1999; Crutcher et al., 2004).

The ratio of mass to magnetic flux in the centre of a core is given
by CM94 in terms of the parameter $\mu$. This is normalised
such that $\mu$$=$1 corresponds to the critical case
where the magnetic field can just support the mass of the core
against collapse (see also Ciolek \& Basu 2000 -- hereafter CB00).
`Model A' quoted above has $\mu$$\sim$0.25.

A somewhat contrasting constraint on the models is provided by
the radial density profiles of the cores observed in mid-infrared
absorption, which show the cores to have very sharp edges
(Bacmann et al., 2000).
The only ambipolar diffusion models which can produce such sharp
edges require a strong magnetic field in the low density ambient
molecular cloud. This requires $\mu$$<<$1. Thus no model can
reproduce all of the observations.

Nonetheless, we can compare the timescales estimated from our data with
those predicted by models that are closer to critical initial conditions.
For example, CB00 modelled the
evolution of L1544, one of our bright cores, with an initial value of
$\mu$=0.8. They predict that the timescale for a core to evolve
from $\sim$5 $\times$ 10$^4$ cm$^{-3}$
to forming a protostar is $\sim$4 $\times$ 10$^5$ years.
This is in very good agreement
with our estimated $\sim$3 $\times$ 10$^5$ years
for the lifetime of a submillimetre core.

CB00 further predict that the timescale for a core to evolve
from $\sim$5 $\times$ 10$^5$ cm$^{-3}$
to forming a protostar is $\sim$9 $\times$ 10$^4$ years.
This is consistent to within a factor of 2 of
our estimated $\sim$1.5 $\times$ 10$^5$ years
for the lifetime of a bright core.
However, we note that the ratio of the two timescales, $t_{submm}$
and $t_{bright}$, is $\sim$4 for the model and only $\sim$2 estimated from
our observations. Hence we appear to be observing
roughly a factor of 2 too many bright cores
relative to the numbers of intermediate cores.

It seems that, for one of these models to match the overall
timescales of prestellar cores, as well as many of the other
observations, the initial conditions must be close to critical
(e.g. Crutcher 1999; Crutcher et al., 2004). However,
initially critical models appear to under-predict
the numbers of bright cores seen.
It is possible that some selection effects in our sample may
account for this, but it is not obvious how such an effect
might be operating.
It is also possible that a more elaborate ambipolar diffusion model that
incorporates the effects of a
non-uniform (turbulent) component to the magnetic
field may fare better in comparison with
the observations.

\begin{figure}
\setlength{\unitlength}{1mm}
\noindent
\begin{picture}(70,60)
\put(-6,-33){\includegraphics{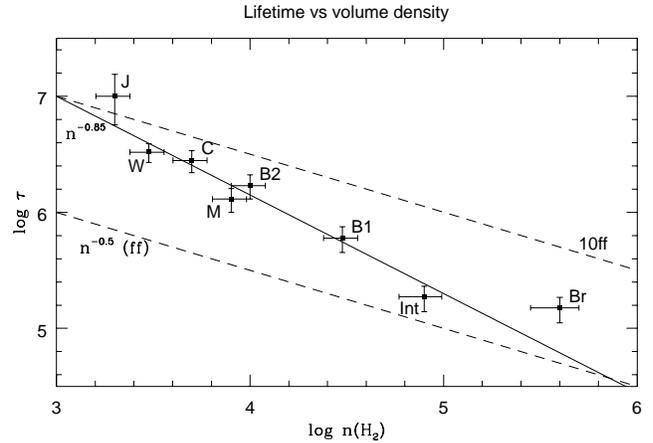}}
\end{picture}
\caption{Plot of inferred starless core lifetime against mean volume
density, adapted from Jessop \& Ward-Thompson (2000), using their original
data with two additional data-points from this paper for the intermediate
and bright core samples, marked as `Int' and `Br' respectively.
The best fit power-law to the original data
of $\tau \propto n({\rm H}_2)^{-0.85}$ is shown as a solid line
and labelled `n$^{-0.85}$'.
Note that the intermediate core data are consistent with this
fit, but the bright core data lie somewhat above the
extrapolation of this line.
The free-fall timescale
($\tau \propto n({\rm H}_2)^{-0.5}$),
as predicted by the `turbulent' models discussed in the text
(e.g. Ballesteros-Paredes et al., 2003),
is shown as the lower
dashed line and labelled `n$^{-0.5}$(ff)'.
A line corresponding to ten times the free-fall time is shown as the
upper dashed line labelled `10ff'. This is roughly equivalent to the
model of cosmic ray ionisaton-regulated star formation
(e.g. Mouschovias 1991).
Other symbols refer to literature data as follows: J -- Jessop \&
Ward-Thompson (2000); W -- Wood et al. (1994); C -- Clemens \& Barvainis
(1988); B1 -- Bourke, Hyland \& Robinson 1995;
B2 -- Bourke et al. (1995); M -- Myers et al. (1983).
See text for further details.}
\end{figure}

\subsection{Comparison with other data-sets}

We can compare our data with data from the literature, and place overall
constraints on all of the models. A similar comparison was
carried out for a number of different data-sets in the literature by
Jessop \& Ward-Thompson (2000). They calculated statistical lifetimes
for a number of data-sets, using the same method as
we have used in this paper. They then plotted the calculated statistical
lifetime against the mean volume density of each sample of cores as
their figure 6. We here reproduce the data from that figure in
our Figure 11.
Remember that each data-point in Figure 11 represents a whole sample
of cores, rather than just a single measurement.
Jessop \& Ward-Thompson (2000) found a best fit to the data of:

\begin{equation}
\tau_{AD} \propto n({\rm H}_2)^{-0.85} ,
\end{equation}

\noindent
with a reduced chi-squared of 1.15. We show this as a solid line on
Figure 11.

We then add to the previous data on Figure 11 the two samples we have studied
here, labelling the intermediate cores as `Int' and the bright cores as `Br'.
We use the mean (FWHM) volume density of each core sample for
consistency with previous surveys, and the lifetimes we calculated
above.

We can also plot on Figure 11 the various model predictions.
The turbulent models discussed above all predict evolution on timescales
roughly equal to the free-fall time. This is shown on Figure 11 as
the lower
dashed line labelled `n$^{-0.5}$~(ff)'. It can be seen that all
groups of cores are longer-lived than the free-fall time.

The magnetic models discussed in the previous section have an
evolutionary timescale
that is the ambipolar diffusion timescale, $\tau_{AD}$.
This is proportional to the ionisation fraction $\chi_i$:

\[ \tau_{AD} \propto \chi_i . \]

\noindent
Hence the manner in which the ionisation of a core is regulated
will strongly affect the timescale on which a core is predicted
to evolve under ambipolar diffusion. Ionisation of the gas can be
caused by both ultra-violet radiation and by cosmic rays. The relation
between ionisation and volume density is usually taken to be a
power-law. For example, Mouschovias (1991) uses
cosmic ray-induced ionisation and finds:

\begin{equation}
\chi_i \propto n({\rm H}_2)^{-0.5},
\end{equation}

\noindent
which leads to:

\begin{equation}
\tau_{AD} \propto n({\rm H}_2)^{-0.5}.
\end{equation}

\noindent
This is shown as the upper dashed line on Figure 11, and labelled `10ff'
(c.f. Ciolek \& Basu 2001).

When ionisation is dominated by UV radiation, and
additional factors are included, such as chemical effects and
the role of multiple charge carriers, the volume density has a slightly
different influence on the recombination rate, and hence the ionisation
fraction. For example, McKee (1989) finds:

\begin{equation}
\chi_i \propto n({\rm H}_2)^{-0.75},
\end{equation}

\noindent
leading to:

\begin{equation}
\tau_{AD} \propto n({\rm H}_2)^{-0.75} ,
\end{equation}

\noindent
which is close to the best-fit power-law in Figure 11.
However, UV ionisation is only prevalent at low column densities,
and for the cores presented here it is not believed to be a
significant effect.

It can be seen that none of the models can fit all of the data.
However, an interesting trend in the data
can be seen on Figure 11. The intermediate core
sample appears to fit with an extrapolation of
the rest of the data, being
roughly consistent with the $n({\rm H}_2)^{-0.85}$ fit.
However, the bright core sample lies distinctly above the extrapolation
of this line.

This is consistent with the trend we saw in the previous
section: if the magnetically regulated (by ionisation) models are
tweaked to fit the low density data (usually by making their
initial conditions close to critical), they then under-estimate
the timescale taken for the cores to evolve through the higher-density
phase to forming a protostar. We again suggest that perhaps
a non-uniform component of the magnetic field (maybe driven by
residual turbulent motions) might provide the additional support
against collapse in the final stages of evolution to protostar formation.

\section{Conclusions}

In this paper we have presented submillimetre continuum data of a large
sample of starless cores. We have detected 29 out of 52 cores at a
wavelength of 850~$\mu$m. Our detection limit corresponds to an A$_V$
of roughly 15. Of the 29 detected cores, 13 were detected at a
peak flux density level of greater than 170~mJy/beam, which corresponds
to an A$_V$ of roughly 50. These 13 cores were designated `bright' cores.
The 16 detected
cores with peak flux density levels less than this value were
designated `intermediate' cores.

The data were combined with previous ISO data and the physical
parameters of the cores were derived.
The temperatures of the cores are all around 10~K.
The central volume densities of the bright cores were found to lie in
the range of 2 $\times$ 10$^5$ -- 2 $\times$ 10$^6$ cm$^{-3}$.
The central volume densities of the intermediate cores were seen to lie in
the range of 5 $\times$ 10$^4$ -- 2 $\times$ 10$^5$ cm$^{-3}$.

The radial density profiles of the bright cores were measured, and seen
to follow the now familiar form
of having a flat central profile that steepens towards the edge.
Detailed comparison of the profiles with that of a Bonnor-Ebert sphere
showed that 6 out of 7 cores had a value of $\xi_{max}$
significantly greater than the critical value of 6.5.
The effective temperature required by the Bonnor-Ebert fitting
also exceeded (by factors of $\sim$2-8) the measured temperature
of the core in every case.
Hence we see that the cores are not critical Bonnor-Ebert spheres.
The masses of the bright cores were seen to have a mean value of
approximately 1.5 times their virial masses, possibly also indicating
that the cores are not supported by internal thermal pressure.

Approximate estimates of core lifetimes were made,
based on statistics of detections
and relative numbers of cores with embedded young stellar objects.
The lifetime of a submillimetre detected core, or prestellar core,
was estimated at
$\sim$3 $\times$ 10$^5$ years, while that of a bright core
was estimated to be $\sim$1.5 $\times$ 10$^5$ years.
None of the current theoretical models were seen to be able to
exactly match all of the observed physical parameters.
Models that regulate collapse magnetically via the ionised component
of the gas were seen to be able to match the timescales at
lower densities, but did not reproduce the relative numbers
of bright and intermediate cores.

\section*{Acknowledgments}

The James Clerk Maxwell Telescope
is operated by the Joint Astronomy Centre, Hawaii, on
behalf of the UK PPARC, the Netherlands NWO, and the Canadian NRC. SCUBA
was built at the Royal Observatory, Edinburgh.
JMK acknowledges PPARC studentship support whilst carrying out
this work.


\begin{thebibliography}{}

\bibitem[\protect\citeauthoryear{{Alves}, {Lada}, \& {Lada}}{{Alves}
   et~al.}{2001}]{all01}
{Alves} J., {Lada} C.~J.,  {Lada} E.~A., 2001, Nature, 409, 159

\bibitem[\protect\citeauthoryear{{Andr{\' e}}}{{Andr{\' e}}}{1994}]{a94}
{Andr{\' e}} P., 1994, in: Montmerle T., Lada C. J., Mirabel F., van Thanh
J. T., eds., `The Cold Universe', Editions Frontieres, Gif-sur-Yvette,
France, p.~179

\bibitem[]{}
Andr\'e P., Belloche A., Hennebelle P., Ward-Thompson D., 2004,
in: Kun M., Eisloeffel J., eds., `Early Stages of Star Formation',
Baltic Astronomy, vol. 13, no. 3, p.~392

\bibitem[]{}
Andr\'e P., Bouwman J., Belloche A., Hennebelle P., 2003,
in: Curry C. L., Fich M., eds.,
`Chemistry as a Diagnostic of Star Formation',
NRC Research Press, p.~127

\bibitem[\protect\citeauthoryear{{Andr{\' e}}, {Ward-Thompson}, \&
   {Barsony}}{{Andr{\' e}} et~al.}{1993}]{awb93}
{Andr{\' e}} P., {Ward-Thompson} D.,  {Barsony} M., 1993, ApJ, 406, 122

\bibitem[\protect\citeauthoryear{{Andr{\' e}}, {Ward-Thompson}, \&
   {Barsony}}{{Andr{\' e}} et~al.}{2000}]{awb00}
{Andr{\' e}} P., {Ward-Thompson} D.,  {Barsony} M., 2000, in: {Mannings}
V., {Boss}
A.~P.,  {Russell} S.~S., eds., {`Protostars and Planets IV'},
\newblock {University of Arizona Press}, p.~59

\bibitem[\protect\citeauthoryear{{Andr{\' e}}, {Ward-Thompson},
\& {Motte}}{{Andr{\' e}}
   et~al.}{1996}]{awm96}
{Andr{\' e}} P., {Ward-Thompson} D.,  {Motte} F., 1996, A\&A, 314, 625, --
Paper~II

\bibitem[\protect\citeauthoryear{{Bacmann} et~al.}{{Bacmann}
   et~al.}{2000}]{bacmann00}
{Bacmann} A., {Andr{\' e}} P., {Puget} J.-L., {Abergel} A., {Bontemps} S.,
   {Ward-Thompson} D., 2000, A\&A, 361, 555

\bibitem[]{}
Ballesteros-Paredes J., Klessen R. S., Vázquez-Semadeni E.,
2003, ApJ, 592, 188

\bibitem[\protect\citeauthoryear{{Beichman} et~al.}{{Beichman}
   et~al.}{1986}]{beich}
{Beichman} C.~A., {Myers} P.~C., {Emerson} J.~P., {Harris} S., {Mathieu} R.,
   {Benson} P.~J.,  {Jennings} R.~E., 1986, ApJ, 307, 337

\bibitem[\protect\citeauthoryear{{Benson} \& {Myers}}{{Benson} \&
   {Myers}}{1989}]{bm89}
{Benson} P.~J.,  {Myers} P.~C., 1989, ApJS, 71, 89

\bibitem[\protect\citeauthoryear{{Bertout}, {Robichon}, \& {Arenou}}{{Bertout}
   et~al.}{1999}]{bra99}
{Bertout} C., {Robichon} N.,  {Arenou} F., 1999, A\&A, 352, 574

\bibitem[\protect\citename{Binney \& Tremaine }1987]{bt}
    Binney, J. and Tremaine S., 1987, Galactic Dynamics, Princeton
    University Press

\bibitem[\protect\citeauthoryear{{Bohlin}, {Savage}, \& {Drake}}{{Bohlin}
   et~al.}{1978}]{bsd78}
{Bohlin} R.~C., {Savage} B.~D.,  {Drake} J.~F., 1978, ApJ, 224, 132

\bibitem[\protect\citeauthoryear{{Bok} \& {McCarthy}}{{Bok} \&
   {McCarthy}}{1974}]{bokm74}
{Bok} B.~J.,  {McCarthy} C.~C., 1974, AJ, 79, 42

\bibitem[\protect\citeauthoryear{{Bok} \& {Reilly}}{{Bok} \&
   {Reilly}}{1947}]{br47}
{Bok} B.~J.,  {Reilly} E.~F., 1947, ApJ, 105, 255

\bibitem[\protect\citename{Bonnor }1956]{bonnor}
    Bonnor W. B., 1956, MNRAS, 116, 351

\bibitem[]{}
Bourke T. L., Hyland A. R., Robinson G., 1995, MNRAS, 276, 1052

\bibitem[]{}
Bourke T. L., Hyland A. R., Robinson G., James S. D., Wright C. M.,
1995, MNRAS, 276, 1067

\bibitem[\protect\citeauthoryear{{Chandrasekhar}}{{Chandrasekhar}}{1939}]
{chandra}
{Chandrasekhar} S., 1939, `An Introduction of the Study of Stellar
Structure', Chicago University Press

\bibitem[\protect\citeauthoryear{{Ciolek} \& {Basu}}{{Ciolek} \&
   {Basu}}{2000}]{cb00}
{Ciolek} G.~E.,  {Basu} S., 2000, ApJ, 529, 925 -- CB00

\bibitem[\protect\citeauthoryear{{Ciolek} \& {Basu}}{{Ciolek} \&
   {Basu}}{2001}]{cb01}
Ciolek G. E., Basu S., 2001, ApJ, 547, 272

\bibitem[\protect\citename{Ciolek }1994]{cm94}
Ciolek G. E., Mouschovias T. C., 1994, ApJ, 425, 142 -- CM94

\bibitem[\protect\citeauthoryear{{Clemens} \& {Barvainis}}{{Clemens} \&
   {Barvainis}}{1988}]{cb88}
{Clemens} D.~P.,  {Barvainis} R., 1988, ApJS, 68, 257

\bibitem[\protect\citeauthoryear{{Codella} et~al.}{{Codella}
   et~al.}{1997}]{cwhb97}
{Codella} C., {Welser} R., {Henkel} C., {Benson} P.~J.,  {Myers} P.~C., 1997,
   A\&A, 324, 203

\bibitem[\protect\citeauthoryear{{Cohen}}{{Cohen}}{1980}]{cohen80}
{Cohen} M., 1980, AJ, 85, 29

\bibitem[\protect\citeauthoryear{{Crutcher}}{{Crutcher}}{1999}]{crutcher99}
{Crutcher} R.~M., 1999, ApJ, 520, 706

\bibitem[]{}
Crutcher R. M., Nutter D. J., Ward-Thompson D., Kirk J. M., 2004, ApJ,
600, 279

\bibitem[\protect\citeauthoryear{{Dame}, {Hartmann}, \& {Thaddeus}}{{Dame}
   et~al.}{2001}]{dht01}
{Dame} T.~M., {Hartmann} D.,  {Thaddeus} P., 2001, ApJ, 547, 792

\bibitem[\protect\citeauthoryear{{de Geus}, {Bronfman}, \& {Thaddeus}}{{de
   Geus} et~al.}{1990}]{gbt90}
{de Geus} E.~J., {Bronfman} L.,  {Thaddeus} P., 1990, A\&A, 231, 137

\bibitem[\protect\citeauthoryear{{Ebert}}{{Ebert}}{1955}]{ebert}
{Ebert} R., 1955, Zeitschrift Astrophysics, 37, 217

\bibitem[\protect\citeauthoryear{{Elias}}{{Elias}}{1978}]{elias78b}
{Elias} J.~H., 1978, ApJ, 224, 857

\bibitem[\protect\citeauthoryear{{Evans} et~al.}{{Evans} et~al.}
{2001}]{ersm01}
{Evans} N.~J., {Rawlings} J.~M.~C., {Shirley} Y.~L.,  {Mundy} L.~G., 2001,
   ApJ, 557, 193

\bibitem[\protect\citeauthoryear{{Foster} \& {Chevalier}}{{Foster} \&
   {Chevalier}}{1993}]{fs93}
{Foster} P.~N.,  {Chevalier} R.~A., 1993, ApJ, 416, 303

\bibitem[\protect\citeauthoryear{{Franco}}{{Franco}}{1989}]{franco89}
{Franco} G.~A.~P., 1989, A\&A, 223, 313

\bibitem[]{}
Frerking M. A., Langer W. D., Wilson R. W., 1982, ApJ, 262, 590

\bibitem[\protect\citeauthoryear{{Goodwin}, {Ward-Thompson}, \&
   {Whitworth}}{{Goodwin} et~al.}{2002}]{gww02}
{Goodwin} S.~P., {Ward-Thompson} D.,  {Whitworth} A.~P., 2002, MNRAS, 330, 769

\bibitem[]{}
Greene T. P., Wilking B. A., Andr\'e, P., Young E. T., Lada C. J.,
1994, ApJ, 434, 614

\bibitem[]{}
Harvey D. W. A., Wilner D. J., Lada C. J., Myers P. C., Alves J. F.,
2003a, ApJ, 598, 1112

\bibitem[]{}
Harvey D. W. A., Wilner D. J., Myers P. C., Tafalla M.,
2003b, ApJ, 597, 424

\bibitem[]{}
Henriksen R., Andr\'e P., Bontemps S., 1997, A\&A, 323, 549

\bibitem[\protect\citename{Holland et al. }1999]{holland}
  Holland W. S., Robson E. I., Gear W. K., Cunningham C. R.,
Lightfoot J. F., Jenness T., Ivison R. J., Stevens J. A., Ade
P. A. R., Griffin M. J., Duncan W. D., Murphy J. A., Naylor D. A.,
1999, MNRAS, 303, 659

\bibitem[\protect\citeauthoryear{{Hotzel} et~al.}{{Hotzel}
   et~al.}{2002}]{hhjmh02}
{Hotzel} S., {Harju} J., {Juvela} M., {Mattila} K.,  {Haikala} L., 2002, A\&A,
   391, 275

\bibitem[\protect\citeauthoryear{Jenness \& Lightfoot}{Jenness \&
   Lightfoot}{1998}]{jenness}
Jenness T.,  Lightfoot J., 1998, Surf -- `Scuba User Reduction Facility
Users Manual', Starlink User Note 216.4, Rutherford Appleton Laboratory,
Didcot, UK

\bibitem[\protect\citeauthoryear{{Jessop} \& {Ward-Thompson}}{{Jessop} \&
   {Ward-Thompson}}{2000}]{jw00}
{Jessop} N.~E.,  {Ward-Thompson} D., 2000, MNRAS, 311, 63

\bibitem[\protect\citeauthoryear{{Jessop} \& {Ward-Thompson}}{{Jessop} \&
   {Ward-Thompson}}{2001}]{jw01}
{Jessop} N.~E.,  {Ward-Thompson} D., 2001, MNRAS, 323, 1025, -- Paper IV

\bibitem[\protect\citeauthoryear{{Jijina}, {Myers}, \& {Adams}}{{Jijina}
   et~al.}{1999}]{jma99}
{Jijina} J., {Myers} P.~C.,  {Adams} F.~C., 1999, ApJS, 125, 161

\bibitem[\protect\citename{Juvela et al. }2002]{juvela}
    Juvela M., Mattila K., Lehtinen K.,
    Lemke D., Laureijs R., Prusti T.,
    2002, A\&A, 382, 583

\bibitem[\protect\citeauthoryear{{Kenyon}, {Dobrzycka},
\& {Hartmann}}{{Kenyon}
   et~al.}{1994}]{kdh94}
{Kenyon} S.~J., {Dobrzycka} D.,  {Hartmann} L., 1994, AJ, 108, 1872

\bibitem[\protect\citeauthoryear{{Kenyon}, {Dobrzycka},
\& {Hartmann}}{{Kenyon}
   et~al.}{1994}]{kh95}
{Kenyon} S.~J.,  {Hartmann} L., 1995, ApJS, 101, 117

\bibitem[\protect\citeauthoryear{{Kessler} et~al.}{{Kessler}
   et~al.}{1996}]{ksetal96}
{Kessler} M.~F., Steinz J. A., Anderegg M. E., Clavel J., Drechsel G.,
Estaria P., Faelker J., Riedinger J. R., Robson A., Taylor B. G.,
Ximenez de Ferran S., 1996, A\&A, 315, L27

\bibitem[]{}
Kirk J. M., 2003, PhD Thesis, Cardiff University

\bibitem[\protect\citeauthoryear{{Lee} \& {Myers}}{{Lee} \&
   {Myers}}{1999}]{lm99}
{Lee} C.~W.,  {Myers} P.~C., 1999, ApJS, 123, 233

\bibitem[\protect\citeauthoryear{{Lehtinen} et~al.}{{Lehtinen}
   et~al.}{2003}]{llmh03}
Lehtinen K., Mattila K., Lemke D., Juvela M., Prusti T., Laureijs R.,
2003, A\&A, 398, 571

\bibitem[\protect\citeauthoryear{{Lemke} et~al.}{{Lemke}
   et~al.}{1996}]{lketal96}
{Lemke} D. et~al., 1996, A\&A, 315, L64

\bibitem[\protect\citeauthoryear{{Mattila}}{{Mattila}}{1979}]{mattila79}
{Mattila} K., 1979, A\&A, 78, 253

\bibitem[\protect\citeauthoryear{{McKee}}{{McKee}}{1989}]{mckee89}
{McKee} C.~F., 1989, ApJ, 345, 782

\bibitem[\protect\citeauthoryear{{Motte}, {Andr{\' e}}, \& {Neri}}{{Motte}
   et~al.}{1998}]{motte98}
{Motte} F., {Andr{\' e}} P.,  {Neri} R., 1998, A\&A, 336, 150

\bibitem[\protect\citeauthoryear{{Motte} et~al.}{{Motte}
   et~al.}{2001}]{motte01}
{Motte} F., {Andr{\' e}} P., {Ward-Thompson} D.,  {Bontemps} S., 2001, A\&A,
   372, L41

\bibitem[\protect\citeauthoryear{{Mouschovias}}{{Mouschovias}}{1991}]{mous91}
{Mouschovias} T.~C., 1991, in: {Lada} C. J.,  {Kylafis} N. D., eds., NATO
ASIC Proc., 342, `The Physics of Star Formation and Early Stellar Evolution',
p.~449

\bibitem[\protect\citeauthoryear{{Murdin} \& {Penston}}{{Murdin} \&
   {Penston}}{1977}]{mp77}
{Murdin} P.,  {Penston} M.~V., 1977, MNRAS, 181, 657

\bibitem[\protect\citeauthoryear{{Myers} \& {Benson}}{{Myers} \&
   {Benson}}{1983}]{mb83}
{Myers} P.~C.,  {Benson} P.~J., 1983, ApJ, 266, 309

\bibitem[\protect\citeauthoryear{{Myers} et al.}{{Myers} et al.}{1988}]
{myers}
Myers P. C., Heyer M., Snell R. L., Goldsmith P. F., 1988, ApJ, 324, 907

\bibitem[\protect\citeauthoryear{{Myers}, {Linke}, \& {Benson}}{{Myers}
   et~al.}{1983}]{mlb83}
{Myers} P.~C., {Linke} R.~A.,  {Benson} P.~J., 1983, ApJ, 264, 517

\bibitem[]{}
Onishi T., Mizuno A., Kawamura A., Fukui Y., 1999, in: Nakamoto T., ed.
`Star Formation 1999', p.~153, Nagoya, Japan

\bibitem[]{}
Ostriker E. C., Gammie C. F., Stone J. M., 1999, ApJ, 513, 219

\bibitem[]{}
Ostriker E. C., Stone J. M., Gammie C. F., 2001, ApJ, 546, 980

\bibitem[]{}
Safier P. N., McKee C. F., Stahler S. W., 1997, ApJ, 485, 660

\bibitem[\protect\citeauthoryear{{Snell}}{{Snell}}{1981}]{snell81}
{Snell} R.~L., 1981, ApJS, 45, 121

\bibitem[\protect\citename{Stamatellos \& Whitworth }2003]{stam}
    Stamatellos D., Whitworth A. P., 2003, A\&A, 407, 941

\bibitem[\protect\citename{Stamatellos \& Whitworth }2003]{stam2}
    Stamatellos D., Whitworth A. P., Andr\'e P., Ward-Thompson D., 2004,
    A\&A, 420, 1009

\bibitem[\protect\citeauthoryear{{Straizys} et~al.}{{Straizys}
   et~al.}{1992}]{sckm92}
{Straizys} V., {Cernis} K., {Kazlauskas} A.,  {Meistas} E., 1992, Baltic
   Astronomy, 1, 149

\bibitem[\protect\citeauthoryear{{Ungerechts} \& {Thaddeus}}{{Ungerechts} \&
   {Thaddeus}}{1987}]{ut87}
{Ungerechts} H.,  {Thaddeus} P., 1987, ApJS, 63, 645

\bibitem[]{}
Ward-Thompson D., 1996, Ap\&SS, 239, 151

\bibitem[\protect\citeauthoryear{{Ward-Thompson}, {Andr{\' e}}, \&
   {Kirk}}{{Ward-Thompson} et~al.}{2002}]{wak02}
{Ward-Thompson} D., {Andr{\' e}} P.,  {Kirk} J.~M., 2002, MNRAS, 329, 257 --
  Paper V

\bibitem[\protect\citeauthoryear{{Ward-Thompson} et~al.}{{Ward-Thompson}
   et~al.}{2000}]{wkcgha00}
{Ward-Thompson} D., {Kirk} J.~M., {Crutcher} R.~M., {Greaves} J.~S., {Holland}
   W.~S.,  {Andr{\' e}} P., 2000, ApJ, 537, L135

\bibitem[\protect\citeauthoryear{{Ward-Thompson}, {Motte}, \&
   {Andr{\' e}}}{{Ward-Thompson} et~al.}{1999}]{wma99}
{Ward-Thompson} D., {Motte} F.,  {Andr{\' e}} P., 1999, MNRAS, 305, 143 --
   Paper III

\bibitem[\protect\citeauthoryear{{Ward-Thompson} et~al.}{{Ward-Thompson}
   et~al.}{1994}]{wsha94}
{Ward-Thompson} D., {Scott} P.~F., {Hills} R.~E.,  {Andr{\' e}} P.,
1994, MNRAS, 268, 276 -- Paper I

\bibitem[\protect\citeauthoryear{{Wheelock} et~al.}{{Wheelock}
   et~al.}{1994}]{issa94}
Wheelock S. L., Gautier T. N., Chillemi J., Kester D., McCallon H., Oken C.,
White J., Gregorich D., Boulanger F., Good J., 1994, `IRAS Sky Survey
Atlas Explanatory Supplement', JPL, Pasadena

\bibitem[\protect\citeauthoryear{{Whitworth} et~al.}{{Whitworth}
   et~al.}{1996}]{wbfw96}
{Whitworth} A.~P., {Bhattal} A.~S., {Francis} N.,  {Watkins} S.~J., 1996,
   MNRAS, 283, 1061

\bibitem[\protect\citeauthoryear{{Whitworth} \& {Summers}}{{Whitworth} \&
   {Summers}}{1985}]{ws85}
{Whitworth} A.,  {Summers} D., 1985, MNRAS, 214, 1

\bibitem[\protect\citeauthoryear{{Whitworth} \& {Ward-Thompson}}{{Whitworth}
 \&
   {Ward-Thompson}}{2001}]{ww01}
{Whitworth} A.~P.,  {Ward-Thompson} D., 2001, ApJ, 547, 317

\bibitem[]{}
Williams J. P., Blitz L., McKee, C. F., 2000, in: Mannings V., Boss A.~P.,
Russell S.~S., eds., `Protostars and Planets IV', University of Arizona
Press, p.~97

\bibitem[]{}
Wood D. O. S., Myers P. C., Daugherty D. A., 1994, ApJS, 95, 457

\bibitem[\protect\citeauthoryear{{Zucconi}, {Walmsley}, \& {Galli}}{{Zucconi}
   et~al.}{2001}]{zwg01}
{Zucconi} A., {Walmsley} C.,  {Galli} D., 2001, A\&A, 376, 650

\bibitem[]{}
Zylka R., Mezger P. G., Ward-Thompson D., Duschl W. J., Lesch, H.,
1995, A\&A, 297, 83

\end{thebibliography}
\end{document}